\documentclass[12pt]{article}

\usepackage[english]{babel}
\usepackage{algpseudocode}

\usepackage{dsfont}
\usepackage[letterpaper,top=2cm,bottom=2cm,left=2.5cm,right=2.5cm,marginparwidth=1.75cm]{geometry}

\usepackage{algorithm}
\usepackage{algpseudocode}

\usepackage{amsmath,amssymb,bm,amsthm}
\usepackage{mathtools}
\usepackage{dsfont}

\mathtoolsset{showonlyrefs=true}
\newtheorem{proposition}{Proposition}
\newtheorem{definition}{Definition}

\newtheorem{theorem}{Theorem}
\newtheorem{corollary}{Corollary}
\newtheorem{lemma}{Lemma}
\usepackage{amsmath}           
\usepackage{mathrsfs}

\DeclareMathOperator\supp{supp}

\usepackage{multicol}
\usepackage{graphicx}
\usepackage{tikz}
\usepackage{pgfplots}
\usetikzlibrary{decorations.pathreplacing,calligraphy}
\usepgfplotslibrary{fillbetween}

\pgfplotsset{compat=newest}
\pgfplotsset{plot coordinates/math parser=false}

\usepackage{amsmath}

\allowdisplaybreaks

\def\bs{\ensuremath\boldsymbol}
\usepackage{dsfont}
\usepackage{graphicx}
\usepackage{enumerate}
\usepackage[colorlinks=true, allcolors=blue]{hyperref}
\definecolor{green}{rgb}{0.0, 0.5, 0.0}

\usepackage{authblk}

\title{Sparse recovery from quadratic equations, part II:\\
hardness and incoherence}
\author{Augustin Cosse\\
\textcolor{blue}{augustin.cosse@univ-littoral.fr}}
\affil{Universit\'e du Littoral C\^ote d'Opale\\
Laboratoire de Math\'ematiques Pures et Appliqu\'ees Joseph Liouville}

\begin{document}
\maketitle

{

\begin{abstract}
We study the square root bottleneck in the recovery of sparse vectors from quadratic equations.  It is acknowledged that a sparse vector $\bs x_0\in \mathbb{R}^n$, $\|\bs x_0\|_0 = k$ can in theory be recovered from as few as $O(k)$ generic quadratic equations but no polynomial time algorithm is known for this task unless $m = \Omega(k^2)$.  This bottleneck was in fact shown in~\cite{soltanolkotabi2019structured} to be essentially related to the initialization of descent algorithms.  Starting such algorithms sufficiently close to the planted signal is known to imply convergence to this signal. In this paper,  we show that as soon as $m\gtrsim \mu_0^{-2}k \vee \mu_0^{-4}$ (up to log factors) where $\mu_0 = \|\bs x_0\|_\infty/\|\bs x_0\|_2$, it is possible to recover a $k$-sparse vector $\bs x_0\in \mathbb{R}^n$ from $m$ quadratic equations of the form $\langle \bs A_i, \bs x\bs x^\intercal\rangle = \langle \bs A_i, \bs x_0\bs x_0^\intercal\rangle + \varepsilon_i $ by minimizing the classical empirical loss.  The proof idea carries over to the phase retrieval setting for which it provides an original initialization that matches the current optimal sample complexity (see e.g.~\cite{cai2023provable}).  In the maximally incoherent regime $\mu_0^{-2}=k$,  and for $m=o(k^2)$ we provide evidence for topological hardness by showing that a property known as the Overlap Gap Property (OGP), which originated in spin glass theory and is conjectured to be indicative of algorithmic intractability when optimizing over random structures, holds for a particular level of overparametrization.  The key ingredient of the proof is a lower bound on the tail of chi-squared random variables which follows from the theory of moderate deviations.
\end{abstract}

}

\section{Introduction}

{We consider the general problem

\begin{align}
\begin{split}
\min_{\bs x} \quad \hat{R}_m(\bs x) = &\frac{1}{m}\sum_{i=1}^m \left(\langle \bs A_i, \bs x\bs x^\intercal\rangle  - \langle \bs A_i, \bs x_0 \bs x_0^\intercal \rangle+\varepsilon_i \right)^2,\\
\text{s.t.} \quad & \|\bs x\|_0 = k.
\end{split}\label{generalProblem00}
\end{align}
We will also consider the following variant of~\eqref{generalProblem00}, known as the phase retrieval problem, in which the sensing operators are subexponential instead of subgaussian 
 \begin{align}
\begin{split}
\min_{\bs x} \quad \hat{R}_m(\bs x) = &\frac{1}{m}\sum_{i=1}^m \left(\langle \bs a_i\bs a_i^\intercal , \bs x\bs x^\intercal\rangle  - \langle \bs a_i\bs a_i^\intercal,  \bs x_0 \bs x_0^\intercal \rangle + \varepsilon_i\right)^2,\\
\text{s.t.} \quad & \|\bs x\|_0 = k.
\end{split}\label{generalProblemPR00}
\end{align}
Problems~\eqref{generalProblem00} and~\eqref{generalProblemPR00} are connected to a number of applications in science, engineering and finance including portfolio optimization~\cite{brodie2009sparse}, subwavelength imaging~\cite{shechtman2011sparsity}, quantum tomography~\cite{shechtman2011sparsity}.  As indicated in~\cite{soltanolkotabi2019structured}, phaseless imaging techniques also play a crucial role in emerging national security applications focusing on the monitoring of electronic devices.  

It is now established that both problems can be solved efficiently either by  semidefinite programming~\cite{li2013sparse}, local iterative algorithms~\cite{cai2016optimal} or even spectral methods~\cite{ma2021spectral, cai2022solving,jaganathan2016phase} when $m = \Omega(k^2)$ The main question is why all of these algorithms seem to fail in the $C_1 k<m<C_2 k^2$ regime while it is known (see for example~\cite{li2013sparse}) that in theory any sparse vector $\bs x_0$ can be recovered from only $\Omega(k)$ generic measurements (corresponding to the degrees of freedom of the unknown). This so-called square root bottleneck has been the focus of a number of research papers over the last few years, the closest to ours being~\cite{lee2017near} where the authors show convergence of a projected power method for~\eqref{generalProblem00} when $\max_i x_0[i]>c$ for some absolute constant $c$ and~\cite{cai2023provable} where the authors show that in the case of problem~\eqref{generalProblemPR00}, recovery of $\bs x_0$ can be ensured for as little as $\mu_0^{-2}k\vee \mu_0^{-4}$ measurements when minimizing an intensity based formulation.  The approach also work for~\eqref{generalProblemPR00} but for convex regularizers only, thus requiring an a priori estimate of the $\ell_1$ norm of $\bs x_0$.

This curiosity has also led to some speculation regarding the origin of this computational hardness.  One plausible explanation being the dual (sparse and rank one) nature of the unknown~\cite{oymak2015simultaneously}, another one being the impossibility to design a sufficiently good initialization~\cite{soltanolkotabi2019structured}.  In a previous line of work (see~\cite{cosse2024sparse}), we suggested a relaxed ``linear+quadratic" formulation   
\begin{align}
   \widehat{R}^\lambda _m(\bs x) = \frac{1}{m}\sum_{i=1}^m \left(\lambda \langle \bs c_i, \bs x\rangle + (1-\lambda)\langle \bs A_i, \bs x\bs x^\intercal \rangle  - b_i^{\varepsilon}\right)^2.\label{problemExternalField01}
\end{align}
where $b_i$ is used to encode the measurements $b_i =\lambda \langle \bs c_i, \bs x_0 \rangle + (1-\lambda) \langle \bs A_i, \bs x_0\bs x_0^\intercal\rangle $.  As tantalizing, as it may appear it is not really easy to find a direct connection between~\eqref{problemExternalField01} and~\eqref{generalProblem00} as such a connection would require a priori knowledge on the support. More specifically, writing~\eqref{generalProblem00} as~\eqref{problemExternalField01} would require to know at least one of the elements from the support of $\bs x_0$ and then normalize by the square of this element.  Such a connection seems to suggest that the knowledge of at least one entry will help improve the initialization of local algorithms when possible.  This is also what seems to be suggested by~\cite{cai2023provable}. 

In what follows, we start by recalling the current benchmarks regarding the minimal number of measurements needed for both problem~\eqref{generalProblem00} and problem~\eqref{generalProblemPR00} and we fill the gaps by providing the missing results when needed.  We then provide geometric/topological evidence for algorithmic hardness by showing that a property based on the disconnectivity of the overlaps of near optimal solutions dubbed \textit{Overlap Gap Property}, which is known to be associated to computational intractability, holds in the regime $m= o(k^2)$. 

We use $\|\bs x\|_p$ to denote the usual $\ell_p$ norm of a vector. $\|\bs X\|$ is used to denote the operator norm of the matrix $\bs X$.  For any random variable $X$, we use $\|X\|_{\psi_p}$ to denote the Orlicz $p$-norm of $X$ (see for exampe~\cite{wellner2013weak} for details).  Given a matrix $\bs A$, we will also use $\bs A_{i*}$ (resp $\bs A_{*j}$) to denote the $i^{\text{th}}$ row (resp.  $j^{\text{th}}$ column of $\bs A$).

\subsection{\label{hardnessIncoherenceSection}Hardness and incoherence}

A number of algorithms have been proposed to solve problems~\eqref{generalProblem00} and~\eqref{generalProblemPR00}, most of them unable to break through the $\Omega(k^2)$ barrier because of (1) a focus on the general setting or (2) a minimization based on the empirical loss defined from intensity as opposed to amplitude measurements~\cite{soltanolkotabi2019structured}.  For the phase retrieval problem~\eqref{generalProblemPR00}, it was in fact shown in~\cite{cai2023provable} that a combination of an amplitude based loss (relying on the absolute value of the measurements) and a focus on sufficiently coherent vectors could lead to the efficient recovery of $\bs x_0$ for as little as $m\gtrsim \mu_0^{-2}k \vee \mu_0^{-4}$ measurements.  Because of the subexponential measurements, as explained in~\cite{soltanolkotabi2019structured}, the success of algorithms based on non-convex regularizers is contingent upon the use of the amplitude based loss.  For problem~\eqref{generalProblem00} on the other hand,  for which it is not clear how to define such a loss, it was shown in~\cite{lee2017near} that the combination of the hard thresholding pursuit algorithm from~\cite{foucart2011hard} followed by a renormalization step could lead to the exact recovery of $\bs x_0$ from $O(k)$ samples when $\|\bs x_0\|_\infty>c \|\bs x_0\|_2$ for some absolute constant $c$ and $O(k^2)$ measurements otherwise.  The result is asymptotic without a clear relation between the accuracy and the number of iterations. 

The first part of this paper bridges the gap between~\cite{lee2017near} and~\cite{cai2023provable}.  We extend the conclusion of~\cite{lee2017near} by providing a more general initialization in the spirit of what is done in~\cite{cai2023provable}.  The initialization that we introduce in this paper however relies on the whole set of measurements, as opposed to~\cite{cai2023provable} where a thresholding is applied.  Because of this difference, the extension to problem~\eqref{generalProblemPR00} requires more advanced (i.e. Talagrand type) inequalities as will be shown in section~\ref{sectionInitializationPR}.  For completeness we then rely on the truncated gradient descent iterations of~\cite{cai2016optimal} to derive an explicit rate of convergence in the regime $m\gtrsim \mu_0^{-2}k \vee \mu_0^{-4}$.  In both the approach of~\cite{lee2017near}  as well as for the truncated gradient descent approach of~\cite{cai2016optimal}, the initialization is key to achieve the optimal sample complexity and can be summarized as follows. We first assemble the estimator $\hat{\bs x}$ defined as
\begin{align}
\hat{\bs x}[\ell] = \frac{1}{m}\sum_{i=1}^m \bs A_i[\ell, \ell] b_i.\label{estimator01}
\end{align}
For such an estimator, we have 
\begin{align}
\mathbb{E}\left\{\hat{\bs x}[\ell]\right\} &= \mathbb{E}\left\{\frac{1}{m}\sum_{i=1}^m A_i[\ell, \ell] \langle \bs A_i, \bs x_0\bs x_0^\intercal\rangle \right\} = x_0^2[\ell].
\end{align}
The sum~\eqref{estimator01} is a sum of chi-squared random variables for which we have
\begin{align}
\left\|\bs A_i[\ell, \ell] \langle \bs A_i, \bs x_0\bs x_0^\intercal \rangle \right\|_{\psi_1}\lesssim \|\bs x_0\|^2.
\end{align}
From this, using an application of Bernstein's inequality (see Proposition~\ref{scalarBernstein}), we get 
\begin{align}
x_0^2[\ell] - \|\bs x_0\|^2\sqrt{\frac{\log(m)}{m}}\leq \hat{\bs x}[\ell] \leq x_0^2[\ell] + \|\bs x_0\|^2\sqrt{\frac{\log(m)}{m}}.
\end{align}
In particular, this shows that the identification of at least one element from the support would require
\begin{align}
\|\bs x_0\|_\infty^2 \geq C \|\bs x_0\|^2\sqrt{\frac{\log(m)}{m}},
\end{align}
which, if we introduce the incoherence $\mu_0$ defined as the ratio between the infinity and the $\ell_2$ norms, $\mu_0 =\|\bs x_0\|_\infty/\|\bs x_0\|_2$ where $1/\sqrt{k}\leq \mu_0\leq 1$, can now read as 
\begin{align}
C\mu_0^{-2}\sqrt{\frac{\log(m)}{m}}\leq 1,
\end{align}
or $m$ sufficiently larger than $\mu_0^{-4}\log(m)$. Note that this does not yet guarantee the recovery of $\bs x_0$. Once we have one element from the support though, in the spirit of~\cite{cosse2024sparse},  we can focus on the corresponding column, from which it becomes easier to get a sufficiently good estimate for $\bs x_0$.  Indeed, once we have a first element, (let us say of index $k^*$), we can define the estimator
\begin{align}
 \hat{S} = \left\{\ell \; s.t \; \left|\frac{1}{m}\sum_{i=1}^m  b_i \bs A_i[\ell,k^*] \right|> \log(m)/m\right\}.
\end{align}
and construct an estimate for $\bs x_0$ based on the leading eigenvector of $\frac{1}{m}\sum_{i=1}^m \mathcal{P}_{\hat{S}} \left(\bs A_i b_i\right)\mathcal{P}_{\hat{S}}$ where $\mathcal{P}_{\hat{S}}$ represents the projector on $\hat{S}$.  The result following from the combination of such an initialization with the approaches of~\cite{cai2016optimal} and~\cite{lee2017near} is summarized by Propositions~\ref{propositionIncoherenceInitializationAndRecovery} and~\ref{propositionTGD} below. 

In tribute to~\cite{lee2017near} we start by considering the following iteration which we label Sparse Power Factorization ({\sf SPF}) for which we prove asymptotic convergence. Assume $\|\bs x_0\|=1$. Then repeat the following steps 
\begin{align}
\begin{split}
\bs x^{(T+1)} &= \text{\sf IHT}(\bs x^{(T)}), \\
\bs x^{(T+1)} &= \frac{\bs x^{(T+1)}}{\|\bs x^{(T+1)}\|}.
\end{split}\tag{\sf{SPF}}\label{SPF}
\end{align}
where {\sf IHT} denotes the iterative Hard Thresholding Algrithm of~\cite{blumensath2009iterative} applied to the linearization of~\eqref{generalProblem00}. Combining~\eqref{SPF} with the initialization discussed above gives our first result.

\begin{proposition}\label{propositionIncoherenceInitializationAndRecovery}
Let $\bs x_0$ denote any $k$-sparse vector from $\mathbb{S}^n$ with incoherence defined as $\mu_0 = \|\bs x_0\|_{\infty}/\|\bs x_0\|_2$ and $b_i = \langle \bs A_i, \bs x_0\bs x_0^\intercal\rangle +\varepsilon_i $ be $m$ quadratic samples in which the sensing matrices $\bs A_i$ are random i.i.d. Gaussian with $\bs A_i[m,n]\sim \mathcal{N}(0,1)$ and the noises $\varepsilon_i$ on the measurements are independent subexponential with $\|\varepsilon_i\|_{\psi_1}\lesssim  \sigma$.  Then as soon as $m\gtrsim k\mu_0^{-2}\vee \mu_0^{-4}$ (up to log factors), the iteration~\eqref{SPF} will be able to recover $\bs x_0$ uniquely and stably.
\end{proposition}

The proof technique used to guarantee the convergence of the iterates in Proposition~\ref{propositionIncoherenceInitializationAndRecovery} relies on a control over the angle between $\bs x^{(T)}$ and $\bs x_0$ and does not provide any explicit relation between the number of iterations and the accuracy of the iterates. Our second result derives such an explicit rate by building on the truncated gradient descent approach of~\cite{cai2016optimal}.  The fact that in the case of problem~\eqref{generalProblem00}, the measurement operator can be shown to satisfy a RIP condition (which is not true anymore in the setting of problem~\eqref{generalProblemPR00}) makes it possible to combine the initialization discussed above with any classical minimization of the conventional $\ell_2$ empirical loss.  Although the result below relies on a simple truncated gradient descent step, a similar extension can be obtained for the iterative hard thresholding approach of~\cite{cai2022sparse} as well as for the truncated amplitude flow of~\cite{wang2017sparse}.  We focus on~\cite{cai2016optimal} as the most interesting approach since it does not require any a priori knowledge on the support of $\bs x_0$. Unlike for the phase retrieval problem, an intensity based formulation is not required. 

We let $\mathcal{T}_{\tau}(\bs x)$ to denote any thresholding operator of the form
\begin{align}
\mathcal{T}_{\tau}(x) =  0\; \text{for $x\in [-\tau, \tau]$ and $|\mathcal{T}_{\tau}(x) - x|\leq \tau$ otherwise}.
\end{align}
From this we introduce the iteration
\begin{align}
\bs x^+ = \mathcal{T}_{\tau}\left(\bs x - \eta \nabla \hat{R}_m(\bs x)\right).\tag{\sf{TGD}}\label{TGD}
\end{align}
When applied to vectors, i.e. $\mathcal{T}_{\tau}(\bs x)$, the action of $\mathcal{T}_{\tau}$ is defined component-wise.  We can now introduce our second result. 

\begin{proposition}\label{propositionTGD}
Let $\bs x_0$ denote any $k$-sparse vector from $\mathbb{S}^n$ with incoherence defined as $\mu_0 = \|\bs x_0\|_{\infty}/\|\bs x_0\|_2$ and $b_i = \langle \bs A_i, \bs x_0\bs x_0^\intercal\rangle +\varepsilon_i $ be $m$ quadratic samples in which the sensing matrices $\bs A_i$ are random i.i.d. Gaussian with $\bs A_i[m,n]\sim \mathcal{N}(0,1)$ and the noises $\varepsilon_i$ on the measurements are independent subexponential with $\|\varepsilon_i\|_{\psi_1}\lesssim \sigma$.  Then as soon as $m\gtrsim (k\mu_0^{-2}\vee \mu_0^{-4})\delta^{-1}$, the truncated gradient descent iterations~\eqref{TGD} will be able to recover $\bs x_0$ uniquely and stably. Moreover,  under appropriate conditions on $\eta$ (the learning rate) and $\tau$, the $T^{th}$ iterate of~\eqref{TGD} obeys
\begin{align}
\left\|\bs x^{(T)} - \bs x_0\right\| &\leq \left(1- \eta\right)^T \delta  + C \sigma \sqrt{\frac{k\log(m)}{m}}.
\end{align}
with probability $1-T/mn^2 - o_m(1)$
\end{proposition}

As indicated above, the initialization used in the proofs of Propositions~\ref{propositionIncoherenceInitializationAndRecovery} and~\ref{propositionTGD} can be transposed to the phase retrieval problem~\eqref{generalProblemPR00}. Although it does not provide any improvement with respect to the current minimal sample complexity derived in~\cite{cai2023provable}, it still exhibits a few differences compared to the initialization discussed in this paper.  We thus view it as being of independent interest and mention it as a side result.  In this second setting, we proceed as follows. We first find the largest entry $k^*$ in the vector $\hat{v}[\ell] = m^{-1}\sum_{i=1}^m b_i \bs a_i^2[\ell]$
\begin{align}
k^* = \underset{k}{\operatorname{argmax}}  \frac{1}{m}\sum_{i=1}^m a^2_i[k] b_i.
\end{align}
Given this entry,  we then consider the estimator
\begin{align}
\hat{S} = \left\{\ell\neq k^* \; s.t \; \left|\frac{1}{m}\sum_{i=1}^m  b_i a_i[k^*]a_i[\ell] \right|> \log(m)/m\right\}.
\end{align}
Finally, the value of $\bs x_0$ on $\hat{S}$ is estimated from the leading eigenvector of the matrix $m^{-1}\sum_{i=1}^m \bs a_i\bs a_i^\intercal$, 
\begin{align}
\left(\bs x^{(0)}\right)_{\hat{S}} \leftarrow \bs v_{\max }\left(\frac{1}{m}\sum_{i=1}^m (\bs a_i)_{\hat{S}}(\bs a_i^\intercal)_{\hat{S}}\right).\label{initializationPR}
\end{align}

\begin{proposition}\label{propositionIncoherenceInitializationAndPhaseRetrieval}
Let $\bs x_0$ denote any $k$-sparse vector from $\mathbb{S}^n$ with incoherence defined as $\mu_0 = \|\bs x_0\|_{\infty}/\|\bs x_0\|_2$ and $b_i = \langle \bs a_i\bs a_i^\intercal , \bs x_0\bs x_0^\intercal\rangle +\varepsilon_i $ be $m$ quadratic samples in which the sensing vectors $\bs a_i$ are random i.i.d. Gaussian with $\bs a_i[r]\sim \mathcal{N}(0,1)$ and the noises $\varepsilon_i$ on the measurements are independent subexponential with $\|\varepsilon_i\|_{\psi_1}\lesssim \sigma$.  Then as soon as $m\gtrsim (k\mu_0^{-2}\vee \mu_0^{-4})\delta^{-1}$,  the estimator $\bs x^{(0)}$ defined from~\eqref{initializationPR} will satisfy $\|\bs x^{(0)} - \bs x_0\|_2\leq \delta$
\end{proposition}

The proofs of Propositions~\ref{propositionIncoherenceInitializationAndRecovery}, \ref{propositionTGD} and~\ref{propositionIncoherenceInitializationAndPhaseRetrieval} are respectively detailed in sections~\ref{sectionProofFirstPropositionCoherence} and~\ref{sectionProofPropositionPR}.  Unlike~\cite{cai2023provable}, the initialization step used in Proposition~\ref{propositionIncoherenceInitializationAndPhaseRetrieval} does not require the estimated support to be of size $k$ and constructs the estimator from the complete set of measurements (without the need to get rid of the measurements that are smaller or larger than $c\|\bs x_0\|$).  Moreover,  unlike~\cite{soltanolkotabi2019structured, cai2023provable}, the iterates~\eqref{TGD} do not require to compute any projection onto a descent cone.

\begin{figure}
\centering
\begin{tikzpicture}
\begin{axis}[width=8cm,
    height=7cm,xmin=0, xmax=10,
    ymin=0, ymax=100, domain=0:10, xlabel={$k$},
    ylabel={$m$}, ylabel style={rotate=-90}, axis line style={line width=2pt}]
\addplot[name path=f, color=black, dashed, line width=1pt]{x^2};
\addplot[name path=g, color=black, line width=1pt]{2*x};
\addplot[name path=h, color=black, line width=1pt]{x^(7/4)};
\node[] at (8,6) {Impossible};
\node[] at (2,70) {Easy};
\node[] at (8.2,80) {$k^2$};
\node[] at (9,25) {$4k$};
\node[] at (8.7,55) {$\mu_0^{-2}k$};
\path[name path=axisv] (axis cs:0,0) -- (axis cs:0,100);
\path[name path=axish] (axis cs:0,0) -- (axis cs:10,0);
\addplot[blue!10] fill between[of=f and h];
\addplot[blue!20] fill between[of=axisv and f];
\addplot[red!20] fill between[of=axish and g];
\end{axis}

\begin{scope}[decoration={calligraphic brace, amplitude=6pt}]
    \draw[thick,decorate] (7,3) -- (7,1.2)
     node[midway,right=1ex, text width=1cm]{\small Conjectured \\ Hard Regime};
 \end{scope}
\end{tikzpicture}
\caption{Representation of the various (conjectured) regimes associated to Problems~\eqref{generalProblem00} and~\eqref{generalProblemPR00}.  The improvement of the paper for problem~\ref{generalProblem00} is shown in light blue.}
\end{figure}
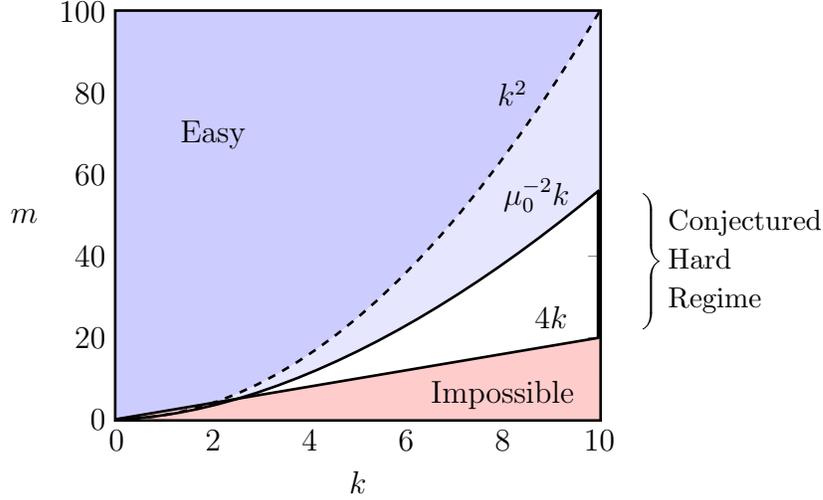


\subsection{\label{topologicalComplexitysection}Topological Complexity Barriers}

In section~\ref{hardnessIncoherenceSection} we tried to characterize as accurately as possible the hard vs easy regimes of problem~\eqref{generalProblem00}.  In this section we focus on the hardest (i.e. fully incoherent) regime for which we provide evidence for algorithmic hardness.  When restricting to fully incoherent vectors, problem~\eqref{generalProblem00} can be equivalently written as the recovery of an unknown $k$-sparse binary vector $\bs x_0\in\mathbb{R}^n$. I.e.
\begin{align}
\begin{split}
\min_{\bs x} \quad  \hat{R}_m(\bs x)& = \frac{1}{m}\sum_{i=1}^m \left(\langle \bs A_i, \bs x\bs x^\intercal \rangle  - \langle \bs A_i, \bs x_0\bs x_0^\intercal \rangle + \varepsilon_i \right)^2,\\
s.t. \quad & \bs x \in \left\{0,1\right\}^n, \quad \|\bs x\|_0 =k.
\end{split}\label{binaryFormulationEmpiricalRisk01}
\end{align} 

A recent line of work from statistical physics focuses on providing formal intuition for the hardness of some optimization problems over random structures by showing that a topological disconnectivity property called \textit{Overlap Gap Property} (OGP) holds for the most difficult instances.  What makes the OGP particularly interesting as a measure of complexity is the fact that there is currently no model known to exhibit some form of algorithmic hardness that does not verify this property~\cite{gamarnik2021overlap}.  Moreover, the OGP is also known to imply the failure of a certain family of local MCMC based algorithms~\cite{arous2023free, gamarnik2019landscape}.  Proving the OGP is not always easy though. In fact it can sometimes be the case that despite evident hardness, because of some bad local behavior in the optimization landscape, this property can only be shown to hold in the so-caled ``impossible regime" or in a limited part of the ``possible but hard" regime.  The beauty of~\cite{gamarnik2019landscape} lies in the introduction of overparametrization as a way to ``smooth" the bad local behavior in the landscape (following a recent trend in deep learning). This in turns makes it possible to extend the range of parameters for which the OGP can be verified. 

In the second part of this paper, in order to derive a topological characterization of the optimization landscape that would be as accurate as possible, we follow~\cite{gamarnik2019landscape} and turn to the following misparametrized version of problem~\eqref{binaryFormulationEmpiricalRisk01}. 
\begin{align}
\begin{split}
\min_{\bs x} \quad  \hat{R}_m(\bs x)& = \frac{1}{m}\sum_{i=1}^m \left(\langle \bs A_i, \bs x\bs x^\intercal \rangle  - \langle \bs A_i, \bs x_0\bs x_0^\intercal \rangle \right)^2,\\
s.t. \quad & \bs x \in \left\{0,1\right\}^n, \quad \|\bs x\|_0 = k'.
\end{split}\label{overparametrizedProblem}
\end{align} 

Note that $\bs x_0$ is still $k$-sparse. Provided that $k'$ is sufficiently close to $k$, it is not difficult to see that the complexities of problems~\eqref{overparametrizedProblem} and~\eqref{binaryFormulationEmpiricalRisk01} should be similar.  In fact a simple exhaustive search would require looking at ${n\choose k}$ candidate solutions in the case of~\eqref{binaryFormulationEmpiricalRisk01} while it would require looking at ${n \choose k'}$ candidate solutions in the case of the misparametrized problem~\eqref{overparametrizedProblem}.  

As pointed out in~\cite{gamarnik2019landscape, arous2023free}, It is however worth noting that not every value of the overparametrization parameter $k'$ is equally useful.  If we let $\bs v$ to denote any $k'$-sparse solution of formulation~\eqref{overparametrizedProblem},  even if we could design a polynomial time algorithm leading to the recovery of $\bs v$, if $\bs v$ does not contain a sufficient amount of information on $\bs x_0$,  the recovery of such a vector is useless in regard to our original objective~\eqref{binaryFormulationEmpiricalRisk01}. Following~\cite{gamarnik2019landscape, arous2023free}, we use the term ``informative" to refer to those values of $k'$ for which a solution of~\eqref{overparametrizedProblem} can be ``boosted" to exact recovery.  Consider the estimate $\hat{x}_{\ell}$ defined from $\bs v$ as 
\begin{align}
\hat{ x}_\ell& = \frac{1}{m}\sum_{i=1}^m b_i \langle (\bs A_i)_{\ell *}, \bs v \rangle ,
\end{align}
where $\bs A_{\ell*}$ is used to denote the $\ell^{th}$ row of the matrix $\bs A$.  The estimate $\hat{ x}_{\ell}$ is a sum of subexponential random variables. Applying the subexponential version of Bernstein's inequality (see for example~\cite{vershynin2018high}, Theorem 2.8.1) and noting that 
\begin{align}
\|b_i \langle (\bs A_i)_{\ell*}, \bs v\rangle \|_{\psi_1} \leq \|\bs x_0\|^2 \sqrt{k'} = k\sqrt{k'}.
\end{align}
If we assume an overlap of size $k$ between the support of $\bs x_0$ and the support of $\bs v$, we also have
\begin{align}
\mathbb{E}\left\{b_i \langle (\bs A_i)_{\ell*}, \bs v\rangle \right\} = x_0[\ell]\langle \bs x_0, \bs v\rangle = kx_0[\ell].
\end{align}
Hence with probability $1-o_m(1)$,
\begin{align}
 kx_0[\ell] - k\sqrt{k'}\sqrt{\frac{\log(m)}{m}}\leq  \hat{x} \leq kx_0[\ell] - k\sqrt{k'}\sqrt{\frac{\log(m)}{m}}.
\end{align}
Since $\bs x_0\in \left\{0,1\right\}$, by thresholding at $1/2$, we can thus guarantee the recovery of the support from $\bs v$ only when $m$ is sufficiently larger than $k'\log(m)$. In the case of problem~\eqref{binaryFormulationEmpiricalRisk01}, much like for the PCA problem discussed in~\cite{arous2023free}, the informative values of the overparametrization parameter $k'$ are thus given by $k\leq k'\leq m/(C\log(m))$ for some sufficiently large absolute constant $C$.  It is now well established that the hard regime is characterized by a sample complexity $m<k^2$ and we can thus focus on the interval $k'\in [k,k^2/C\log(k)]$. Such an interval not only implies that any solution of~\eqref{overparametrizedProblem} sharing a complete overlap with $\bs x_0$ can be used to recover $\bs x_0$ but it also implies that if we can show the existence of an energy barrier for some overlap $\ell<k$, then, as a purely random guess would give an overlap $kk'/n\ll k$,  we are also guaranteed that no naive approach should be able to recover the planted signal~\cite{arous2023free}.

We recall the definition of the $k'$-Overlap Gap Property from~\cite{gamarnik2019landscape,arous2023free}

\begin{definition}[$k'$-OGP]
Problem~\eqref{binaryFormulationEmpiricalRisk01} exhibits the $k'$-Overlap Gap Property ($k'$-OGP) if there exists $\zeta_{1,n}, \zeta_{2,n}\in \left\{1, \ldots, k\right\}$ with $\zeta_{1,n}<\zeta_{2,n}$ and some $r_n\in \mathbb{R}$ such that
\begin{enumerate}[(i)]
\item There exist $\bs v, \bs w \in \left\{0,1\right\}^n$ with $\langle \bs v, \bs x_0\rangle\leq \zeta_{1,n}$ and $\langle \bs w, \bs x_0\rangle \geq \zeta_{2,n}$ as well as $\hat{R}_m(\bs v)\vee \hat{R}_m(\bs w)\leq r_n$

\item For any $\bs v\in \left\{0,1\right\}^n$ with $\hat{R}_m(\bs v)\leq r_n$, it holds that $\langle \bs v, \bs x_0\rangle \leq \zeta_{1,n}$ or $\langle \bs v, \bs x_0\rangle \geq \zeta_{2,n} $ with high probability at $k, n\rightarrow \infty$. 
\end{enumerate}
\end{definition}

As indicated in~\cite{arous2023free,gamarnik2017high,gamarnik2021overlap}, the OGP can be shown to imply the failure of certain ``local" algorithms.  If an algorithm iteratively updates the unknown vector by adding or removing entries in a way that always improves the objective,  then that algorithm will be stuck if initialized in a region where the overlap is sufficiently large because it will never be able to ``climb" the energy barrier and escape outside of the region in which it was initialized.  In fact it is known (see~\cite{arous2023free}) that if the $k'$-OGP holds with $\zeta_{1,n} - \zeta_{2,n}>\Delta$,  any local algorithm that relies on the update of $\Delta$ coordinates at each step is bound to fail.  We are now ready to state the main result of this section. 

\begin{theorem}\label{theoremMainOGP}
let $k'$ be informative, i.e. $k\leq k'\leq m/\log(m)$ also assume that $m<k^2/C$ for some sufficiently large absolute constant $C$.  Then for $k',m, n$ sufficiently large with $k,k' = o(m)$ as well as
\begin{align}
 k' \leq m^{1/3}k^{2/3}{\log^{1/3}(n)} \wedge k m^{1/4} /\log(n)
\end{align} 
the optimization problem~\eqref{binaryFormulationEmpiricalRisk01} exhibits the $k'$-Overlap Gap Property with $\zeta_1 - \zeta_2 = O((k')^{3/2}/\sqrt{m})$
\end{theorem}

The proof of Theorem~\ref{theoremMainOGP} can be found in section~\ref{overlapGapPropertyProof}.  It relies on a combination between the second moment method of~\cite{gamarnik2017high,arous2023free} (for a certain level of overparametrization),  the extension of the classical Gaussian concentration inequality for Lipschitz functions and the use of a moderate deviation principle for chi-squared random variables.


\section{\label{sectionProofFirstPropositionCoherence}Proof of Propositions~\ref{propositionIncoherenceInitializationAndRecovery} and~\ref{propositionTGD}}

\subsection{\label{initializationRankOneRecovery}Initialization}

%

In what follows, we assume that $x_0[1]$ is the largest entry in $\bs x_0$ so that $\bs x_0^2[1] = \|\bs x_0\|_{\infty}^2$. We let
\begin{align}
\bs v^{(0)}& = \frac{1}{m}\sum_{i=1}^m b_i \bs A_i[1, 1] \bs e_1.\label{sumEstimatex1}
\end{align}
As a result we have $\mathbb{E}\left\{\bs v^{(0)}\right\} = x_0^2[1]\bs e_1$. Moreover, since the sum~\eqref{sumEstimatex1} is a sum of subexponential variables, noting that 
\begin{align}
\left\|\langle \bs A_i, \bs x_0 \bs x_0^{\intercal}\rangle \bs A_i[1,1] \right\|_{\psi_1} \leq \|\bs x_0\|^2,
\end{align}
and applying Proposition~\ref{scalarBernstein} gives 
\begin{align}
\bs e_1\left(x_0^2[1]- \|\bs x_0\|^2\sqrt{\frac{\log(m)}{m}}\right) \leq \bs v^{(0)} \leq \left(x_0^2[1] + \|\bs x_0\|^2\sqrt{\frac{\log(m)}{m}}\right)\bs e_1.\label{boundOnFirstEntryInitialization}
\end{align}
Substituting $\|\bs x_0\|^2 = \mu_0^{-2}\|\bs x_0\|_\infty^2$ in~\eqref{boundOnFirstEntryInitialization} gives
\begin{align}
\bs e_1 \left(x_0^2[1] - x_0^2[1]\mu_0^{-2}\sqrt{\frac{\log(m)}{m}}\right)\leq \bs v^{(0)} \leq  \left(x_0^2[1] + x_0^2[1]\mu_0^{-2}\sqrt{\frac{\log(m)}{m}}\right)\bs e_1.\label{concentrationX01}
\end{align}
From this we have $x_0^2[1] (1-\delta) \leq \bs v^{(0)}[1] \leq x_0^2[1] (1+\delta)$ for any constant $\delta$ as soon as $m\gtrsim \mu_0^{-4}\log(m)\delta^{-1}$. We define $\bs x^{(0)}[1]$ as the square root of $\bs v^{(0)}[1]$. I.e. $\bs x^{(0)}[1] = \sqrt{\bs v^{(0)}}[1]$. 

Now consider the vector $\bs y_0 = \bs x_0[1]\bs z_0$ where $\bs z_0$ contains all the entries of $\bs x_0$ except for $\bs x_0[1]$.  That is to say $\bs z_0 = [0, \bs x_0[2], \ldots, \bs x_0[n]]$. We can guess the entries of $\bs y_0$ by looking at the estimator
\begin{align}
\hat{\bs y} = \frac{1}{m}\sum_{i=1}^m b_i (\bs A_i)_{*1}.
\end{align}
Using Proposition~\ref{scalarBernstein}, we again have 
\begin{align}
y_0[\ell] - \sqrt{\frac{\log(m)}{m}}\|\bs x_0\|^2 \leq  \hat{y}_0[\ell] \leq y_0[\ell]+ \sqrt{\frac{\log(m)}{m}}\|\bs x_0\|^2
\end{align}
We define the vector $\hat{\bs w}$ by only retaining those entries in $\hat{\bs y}_0$ that are sufficiently larger than $\sqrt{\log(m)/m}\|\bs x_0\|^2$, i.e.
\begin{align}
\hat{\bs w}[\ell] = \left\{\begin{array}{ll}
\hat{y}_0[\ell] & |\hat{y}_0[\ell]|>C\sqrt{\log(m)/m}\|\bs x_0\|^2\\
0 &\text{otherwise}.
\end{array}\right.
\end{align}
where $C$ is some absolute constant. Note that $\hat{\bs w}$ is an estimator for the vector $\bs x_0[1]\bs z_0$. In particular, if we let $\hat{S} = \supp \hat{\bs w}_0$,  we have $\hat{S}\subseteq \supp(\bs x_0)$. Moreover, the error $\hat{\bs w}- \bs x_0[1]\bs z_0$ made by the estimator $\hat{\bs w}$ outside $\hat{S}$ obeys
\begin{align}
\|\left(\hat{\bs w} - \bs x_0[1]\bs z_0\right)_{\hat{S}^c}\| = \|(\bs x_0[1]\bs z_0)_{\hat{S}^c}\|\leq \sqrt{k}\sqrt{\frac{\log(m)}{m}}\|\bs x_0\|^2
\end{align}
which gives 
\begin{align}
\|(\bs z_0)_{\hat{S}^c}\|\leq \mu_0^{-1} \|\bs x_0\|^2\sqrt{\frac{k\log(m)}{m}}.\label{estimatorZ0}
\end{align}
On the estimated support,\eqref{concentrationX01} clearly gives, as soon as $m\gtrsim\mu_0^{-4}\log(m)$, 
\begin{align}
|x_0[1] - \hat{x}_0[1]|\leq\sqrt{ |x_0^2[1] - \hat{x}^2_0[1]|}\leq \sqrt{\delta x_0^2[1]}\leq \sqrt{\delta} \|\bs x_0\|_\infty
\end{align}
for the rest of the support, $\hat{S}' = \hat{S}\setminus (1,1)$ we define $(\hat{\bs x})_{\hat{S}'}$ from the top eigenvector of 
\begin{align}
\frac{1}{m}\sum_{i=1}^m \left((\bs A_i + \bs A_i^\intercal)/2\right)_{\hat{S}\times \hat{S}} b_i = \frac{1}{m}\sum_{i=1}^m \left((\bs A_i + \bs A_i^\intercal)/2\right)_{\hat{S}\times \hat{S}} \langle \bs A_i, \bs x_0\bs x_0^\intercal \rangle.
\end{align}
Using Proposition~\ref{BernsteinProposition} together with $\hat{S}\subseteq \supp(\bs x_0)$ with $|\supp(\bs x_0)| = k$ and noting that 
\begin{align}
\sigma^2 &\lesssim \left\|\mathbb{E}\left\{\frac{1}{m}\sum_{i=1}^m (\bs A_i)_{\hat{S}\times \hat{S}}(\bs A_i)_{\hat{S}\times \hat{S}}^* |\langle (\bs A_i)_{\hat{S}\times \hat{S}}, \bs x_0\bs x_0^\intercal \rangle |^2\right\}\right\|\\
&\lesssim \left\|k\bs I \|\bs x_0\|^4 + \|\bs x_0\|^2 \bs x_0\bs x_0^\intercal \right\|\leq \|\bs x_0\|^4 + k\|\bs x_0\|^2
\end{align}
as well as 
\begin{align}
\left\|(\bs A_i)_{\hat{S}\times \hat{S}} \langle \bs A_i, \bs x_0\bs x_0^\intercal\rangle \right\|_{\psi_1}\lesssim \sqrt{k}\|\bs x_0\|^2
\end{align}
we have with probability at least $1-o_m(1)$
\begin{align}
\left\|\frac{1}{m}\sum_{i=1}^m ((\bs A_i + \bs A_i^\intercal)/2)_{\hat{S}\times \hat{S}} b_i - (\bs x_0)_{\hat{S}}(\bs x_0)_{\hat{S}}^\intercal\right\|\leq \delta \|\bs x_0\|^2\label{boundOperatorNormAiAitranspose}
\end{align}
as soon as $m\gtrsim \delta^{-1} (\|\bs x_0\|^2 + k \vee \sqrt{k})$. Using Davis-Kahan `sin $\theta$' theorem (see e.g. Corollary 3 in~\cite{yu2015useful}) and letting $\hat{\bs x}$ to denote the normalized top eigenvector of the matrix $m^{-1}\sum_{i=1}^m (\bs A_i + \bs A_i^\intercal) b_i/2$
\begin{align}
\min\left\{\left\|\frac{(\bs x_0)_{\hat{S}}}{\|(\bs x_0)_{\hat{S}}\|} - \hat{\bs x}_{\hat{S}} \right\|, \left\|\frac{(\bs x_0)_{\hat{S}}}{\|(\bs x_0)_{\hat{S}}\|} + \hat{\bs x}_{\hat{S}}\right\|\right\}\leq 2^{3/2} \delta \|\bs x_0\|/\|(\bs x_0)_{\hat{S}}\|\label{afterDavisKahan}
\end{align}
Using~\eqref{estimatorZ0}, we also have
\begin{align}
\|(\bs x_0)_{\hat{S}^c}\| = \|\bs x_0 -(\bs x_0)_{\hat{S}}\|\leq \|\bs x_0\|\sqrt{\frac{\mu_0^{-2}k\log(m)}{m}} + \|\bs x_0\|\left(\frac{\mu_0^{-4}\log(m)}{m}\right)^{1/4}\label{x0outsideSw}
\end{align}
which gives 
\begin{align}
\|(\bs x_0)\|\left(1 - \sqrt{\frac{\mu_0^{-2}k\log(m)}{m}} - \left(\frac{\mu_0^{-4}\log(m)}{m}\right)^{1/4}\right)\leq \|(\bs x_0)_{\hat{S}}\|
\end{align}
Substituting in~\eqref{afterDavisKahan}, gives 
\begin{align}
\min\left\{\left\|\frac{(\bs x_0)_{S}}{\|(\bs x_0)_{\hat{S}}\|} - \hat{\bs x}_{\hat{S}} \right\|, \left\|\frac{(\bs x_0)_{S}}{\|(\bs x_0)_{\hat{S}}\|} + \hat{\bs x}_{\hat{S}}\right\|\right\}\leq 2^{3/2} \delta (1-\delta')^{-1}\label{boundSupport01}
\end{align}
where
\begin{align}
\delta' =\sqrt{\frac{\mu_0^{-2}k\log(m)}{m}}+\left(\frac{\mu_0^{-4}\log(m)}{m}\right)^{1/4}
\end{align}
Note that~\eqref{boundOperatorNormAiAitranspose} also implies 
\begin{align}
 (1-\delta)\|(\bs x_0)_{\hat{S}}\|^2 \leq \phi^2 = \left\|\frac{1}{m}\sum_{i=1}^m \left((\bs A_i + \bs A_i^\intercal)/2\right)_{\hat{S}\times \hat{S}} b_i\right\| \leq (1+\delta)\|(\bs x_0)_{\hat{S}}\|^2\label{sandwichForphi}
\end{align}
Using~\eqref{boundSupport01} we have
\begin{align}
\min\left\{\left\|(\bs x_0)_{\hat{S}} - \hat{\bs x}_{\hat{S}} \|(\bs x_0)_{\hat{S}}\|\right\| , \left\|(\bs x_0)_{\hat{S}} + \hat{\bs x}_{\hat{S}} \|(\bs x_0)_{\hat{S}}\|\right\|\right\}\leq \delta'' \|(\bs x_0)_{\hat{S}}\|
\end{align}
Noting that 
\begin{align}
\left\|(\bs x_0)_{\hat{S}} - \hat{\bs x}_{\hat{S}} \|(\bs x_0)_{\hat{S}}\|\right\|& = \left\|(\bs x_0)_{\hat{S}} - \hat{\bs x}_{\hat{S}} \phi + \hat{\bs x}_{\hat{S}} \phi - \hat{\bs x}_{\hat{S}} \|(\bs x_0)_{\hat{S}}\|\right\|\\
\left\|(\bs x_0)_{\hat{S}} + \hat{\bs x}_{\hat{S}} \|(\bs x_0)_{\hat{S}}\|\right\|& = \left\|(\bs x_0)_{\hat{S}} + \hat{\bs x}_{\hat{S}} \phi - \hat{\bs x}_{\hat{S}} \phi + \hat{\bs x}_{\hat{S}} \|(\bs x_0)_{\hat{S}}\|\right\|
\end{align} 
Hence
\begin{align}
&\min\left\{\left\|(\bs x_0)_{\hat{S}} - \hat{\bs x}_{\hat{S}} \phi\right\| -  \left\|\hat{\bs x}_{\hat{S}}\right\| \left| \phi - \|(\bs x_0)_{\hat{S}}\|\right|, \left\|(\bs x_0)_{\hat{S}} + \hat{\bs x}_{\hat{S}} \phi\right\| -  \left\|\hat{\bs x}_{\hat{S}}\right\| \left| \phi - \|(\bs x_0)_{\hat{S}}\|\right|\right\}\\
&\leq \min \left\{\left\|(\bs x_0)_{\hat{S}} - \hat{\bs x}_{\hat{S}} \|(\bs x_0)_{\hat{S}}\|\right\|, \left\|(\bs x_0)_{\hat{S}} + \hat{\bs x}_{\hat{S}} \|(\bs x_0)_{\hat{S}}\|\right\|\right\}\\
&\leq \delta'' \|(\bs x_0)_{\hat{S}}\|
\end{align}
from which we can deduce
\begin{align}
\min \left\{\left\|(\bs x_0)_{\hat{S}} - \hat{\bs x}_{\hat{S}} \phi\right\|, \left\|(\bs x_0)_{\hat{S}} + \hat{\bs x}_{\hat{S}} \phi\right\| \right\}\leq  \delta'' \|(\bs x_0)_{\hat{S}}\| +  \left\|\hat{\bs x}_{\hat{S}}\right\| \left| \phi - \|(\bs x_0)_{\hat{S}}\|\right|.
\end{align}
Using~\eqref{sandwichForphi},  and noting that
\begin{align}
\left|\phi - \|(\bs x_0)_{\hat{S}}\|\right|\leq \sqrt{|\phi^2 - \|(\bs x_0)_{\hat{S}}\|^2 |} \leq \delta^{1/2} \|(\bs x_0)_{\hat{S}}\|
\end{align}
we finally get 
\begin{align}
\min \left\{\left\|(\bs x_0)_{\hat{S}} - \hat{\bs x}_{\hat{S}} \phi\right\|, \left\|(\bs x_0)_{\hat{S}} + \hat{\bs x}_{\hat{S}} \phi\right\| \right\}&\leq  \delta'' \|(\bs x_0)_{\hat{S}}\| +  \left\|\hat{\bs x}_{\hat{S}}\right\| \delta^{1/2}\|(\bs x_0)_{\hat{S}}\|\\
&\leq \delta'' \|(\bs x_0)_{\hat{S}}\| +   \delta^{1/2}\|(\bs x_0)_{\hat{S}}\|\\
&\leq o_m(\|(\bs x_0)_{\hat{S}}\|) \leq o_m(\|\bs x_0\|)
\end{align}
as soon as $m\gtrsim \log(m)\mu_0^{-2}k \vee \log(m)\mu_0^{-4}$. Combining this with~\eqref{x0outsideSw}, we see that if we choose the estimator $\hat{\bs w}'$ defined as
\begin{align}
\hat{\bs w}' = \left\{\begin{array}{ll}
\hat{\bs x}[\ell]& \ell \in \hat{S}\\
 0& \ell \in \hat{S}^c
\end{array}\right.
\end{align}
we get $\|\hat{\bs w}' - \bs x_0\|\leq \delta \|\bs x_0\|$ as soon as $m\gtrsim \delta^{-1}(\mu_0^{-2}k\log(m)\vee \mu_0^{-4})$.

\subsection{\label{convergenceSPF}Convergence of~\eqref{SPF}}

By analogy with~\cite{soltanolkotabi2019structured}, we use $\hat{R}^A$ to denote the linearized empirical loss
\begin{align}
\hat{R}^A_m(\bs x; \tilde{\bs x}) &  = \frac{1}{m}\sum_{i=1}^m \left(\langle \tilde{\bs a}_i, \bs x\rangle - \langle \bs A_i, \bs x_0\bs x_0^\intercal\rangle  +\varepsilon_i \right)^2\\
& = \frac{1}{m}\sum_{i=1}^m \left(\langle \tilde{\bs a}_i, \bs x\rangle - \langle \tilde{\bs x}^\intercal \bs A_i, \bs x_0\rangle +\langle \left(\bs x_0 - \tilde{\bs x}\right)\bs A_i, \bs x_0\rangle   +\varepsilon_i \right)^2
\end{align}
where $\tilde{\bs a}_i = (\bs x^{(T-1)})^\intercal \bs A_i$ for some estimate $\bs x^{(T-1)}$.  In particular, if we decompose $\bs x_0$ as $\bs x_0 = \mathcal{P}_{\bs x^{(T-1)}}\bs x_0 + \mathcal{P}^\perp_{\bs x^{(T-1)}}\bs x_0  = \langle \bs x^{(T-1)} , \bs x_0\rangle \bs x^{(T-1)} + \bs \zeta $ where $\mathcal{P}_{\bs x}$ (resp. $\mathcal{P}_{\bs x}^\perp$) denote the orthogonal projectors onto $\bs x$ (resp. its orthogonal complement)  and let $\bs z = \mathcal{P}_{\bs x^{(T-1)}}\bs x_0$, we have 
\begin{align}
\hat{R}^A_m(\bs x; \tilde{\bs x}) &  = \frac{1}{m}\sum_{i=1}^m \left(\langle \tilde{\bs a}_i, \bs x\rangle - \langle \bs z^\intercal \bs A_i, \bs x_0\rangle +\langle \left(\bs x_0 - \bs z\right)\bs A_i, \bs x_0\rangle   +\varepsilon_i \right)^2\\
& =  \frac{1}{m}\sum_{i=1}^m \left(\langle \tilde{\bs a}_i, \bs x\rangle - \langle \langle \bs x_{T-1}, \bs x_0\rangle \bs x_{T-1}^\intercal \bs A_i, \bs x_0\rangle +\langle \bs \zeta^\intercal\bs A_i, \bs x_0\rangle   +\varepsilon_i \right)^2\\
& =  \frac{1}{m}\sum_{i=1}^m \left(\langle \tilde{\bs a}_i, \bs x\rangle - \langle \bs x_{T-1}^\intercal \bs A_i- \langle \bs x_{T-1}, \bs x_0\rangle^{-1}\bs \zeta^\intercal\bs A_i,  \langle \bs x_{T-1}, \bs x_0\rangle\bs x_0\rangle   +\varepsilon_i \right)^2\\
& =  \frac{1}{m}\sum_{i=1}^m \left(\langle \tilde{\bs a}_i, \bs x\rangle - \langle \tilde{\bs a}_i - \bs v_i,  \langle \bs x_{T-1}, \bs x_0\rangle\bs x_0\rangle   +\varepsilon_i \right)^2\label{linarizedLossDecompositionPPperp}
\end{align}
where we defined $\bs v_i$ as $\bs v_i = \langle \bs x_{T-1}, \bs x_0\rangle^{-1}\bs \zeta^\intercal \bs A_i $. From this, the hard thresholding iterations are given by
\begin{align}
\bs x^+& = \mathcal{H}_k\left(\bs x - \eta \nabla \hat{R}^A_m(\bs x)\right)
\end{align}
where $\mathcal{H}_k(\bs x)$ denotes the hard thresholding operator that sets all but the largest $k$ elements of $\bs x$ to zero. In order to control these iterations, we will use the notion of restricted isometry constant recalled below
\begin{definition}
A linear map $\mathcal{A}:\mathbb{C}^{n_1\times n_2}\mapsto \mathbb{C}^m$ satisfies the rank-$r$ $k$-sparse restricted isometry property with isometry constant $\delta$ if 
\begin{align}
(1-\delta) \|\bs X\|_F^2 \leq \|\mathcal{A}(\bs X)\|_2^2 \leq (1+\delta) \|\bs X\|_F^2
\end{align}
for all $\bs X\in \mathbb{C}^{n\times n}$ such that $\text{rank}(\bs X)\leq r$ and $\|\bs X\|_{0,2}, \|\bs X^\intercal\|_{0,2}\leq k$. 
\end{definition}

We follow~\cite{lee2017near} and define the matrices $F(\bs y)\in \mathbb{C}^{m\times n_1}$ and $G(\bs x)\in \mathbb{C}^{m\times n_2}$ as
\begin{align}
F(\bs y) = \frac{1}{\sqrt{m}} \left[\begin{array}{c}
\bs y^\intercal \bs A_1 \\
\vdots \\
\bs y^\intercal \bs A_m 
\end{array}\right], \quad G(\bs x) = \frac{1}{\sqrt{m}}\left[\begin{array}{c}
 \bs x\bs A_1^\intercal \\
\vdots \\
\bs x \bs A_m^\intercal 
\end{array}\right]\label{definitionFandG}
\end{align}
The restricted isometry constant of $\tilde{\bs A} = [\tilde{\bs a}_1, \ldots \tilde{\bs a}_m]^\intercal$ can be derived through Lemma~\ref{lemmaRIP2} (Lemma B.1 in~\cite{lee2017near}) and noting that 
\begin{align}
F^\intercal(\bs y)F(\bs y) = \frac{1}{m}\sum_{i=1}^m  \bs A_i^\intercal \bs y \bs y^\intercal \bs A_i, 
\end{align}
Using this with Lemma~\ref{lemmaRIP2} gives for all subset $J$ with $|J|\leq 3k$,
\begin{align}
\left\|\Pi_{J}\left(\frac{1}{m}\sum_{i=1}^m \tilde{\bs a}_i \tilde{\bs a}_i^\intercal - \|\bs y\|^2\bs I\right)\Pi_{J}\right\|\leq \delta \|\bs y\|^2
\end{align}
which in turns implies for every $3k$-sparse vectors $\bs x$
\begin{align}
(1-\delta)\|\bs y\|^2 \|\bs x\|^2 \leq \frac{1}{m}\sum_{i=1}^m |\langle \tilde{\bs a}_i, \bs x\rangle |^2 \leq (1+\delta )\|\bs y\|^2 \|\bs x\|^2
\end{align}
In particular if we take $\bs y = \bs x^{(T-1)}$ with $\|\bs x^{(T-1)}\|=1$ we see that the map $\bs \Phi: \bs x \rightarrow \left\{\langle \tilde{\bs a}_i, \bs x\rangle \right\}_{i=1}^m $ satisfies the restricted isometry property with constant $\delta$ (note that $\delta$ can be made sufficiently small by taking $m$ sufficiently larger than $k$).  Using this with $w_i = \langle  \bs \zeta^\intercal \bs A_i , \bs x_0  \rangle+ \varepsilon_i $ and following the steps in~\cite{blumensath2009iterative} (see the proof of Corollary 1), if we assume that $\bs x^{(T)}$ is obtained after $L$ iterations of {\sf IHT} applied to problem~\eqref{linarizedLossDecompositionPPperp} with $\bs x^{T-1}$ following from the previous iteration of~\eqref{SPF}, we have
\begin{align}
\left\|\bs x^{(T)} - \langle \bs x^{(T-1)}, \bs x_0\rangle \bs x_0 \right\| & = \left\|\bs x^{(L)} - \langle \bs x^{(T-1)}, \bs x_0\rangle \bs x_0 \right\| \\
&\leq \frac{1}{2}\left\|\bs x^{(L-1)} - \langle \bs x^{(T-1)}, \bs x_0\rangle \bs x_0 \right\| + 2 \left\|\bs \Phi_{\Gamma_L}^\intercal \bs w\right\|\\
&\leq \left(\frac{1}{2}\right)^{T} \|\bs x_0 \langle \bs x_T ,\bs x_0\rangle\| + 2\sum_{\ell=0}^L \left(\frac{1}{2}\right)^L \left\|\bs \Phi^\intercal_{\Gamma_{L - \ell}}\bs w\right\|\\
&\leq 2^{-L} \|\langle \bs x_0, \bs x_T\rangle \bs x_0 \| + 4\sup_{|\Gamma|\leq 2k} \left\|\bs \Phi^\intercal_{\Gamma} \bs w\right\|\label{boundIHTperturbed01}
\end{align}
where $\Gamma_L$ denotes the support of $\bs x^{(L)}$.  Applying Lemma~\ref{lemmaRIP2} we have for any $|\Gamma|\leq 3k$
\begin{align}
\left\|\frac{1}{m}\sum_{i=1}^m \left(\tilde{\bs a}_i\right)_{\Gamma} \langle \bs \zeta^\intercal \bs A_i, \bs x_0\rangle - \langle \bs x_0, \bs \zeta\rangle \mathcal{P}_{\Gamma}\bs I \bs x_0 \right\|\leq \delta \|\bs \zeta\| \|\bs x_{T-1}\| \|\bs x_0\|
\end{align}
for all $\bs \zeta, \bs x^{(T-1)}$.  Substituting this in~\eqref{boundIHTperturbed01} gives after $L$ iterations of {\sf IHT}, noting that $\|\bs \Phi^\intercal \bs \varepsilon\|\lesssim \|\bs \varepsilon\|$, 
\begin{align}
\left\|\bs x^{(T)} - \langle \bs x^{(T-1)}, \bs x_0\rangle \bs x_0 \right\| &\leq 2^{-L} \|\langle \bs x_0, \bs x_T\rangle  \bs x_0\| + 4 \|\bs \varepsilon\| + 4\delta' \|\bs \zeta\|\|\bs x_{T-1}\|
\end{align}
Following the proof of Corollary 8.3 in~\cite{lee2017near}, we let $\theta_{T}$ to denote the angle between $\bs x_0$ and $\bs x^{(T)}$. From this, 
\begin{align}
\sin \theta_T & = \left\|\mathcal{P}^\perp_{(\bs x^{(T)})}\bs x_0\right\|\\
& = \frac{\left\|\mathcal{P}^\perp_{(\bs x^{(T)})}\bs x_0 \langle \bs x_0, \bs x^{(T-1)}\rangle \right\|}{|\langle \bs x_0, \bs x^{(T-1)}\rangle |}\\
&\leq \frac{\|\bs x_0 \langle \bs x_0, \bs x^{(T-1)}\rangle - \bs x^{(T)} \|}{|\langle \bs x_0, \bs x^{(T-1)}\rangle |}\\
&\leq \frac{2^{-L} + 4\|\bs \varepsilon \|+4\delta' \sin \theta_{T-1}}{\cos \theta_{T-1}}
\end{align}
as a result we have the relation
\begin{align}
\sin \theta_{T}\leq \left(2^{-L} + 4\|\bs \varepsilon\|\right)\sec \theta_{T-1} + 4\delta' \tan \theta_{T-1}.\label{relationThetaTthetaTminus1}
\end{align}
Using the proof of Lemma 8.7 in~\cite{lee2017near}, we define the functions
\begin{align}
f(\theta) &= \sin^{-1}\left(\left(2^{-L} + 4\|\bs \varepsilon\|\right)\sec \theta + 4\delta' \tan \theta\right)\\
f_\nu(\theta) &= \sin^{-1}\left(\left(2^{-L} + 4\nu\right)\sec \theta + 4\delta' \tan \theta\right)
\end{align}
For $\|\bs \varepsilon \|\leq \nu$. Both $f(\theta)$ and $f_\nu(\theta)$ are monotonic increasing and convex on $\theta\in [0, \pi/2)$.  We can thus write $\theta\leq f(\theta)$ on $[0, \theta_{\inf}]\cup [\theta_{\sup},\pi/2)$ (resp. $\theta\geq  f_\nu(\theta)$ on $[0, \theta_{\nu,\inf}]\cup [\theta_{\nu,\sup},\pi/2)$) and $\theta\geq f(\theta)$ on $[\theta_{\inf}, \theta_{\sup}]$ (resp.  $\theta\geq f_{\nu}(\theta)$ on $[\theta_{\nu,{\inf}}, \theta_{\nu,{\sup}}]$).  Moreover, since $f_\nu(\theta)\leq f(\theta)$, we have $[\theta_{\nu,\inf}, \theta_{\nu,\sup}]\subseteq [\theta_{\inf}, \theta_{\sup}]$.  To have $\pi/4 \in [\theta_{\nu,\inf}, \theta_{\nu,\sup}]$, we need
\begin{align}
\sin \pi/4 = \frac{\sqrt{2}}{2} \geq \left((2^{-L} + 4\nu)\frac{2}{\sqrt{2}} + 4\delta' \right)
\end{align}
which can always be achieved for $K$ sufficiently large, $\nu, \delta'$ sufficiently small.  Using the conclusion of section~\ref{initializationRankOneRecovery}, 
\begin{align}
\|\bs x^{(0)} - \bs x_0\|^2 =  2(1-\cos \theta_0)  \leq \frac{\mu_0^{-2} k}{m}
\end{align}
as soon as $\nu, \delta'$ are sufficiently small and $k$ is sufficiently large, if we take
\begin{align}
\frac{\mu_0^{-2} k}{m}\leq \frac{1}{2}
\end{align}
we have $\theta_0\leq \theta_{\sup, \nu}$ and the discussion above together with~\eqref{relationThetaTthetaTminus1} imply
\begin{align}
\theta_{1} \leq f(\theta_0)< \theta_0 
\end{align}
The strict inequality follows from the monotonicity of the function $f(\theta)$ and the fact that $\theta_0<\theta_{\sup}$.  Applyign this recursively, we can thus conclude that the iterates $\bs x^{T}$ for $L$ sufficiently large and $m$ sufficiently larger than $\mu_0^{-2}k\vee \mu_0^{-4}$ converge to the interval $[0, \theta_{\inf}]$.  On the other hand, following~\cite{lee2017near} we have
\begin{align}
\theta_{\inf} = \operatorname{argmin}\left\{\theta| \theta\geq f_\nu(\theta)\right\}\leq  \operatorname{argmin}\left\{\theta| \theta\geq \sin^{-1}\left(\frac{1}{\cos\theta_{\sup}}(2^{-k} + 4\|\varepsilon\|)  + 4\delta' \sin \theta\right)\right\} = \tilde{\theta}
\end{align}
Using
\begin{align}
\sin \tilde{\theta} (\cos \theta_{\sup} - 4\delta') = (2^{-L} + 4\nu)
\end{align}
we get 
\begin{align}
\theta_{\inf} \leq \frac{ (2^{-L} + 4\nu)}{(\cos \theta_{\sup} - 4\delta') }\leq \frac{ (2^{-L} + 4\nu)}{(\sqrt{2}/2 - 4\delta') }
\end{align}
For $\nu, \delta'$ sufficiently small and $L$ sufficiently large,  we can thus make $\theta_{\inf}$ arbitrarily small.

\subsection{\label{convergenceRankOneSparseRecovery}Convergence of~\eqref{TGD} (Non asymptotic)}

We use $\bs x^+$ to denote the update
\begin{align}
\bs x^+ = \mathcal{T}_{\eta\tau(\bs x)}\left(\bs x - \eta \left(\nabla \hat{R}_m(\bs x)\right)\right).
\end{align}
Let us assume for now that $\supp \bs x, \supp \nabla \hat{R}_m \subseteq S$ (note that this is always true for the initialization $\bs x^{(0)}$). We can then write $\bs x^+$ as
\begin{align}
\bs x^+ = \bs x - \eta \left(\nabla \hat{R}_m(\bs x)\right)_S + \eta\tau(\bs x) \bs v
\end{align}
where $\supp \bs v\subseteq S$ and $\|\bs v\|_\infty\leq 1$. Moreover, 
\begin{align}
\left(\nabla \hat{R}_m(\bs x)\right)_S = \frac{2}{m}\sum_{i=1}^m \left(\langle \bs A_i, \bs x\bs x^\intercal \rangle - \langle \bs A_i, \bs x_0\bs x_0^\intercal\rangle + \varepsilon_i  \right)\left(\bs A_i\bs x + \bs A_i^\intercal \bs x\right)_S
\end{align}
We follow~\cite{cai2016optimal} and let 
\begin{align}
\left(\nabla \hat{R}_m(\bs x)\right)_S &= \frac{1}{m}\sum_{i=1}^m \left(\langle \bs A_i, \bs x \bs x^\intercal\rangle - b_i\right) \left(\bs A_i \bs x + \bs A_i^\intercal \bs x\right)_S+ \frac{1}{m}\sum_{i=1}^m \varepsilon_i \left(\bs A_i \bs x + \bs A_i^\intercal \bs x\right)_S\\
& = \bs A + \bs E 
\end{align}
Letting $\bs h = \bs x - \bs x_0$, we further have
\begin{align}
\left\|\bs x^+ - \bs x_0\right\|&\leq \left\|\bs x - \bs x_0 - \eta \left(\nabla\hat{R}_m(\bs x)\right)_S\right\| + \eta\tau(\bs x) \|\bs v\|\\
&\leq \|\bs h - \eta\bs A\| + \eta \left\| \bs E\right\| + \eta\tau(\bs x)\sqrt{k}.\label{decompositiondeviationBetweenIterate01}
\end{align}

Note that for any vector $\bs u\in \mathbb{R}^n$, we have 
\begin{align}
\langle \bs u, \nabla \hat{R}_m(\bs x)\rangle& = \frac{1}{m}\sum_{i=1}^m \langle \bs A_i\bs x + \bs A_i^\intercal \bs x, \bs u\rangle \left(\langle \bs A_i, \bs x\bs x^\intercal\rangle - b_i \right)
\end{align}
Using $\langle \bs A_i, \bs x\bs x^\intercal\rangle  = \langle \bs A_i, \bs x_0\bs x_0^\intercal\rangle + \langle \bs h, \left(\bs A_i + \bs A_i^\intercal,\right) \bs x_0\rangle + \langle\bs A_i, \bs h\bs h^\intercal \rangle $
From this, we have 
\begin{align}
\langle \bs u, \nabla\hat{R}_m(\bs x)\rangle& = \left(\langle \bs u, (\bs A_i + \bs A_i^\intercal)\bs h\rangle + \langle \bs u, (\bs A_i + \bs A_i^\intercal)\bs x_0 \rangle  \right) \left(\langle \bs A_i, \bs x\bs x^\intercal \rangle - b_i \right)\\
& = \left(\langle \bs u, (\bs A_i + \bs A_i^\intercal)\bs h\rangle + \langle \bs u, (\bs A_i + \bs A_i^\intercal)\bs x_0\rangle  \right) \left(\langle\bs h, (\bs A_i + \bs A_i^\intercal)\bs x_0 \rangle + \langle \bs A_i, \bs h\bs h^\intercal\rangle  \right) \\
\begin{split}
& = \langle \bs u, (\bs A_i + \bs A_i^\intercal)\bs h\rangle \langle \bs h, \left(\bs A_i+ \bs A_i^\intercal \right)\bs x_0\rangle\\
& + \langle \bs u, (\bs A_i + \bs A_i^\intercal)\bs x_0\rangle \langle \bs h, \left(\bs A_i + \bs A_i^\intercal\right)\bs x_0\rangle\\
& + \langle \bs u, (\bs A_i + \bs A_i^\intercal)\bs h\rangle \langle \bs A_i, \bs h\bs h^\intercal\rangle + \langle \bs u, (\bs A_i + \bs A_i^\intercal)\bs x_0\rangle \langle \bs A_i, \bs h\bs h^\intercal\rangle \end{split} \label{decompositionGradient001}\\
& = T_1+ T_2 +    T_3
\end{align}
where 
\begin{align}
T_1 & =  \frac{1}{m} \sum_{i=1}^m \bs u\bs A_i^\intercal \bs h\bs h^\intercal \bs A_i \bs x_0+ \frac{1}{m}\sum_{i=1}^m \bs u \bs A_i \bs h\bs h^\intercal \bs A_i^\intercal \bs x_0\\
&+ \frac{2}{m} \sum_{i=1}^m \bs u^\intercal \bs A_i\bs h \bs h^\intercal \bs A_i^\intercal \bs h +\frac{2}{m} \sum_{i=1}^m \bs u^\intercal \bs A_i \bs x_0\bs h^\intercal \bs A_i^\intercal \bs h\\
T_2& = \frac{1}{m}\sum_{i=1}^m \bs u^\intercal \bs A_i \bs h\bs h^\intercal \bs A_i \bs x_0 +  \frac{1}{m}\sum_{i=1}^m\bs u^\intercal \bs A_i^\intercal \bs h\bs h^\intercal \bs A_i^\intercal \bs x_0 \\
T_3& = \frac{1}{m}\sum_{i=1}^m \bs u\bs A_i \bs x_0 \bs h^\intercal \bs A_i^\intercal \bs x_0 + \frac{1}{m}\sum_{i=1}^m \bs u\bs A_i^\intercal \bs x_0\bs h^\intercal \bs A_i \bs x_0 \\
&+ \frac{1}{m}\sum_{i=1}^m \bs u^\intercal \bs A_i \bs x_0\bs h^\intercal \bs A_i \bs x_0 + \frac{1}{m}\sum_{i=1}^m\bs u^\intercal \bs A_i^\intercal \bs x_0\bs h^\intercal \bs A_i^\intercal \bs x_0\\
\end{align}
For $T_1$,  using $F_1$ and $F_2$ to denote the operators 
\begin{align}
F_1(\bs x) = \left[\begin{array}{c}
\bs x^\intercal \bs A_1^\intercal \\
\vdots \\
\bs x^\intercal \bs A_m^\intercal \\
\end{array}\right], \quad F_2(\bs x) = \left[\begin{array}{c}
\bs x^\intercal \bs A_1\\
\vdots \\
\bs x^\intercal \bs A_m\\
\end{array}\right]
\end{align}
and using Lemma~\ref{lemmaRIP2},  we have for every $\bs h$ with $|\supp (\bs h)|\leq 3k$
\begin{align}
|T_1|&\leq  2\|\bs h\|^2 (1+\delta) \|\bs x_0\| \|\bs u\| \\
&+2 (1+\delta) \|\bs h\|^3 \|\bs u\| + 2\left(|\langle \bs x_0, \bs h\rangle | + \|\bs x_0\|\|\bs h\|\right) \|\bs h\|\|\bs u\|\label{boundT1Cai}
\end{align}
where $\delta$ is the rank-$2$, $(3k, 3k)$-RIP constant of the operator $\mathcal{A}:\bs X \mapsto \left\{\langle \bs A_i, \bs X\rangle\right\}_{i=1}^m $. For $T_2$, adapting the proof of Lemma B.1 in~\cite{lee2017near}, we first note that 
\begin{align}
\mathbb{E}\left\{ \frac{1}{m}\sum_{i=1}^m \bs A_i \langle \bs A_i^\intercal , \bs X\rangle\right\}  = \bs X^\intercal 
\end{align}
If we let $\mathcal{T}$ to denote the operator $\mathcal{T}:\bs X \mapsto \bs X^\intercal$,  for any subset $J$ with $|J|\leq 3k$ and vectors $\bs y, \bs \zeta$ with $\|\bs y\| = \|\bs \zeta\|=1$, we have 
\begin{align}
\langle \bs u\bs v^\intercal, \left(\Pi_{J}\otimes \bs y\bs y^\intercal \right) \mathcal{T} \left(\Pi_{J}\otimes \bs \zeta \bs \zeta^\intercal \right) (\tilde{\bs u}\tilde{\bs v}^\intercal) \rangle & = \langle \Pi_{J} \bs u \bs v^\intercal \bs y\bs y^\intercal,  \bs \zeta \left(\Pi_{J} \tilde{\bs u}\right)^\intercal \rangle \langle \tilde{\bs v}, \bs \zeta\rangle\\
& = \langle \tilde{\bs v}, \bs \zeta\rangle \langle \Pi_{J}\bs u, \bs \zeta \rangle \langle \Pi_{J} \tilde{\bs u}, \bs y\rangle \langle \bs y, \bs v\rangle \label{expressionTTildeOperator}    
\end{align}
On the other hand, if we let $\mathcal{P}_{\bs y} = \bs y\bs y^\intercal$ to denote the orthogonal projector onto the span of $\bs y$, we have
\begin{align}
\frac{1}{m}\sum_{i=1}^m \left(\Pi_{J} \bs u\right)^\intercal \bs A_i \bs y \bs \zeta^\intercal \bs A_i \left( \Pi_{J} \tilde{\bs u}\right) \langle \bs y, \bs v \rangle \langle \bs \zeta, \tilde{\bs v}\rangle  & = \frac{1}{m} \sum_{i=1}^m \langle \bs A_i, \Pi_{J} \bs u \bs y^\intercal \rangle \langle \bs y, \bs v\rangle \langle \bs A_i^\intercal, \Pi_{J} \tilde{\bs u}\bs \zeta^\intercal \rangle \langle \bs \zeta, \tilde{\bs v}\rangle  \\
& = \frac{1}{m}\sum_{i=1}^m \langle \bs A_i, \left(\Pi_{J}\otimes \bs y\bs y^\intercal\right) \bs u\bs v^\intercal\rangle \langle\bs A_i^\intercal, \left(\Pi_{J} \otimes \bs \zeta\bs \zeta^\intercal \right)\tilde{\bs u}\tilde{\bs v}^\intercal\rangle  \\
& = \langle \left(\Pi_{J} \otimes \bs y\bs y^\intercal \right) \bs u\bs v^\intercal, \frac{1}{m}\sum_{i=1}^m \bs A_i \langle \bs A_i^\intercal,\left(\Pi_{J} \otimes \bs \zeta\bs \zeta^\intercal\right)\tilde{\bs u}\tilde{\bs v}^\intercal \rangle\rangle\\
& = \langle \bs u\bs v^\intercal , \left(\Pi_{J} \otimes \bs y\bs y^\intercal\right)\frac{1}{m} \sum_{i=1}^m \bs A_i\langle \bs A_i^\intercal, \left(\Pi_{J} \otimes \bs \zeta\bs \zeta^\intercal\right)\tilde{\bs u}\tilde{\bs v}^\intercal\rangle \rangle   \\
& = \langle \bs u\bs v^\intercal, \left(\Pi_{J}\otimes \mathcal{P}_{\bs y}\right) \frac{1}{m}\sum_{i=1}^m \bs A_i \langle \bs A_i^\intercal, \left(\Pi_{J} \otimes \mathcal{P}_{\bs \zeta}\right)\tilde{\bs u}\tilde{\bs v}^\intercal \rangle \rangle \\
& =\langle \bs u\bs v^\intercal,  \left(\Pi_{J} \otimes \mathcal{P}_{\bs y}\right) \mathcal{A}^*\tilde{\mathcal{A}}(\Pi_{J}\otimes \mathcal{P}_{\bs \zeta}) \tilde{\bs u}\tilde{\bs v}^\intercal \rangle \label{expressionAATildeOperator}
\end{align}
In the last line, we define the operator $\tilde{\mathcal{A}}$ as $\tilde{\mathcal{A}}:\bs X\mapsto \tilde{\mathcal{A}}(\bs X) = \langle \bs A_i^\intercal, \bs X\rangle $. Combining~\eqref{expressionAATildeOperator} with~\eqref{expressionTTildeOperator}, we thus get
\begin{align}
&\frac{1}{m}\sum_{i=1}^m \left(\Pi_{J} \bs u\right)^\intercal \bs A_i \bs y \bs \zeta^\intercal \bs A_i\left( \Pi_{J} \tilde{\bs u} \right)\langle \bs y, \bs v \rangle \langle \bs \zeta, \tilde{\bs v}\rangle - \langle \tilde{\bs v}, \bs \zeta\rangle \langle\left( \Pi_{J }\bs u\right), \bs \zeta \rangle \langle \left(\Pi_{J} \tilde{\bs u}\right), \bs y\rangle \langle \bs y, \bs v\rangle\\
& = \langle \bs u\bs v^\intercal,  \left(\Pi_{J} \otimes \mathcal{P}_{\bs y}\right) \left(\mathcal{A}^*\tilde{\mathcal{A}} - \mathcal{T}\right)(\Pi_{J}\otimes \mathcal{P}_{\bs \zeta} )\tilde{\bs u}\tilde{\bs v}^\intercal \rangle\label{relationAiAiandRIP}
\end{align}
Taking $\tilde{\bs v} = \bs \zeta$ and $\bs v= \bs y$, we get 
\begin{align}
\begin{split}
&\frac{1}{m}\sum_{i=1}^m \left(\Pi_{J} \bs u\right)^\intercal \bs A_i \bs y \bs \zeta^\intercal \bs A_i \Pi_{J} \tilde{\bs u}\|\bs y\|^2 \|\bs \zeta\|^2 - \|\bs \zeta\|^2 \langle \Pi_{J}\bs u, \bs \zeta \rangle \langle \Pi_{J} \tilde{\bs u}, \bs y\rangle \|\bs y\|^2\\
=&\frac{1}{m}\sum_{i=1}^m \left(\Pi_{J} \bs u\right)^\intercal \bs A_i \bs y \bs \zeta^\intercal \bs A_i \Pi_{J} \tilde{\bs u}-  \langle \Pi_{J }\bs u, \bs \zeta \rangle \langle \Pi_{J} \tilde{\bs u}, \bs y\rangle \|\bs y\|^2\|\bs \zeta\|^2\\
=& \langle \bs u\bs y^\intercal,  \left(\Pi_{J} \otimes \mathcal{P}_{\bs y}\right) \left(\mathcal{A}^*\tilde{\mathcal{A}} - \mathcal{T}\right)(\Pi_{J}\otimes \mathcal{P}_{\bs \zeta} )\tilde{\bs u}\bs \zeta^\intercal \rangle
\end{split}\label{lowerBoundT3detailWithoutAbsoluteValue}
\end{align}
From this, for every $\bs y, \bs \zeta$ with unit norm, we have 
\begin{align}
\left|\frac{1}{m}\sum_{i=1}^m \left(\Pi_{J} \bs u\right)^\intercal \bs A_i \bs y \bs \zeta^\intercal  \bs A_i \Pi_{J} \tilde{\bs u}\right| &\leq \|\bs u\| \|\tilde{\bs u}\| \\
& + \langle \bs u\bs y^\intercal,  \left(\Pi_{J} \otimes \mathcal{P}_{\bs y}\right) \left(\mathcal{A}^*\tilde{\mathcal{A}} - \mathcal{T}\right)(\Pi_{J}\otimes \mathcal{P}_{\bs \zeta} )\tilde{\bs u}\bs \zeta^\intercal \rangle\label{intermediateResultAAtilde}  
\end{align}
The operator $\mathcal{A}^*\tilde{\mathcal{A}}$ is subexponential.  Moreover, from standard results on subexponential random matrices (see e.g.~\cite{foucartmathematical}), we have
\begin{align}
\left|\langle \bs Y, \mathcal{A}\tilde{\mathcal{A}}(\bs X)\rangle - \langle \bs Y, \bs X^\intercal\rangle \right|< \delta \|\bs X\|_F\|\bs Y\|_F\label{concentrationOfOperatorAAtilde}
\end{align}
with probability $1-e^{-Cm + C'k}$ for all doubly $3k$-sparse and rank two matrices $\bs X$ and $\bs Y$.  Combining this with~\eqref{relationAiAiandRIP}, and substituting in~\eqref{decompositionGradient001}, 
\begin{align}
|T_2|\leq2 \|\bs h\|^2 \|\bs u\|\|\bs x_0\| (1+\delta) \label{definitionT2}
\end{align}
Finally for $T_3$ using the discussion above we have 
\begin{align}
4\|\bs x_0\|^2 \|\bs u\| \|\bs h\| (1-\delta)  \leq |T_3|\leq 4\|\bs x_0\|^2 \|\bs u\| \|\bs h\| (1+\delta)\label{upperLowerBoundT3}
\end{align}
Using~\eqref{upperLowerBoundT3},~\eqref{definitionT2} and~\eqref{boundT1Cai}, together with $\|\bs h\|$ sufficiently small,  we can derive the following local smoothness condition
\begin{align}
\left\|\nabla \hat{R}_m(\bs x)\right\|^2& \leq \left(\sup_{\|\bs u\|\leq 1} \langle \nabla \hat{R}_m(\bs x), \bs u\rangle \right)^2 \\
&\lesssim 4\|\bs h\|^4 \|\bs x_0\|^2 (1+\delta)^2 + 16 \|\bs x_0\|^4 \|\bs h\|^2 (1+\delta)^2\\
& + \|\bs h\|^4 (1+\delta)^2  + 4 \|\bs h\|^4 (1+\delta)^2 \|\bs x_0\|^2\\
& + 4(1+\delta)^2 \|\bs h\|^6 + 4 \|\bs x_0\|^2 \|\bs h\|^4 (1+\delta)^2 + 8\|\bs x_0\|^2 \|\bs h\|^4 \\
&\leq  20\|\bs h\|^2\label{smoothnessupperBound}
\end{align}
To derive a corresponding lower bound on the curvature, we use~\eqref{lowerBoundT3detailWithoutAbsoluteValue}, taking $\bs y = \tilde{\bs u} = \bs x_0$, $\bs \zeta = \bs u = \bs h$ together with~\eqref{concentrationOfOperatorAAtilde} and combining this with the result of Lemma~\ref{lemmaRIP2}, as well as the the upper bounds on $T_1$ and $T_2$ (taking $\|\bs h\|<1/32$, $\delta<0.1$ for example) we get 
\begin{align}
\langle \bs h, \nabla \hat{R}_m(\bs x)\rangle \geq  4(1-\delta)\|\bs h\|^2 - 0.3\|\bs h\|^2 \label{curvatureLowerBound01}
\end{align}
Substituting~\eqref{curvatureLowerBound01} and~\eqref{smoothnessupperBound} in~\eqref{decompositiondeviationBetweenIterate01}, we can write 
\begin{align}
\left\|\bs h - \eta \bs A\right\|^2 \leq \|\bs h\|^2 (1 - 8(1-\delta)\eta + 0.3\eta ) + 20\eta^2 \|\bs h\|^2\leq   \|\bs h\|^2 (1 -5\eta) \label{boundDeviationhminusetaA}
\end{align}
as soon as $\eta <1/20$. 

To control the third term in~\eqref{decompositiondeviationBetweenIterate01},  for $S = \supp(\bs x_0)$, we assume that $\bs x^{(n)}$ is independent of $(\bs A_i)_{S^c\times S }$ and $(\bs A_i)_{S\times S^c }$ (this is true for $\bs x^{(0)}$ by construction as we take $\hat{S}\subset \supp(\bs x_0)$ and define $\bs x^{(0)}$ from the truncated eigenvalue of decomposition of $m^{-1}\sum_{i=1}^m (\bs A_i + \bs A_i^\intercal)b_i$ on $\hat{S}\times \hat{S}$). Using the expression of the gradient, we can write for any $k\in S^c$, 
\begin{align}
\left|\nabla \hat{R}_m(\bs x)\right|_k& = \frac{1}{m}\sum_{i=1}^m \left(\langle \bs A_i, \bs x\bs x^\intercal \rangle  - b_i\right) \langle (\bs A_i)_{k*} + (\bs A_i)_{*k}, \bs x\rangle \label{gradientboundOutsideSupport}
\end{align}
where $\bs A_{k*}$ (resp. $\bs A_{*k}$) is used to denote the $k^{th}$ row (resp $k^{th}$ column) of the matrix $\bs A$.  Since we assume that $\hat{\bs x}$ only depends on $(\bs A_i)_{S\times S}$, when conditionning on the $(\bs A_i)_{S\times S}$, the sum on the RHS of~\eqref{gradientboundOutsideSupport} can be viewed as a sum of independent random variables. Letting $\rho = (S\times S^c) \cup (S^c\times S)$, and noting that  
\begin{align}
\mathbb{E}_{(A_i)_{\rho}} \left\{\left(\langle \bs A_i, \bs x\bs x^\intercal\rangle - \langle \bs A_i,\bs x_0\bs x_0^\intercal \rangle  \right) \langle (\bs A_i)_{k*} + (\bs A_i)_{*k}, \bs x\rangle |(\bs A_i)_{S\times S}\right\} = 0
\end{align}
as well as 
\begin{align}
\left\|\left(\langle \bs A_i, \bs x\bs x^\intercal\rangle - \langle \bs A_i, \bs x_0\bs x_0^\intercal \rangle  \right) \langle (\bs A_i)_{k*} + (\bs A_i)_{*k}, \bs x\rangle \right\|_{\psi_2} \left|\lesssim \left(\langle \bs A_i, \bs x\bs x^\intercal\rangle - \langle \bs A_i, \bs x_0\bs x_0^\intercal \rangle  \right)\right| \sqrt{4}\|\bs x\|
\end{align}
Using standard subgaussian concentration results (e.g. Hoeffding's inequality, see~\cite{wainwright2019high} Proposition 2.5 or~\cite{vershynin2018high} Theorem 2.6.2) we can then write 
\begin{align}
&P\left(\left|\sum_{i=1}^m \left(\langle \bs A_i, \bs x\bs x^\intercal\rangle - \langle \bs A_i, \bs x_0\bs x_0^\intercal \rangle  \right) \langle (\bs A_i)_{*k} +(\bs A_i)_{k*}, \bs x\rangle \right|> t\right)\\
& \leq \exp \left(- \frac{ t^2}{4\sum_{i=1}^m \left|\langle \bs A_i, \bs x\bs x^\intercal\rangle - \langle \bs A_i, \bs x_0\bs x_0^\intercal\rangle  \right|^2 \|\bs x\|^2}\right)
\end{align}
Taking $t = \sqrt{\sum_{i=1}^m \left|\langle \bs A_i, \bs x\bs x^\intercal  \rangle  - \langle \bs A_i, \bs x_0\bs x_0^\intercal \rangle \right|^2 \|\bs x\|^2\log((mn)^2)}$ we get 
\begin{align}
&\left|\frac{1}{m} \sum_{i=1}^m \left(\langle \bs A_i, \bs x\bs x^\intercal\rangle - \langle \bs A_i, \bs x_0\bs x_0^\intercal\rangle  \right) \langle (\bs A_i)_{*k} + (\bs A_i)_{k*}, \bs x\rangle \right|\\
&\leq \frac{\sqrt{\log(m^2 n^2)}}{m} \sqrt{\sum_{i=1}^m \left(\langle \bs A_i, \bs x\bs x^\intercal\rangle - \langle \bs A_i, \bs x_0\bs x_0^\intercal\rangle  \right)^2 \|\bs x\|^2}
\end{align}
with probability $1-1/(mn)^2$.  

As a result, if we take 
\begin{align}
\tau(\bs x)  = \sqrt{\frac{C\log(mn)}{m^2} \sum_{i=1}^m \left(\langle \bs A_i, \bs x\bs x^\intercal\rangle - \langle \bs A_i, \bs x_0\bs x_0^\intercal\rangle  \right)^2 \|\bs x\|^2}
\end{align}
for some sufficiently large constant $C$ we have
\begin{align}
\left|\nabla \hat{R}_m(\bs x)\right|_k <\tau(\bs x)\label{boundonGradientOutsideSupp}
\end{align}
with probability $1- 1/(m^2n)$ for every $k\in S^c$.  To get~\eqref{boundonGradientOutsideSupp} we had to condition on $\bs x$ (in particular assuming that $\text{supp}(\bs x)\subseteq S$ so that the inequality above only holds for a fixed $\bs x$).

For the third term in~\eqref{decompositiondeviationBetweenIterate01}, we can now write
\begin{align}
\sqrt{k}\tau(\bs x) = \sqrt{\frac{C k \log(mn)}{m^2} \sum_{i=1}^m \left( \langle \bs A_i, \bs x\bs x^\intercal \rangle - \langle \bs A_i, \bs x_0\bs x_0^\intercal \rangle  \right)^2 \|\bs x\|^2 }
\end{align}
using $\|\bs x\|\leq \|\bs h\|+ \|\bs x_0\|\lesssim (1+\delta)$ as soon as $m\gtrsim \mu_0^{-2} k \vee \mu_0^{-4}$ (up to log factors) we have 
\begin{align}
\left|\sqrt{k}\tau(\bs x)\right|\leq \sqrt{\frac{C k \log(mn)}{m^2} \sum_{i=1}^m \left(\langle \bs A_i, \bs x\bs x^\intercal\rangle - \langle \bs A_i , \bs x_0\bs x_0^\intercal \rangle  \right)^2}\label{boundsqrtktau01}
\end{align}
Then using 
\begin{align}
\left(\langle \bs A_i, \bs x\bs x^\intercal \rangle - \langle \bs A_i, \bs x_0\bs x_0^\intercal \rangle  \right) = \langle \bs A_i, \bs h\bs h^\intercal\rangle + \langle \bs A_i+ \bs A_i^\intercal, \bs h\bs x_0^\intercal \rangle 
\end{align}
we can write  
\begin{align}
\sum_{i=1}^m \left(\langle \bs A_i, \bs x\bs x^\intercal\rangle - \langle \bs A_i, \bs x_0\bs x_0^\intercal\rangle  \right)^2 \lesssim \sum_{i=1}^m \langle \bs A_i, \bs h\bs h^\intercal \rangle^2 +\langle \bs A_i +\bs A_i^\intercal, \bs h\bs x_0^\intercal \rangle^2  
\end{align}

\begin{align}
\frac{1}{m}\sum_{i=1}^m \langle \bs A_i, \bs h\bs h^\intercal \rangle  = \frac{1}{m}\sum_{i=1}^m  \bs h^\intercal \bs A_i \bs h\bs h^\intercal \bs A_i \bs h
\end{align}
as well as
\begin{align}
\frac{1}{m}\sum_{i=1}^m \langle \bs A_i +\bs A_i^\intercal, \bs h\bs x_0^\intercal \rangle & = \frac{1}{m}\sum_{i=1}^m \bs h^\intercal \bs A_i \bs x_0 \bs h^\intercal \bs A_i \bs x_0 +\frac{1}{m}\sum_{i=1}^m\bs h^\intercal \bs A_i \bs x_0 \bs h^\intercal \bs A_i^\intercal \bs x_0 \\
&+\frac{1}{m}\sum_{i=1}^m \bs h^\intercal \bs A_i^\intercal \bs x_0 \bs h^\intercal \bs A_i^\intercal \bs x_0 +\frac{1}{m}\sum_{i=1}^m\bs h^\intercal \bs A_i^\intercal \bs x_0\bs h^\intercal \bs A_i \bs x_0
\end{align}
Using this with Lemma~\ref{lemmaRIP2} as well as~\eqref{intermediateResultAAtilde} and~\eqref{concentrationOfOperatorAAtilde}, we can write
\begin{align}
\frac{1}{m}\sum_{i=1}^m \left(\langle \bs A_i, \bs x\bs x^\intercal\rangle -  \langle \bs A_i, \bs x_0\bs x_0^\intercal \rangle \right)^2 &\lesssim \frac{1}{m} \sum_{i=1}^m \langle \bs A_i, \bs h\bs h^\intercal\rangle^2 + \frac{1}{m}\sum_{i=1}^m \langle \bs A_i + \bs A_i^\intercal, \bs h\bs x_0^\intercal \rangle^2 \\
&\leq\delta\left( \|\bs h\|^4 +2\|\bs h\|^2 \|\bs x_0\|^2 +2\langle \bs h, \bs x_0\rangle^2 \right)\\
& \leq \delta\left( \|\bs h\|^4 + \|\bs h\|^2 \|\bs x_0\|^2\right)
\end{align}
with probability $1-e^{-C\delta m + Ck}$. Combining this with~\eqref{boundsqrtktau01} gives 
\begin{align}
|\sqrt{k}\tau(\bs x)| \lesssim \sqrt{\frac{C k \log(mn)\log(m)}{m}}\left( \|\bs h\|^2 + \|\bs h\|\|\bs x_0\|\right)
\end{align}
with probability $1-e^{-C\delta m + Ck} - 1/(mn^2)$ (assuming that $\nabla \hat{R}_m$ is supported on $S$).  Combining this with~\eqref{boundDeviationhminusetaA} and substituting in~\eqref{decompositiondeviationBetweenIterate01}, we get 
\begin{align}
\left\|\bs x^+ - \bs x_0\right\| &\leq  \|\bs h\|\sqrt{1- 5\eta} + \eta \|\bs E\| + \eta \sqrt{\frac{k\log(mn)\log(m)}{m}} \|\bs h\|\\
&\leq \left(1- 5\eta/2 + \eta^2\sqrt{8}/4 \right) \|\bs h\| + \eta \|\bs E\| + \eta \sqrt{\frac{Ck\log(mn)\log(m)}{m}}\|\bs h\|
\end{align}
The last line follows from a Taylor expansion of $\sqrt{1-x}$ around $x= 0$. Assuming $\eta \sqrt{8}/4 < 1/2$ as well as $\sqrt{\frac{Ck\log(mn)\log(m)}{m}}<1/2$ we have 
\begin{align}
\left\|\bs x^+ - \bs x_0\right\| &\leq \left(1- \eta\right) \|\bs x - \bs x_0\| + \eta \|\bs E\|\label{boundConvergenceAlmostFinal}
\end{align}
For the last term in~\eqref{boundConvergenceAlmostFinal}, we note that 
\begin{align}
\left\|\frac{1}{m}\sum_{i=1}^m \varepsilon_i \mathcal{P}_{S}\left(\bs A_i + \bs A_i^\intercal\right)\bs x\right\|\leq \left\|\frac{1}{m}\sum_{i=1}^m \varepsilon_i \mathcal{P}_{S}\left(\bs A_i + \bs A_i^\intercal\right)\mathcal{P}_S\right\| \|\bs x\|
\end{align}
On the other hand, using 
\begin{align}
\left\|\frac{1}{m}\mathbb{E}\sum_{i=1}^m \varepsilon_i^2 \mathcal{P}_S\left(\bs A_i + \bs A_i^\intercal \right) \mathcal{P}_S^* \mathcal{P}_S \left(\bs A_i + \bs A_i^\intercal\right)\mathcal{P}_S\right\| &\lesssim \left\|\frac{1}{m}\sum_{i=1}^m \varepsilon_i^2 \bs I\right\| + \left\|\frac{1}{m}\sum_{i=1}^m \varepsilon_i^2 k \bs I\right\| \\
&\lesssim \frac{k}{m} \left|\sum_{i=1}^m \varepsilon_i^2 \right|
\end{align}
using an application of Proposition~\ref{BernsteinProposition} gives
\begin{align}
\left\|\frac{1}{m}\sum_{i=1}^m \varepsilon_i \mathcal{P}_S \left(\bs A_i + \bs A_i^\intercal\right) \mathcal{P}_S \right\|\leq \sqrt{\frac{k\log(m)(1/m)\sum_{i=1}^m \varepsilon_i^2}{m}}\label{boundNoisyTerm01}
\end{align}
On the other hand,  letting $\sigma = \max_{1\leq i\leq m} \|\varepsilon_i\|_{\psi_1}$, Lemma A.7 in~\cite{cai2016optimal} (Chebyshev's inequality) gives
\begin{align}
\left|\frac{1}{m}\sum_{i=1}^m \varepsilon_i^2 \right| \leq  C\sigma^2, \quad w.p\quad 1-3/m\label{boundSubexponentialNoiseCai}
\end{align}
Combining~\eqref{boundSubexponentialNoiseCai} with~\eqref{boundNoisyTerm01} by means of a union bound, we finally get 
\begin{align}
\left\|\frac{1}{m}\sum_{i=1}^m \varepsilon_i \mathcal{P}_S \left(\bs A_i + \bs A_i^\intercal\right) \bs x\right\| \leq \sigma \sqrt{\frac{k\log(m)}{m}}\label{boundEterm}
\end{align}
with probability $1-4/m$.  Combining~\eqref{boundEterm} with~\eqref{boundConvergenceAlmostFinal} gives 
\begin{align}
\left\|\bs x^+ - \bs x_0\right\| &\leq \left(1- \eta\right) \|\bs x - \bs x_0\| + \eta \sigma \sqrt{\frac{k\log(m)}{m}}
\end{align}
For any $T$ such that $T\eta\lesssim 1$ we have 
\begin{align}
\left\|\bs x^{(T)} - \bs x_0\right\| &\leq \left(1- \eta\right)^T \delta  + C \sigma \sqrt{\frac{k\log(m)}{m}}\label{finalBoundIteratesTruncatedGradientDescent}
\end{align}
where $C$ is an absolute constant. Since all of our concentration results are defined based on the measurement operators $\bs A_i$ and for all $\bs x$'s except for~\eqref{boundonGradientOutsideSupp} which is conditioned on $\bs x$.  Relation~\eqref{finalBoundIteratesTruncatedGradientDescent} holds with probability at least $1-o_m(1) - T/(m^2n)$.

\section{\label{sectionProofPropositionPR}Proof of Proposition~\ref{propositionIncoherenceInitializationAndPhaseRetrieval}}

\subsection{\label{sectionInitializationPR}Initialization}

Once we have the largest entry, we consider the estimator 
\begin{align}
\hat{v}[\ell] = \frac{1}{m} \sum_{i=1}^m  b_i a_i[1]a_{i}[\ell] = \frac{1}{m} \sum_{i=1}^m \langle \bs a_i, \bs x_0\rangle^2 a_i[1]a_{i}[\ell]  \label{supportEstimatorPR01}
\end{align}
for any $\ell\neq 1$.  


We start by considering the case $\ell\in S$, letting $\bs a_{i\setminus \left\{1, \ell\right\}}$ to denote the vector $\bs a_i$ in which the entries $1$ and $\ell$ have been set to zero and decoupling, we get
\begin{align}
\frac{1}{m}\sum_{i=1}^m |\langle \bs a_i, \bs x\rangle |^2 a_i[1]a_i[\ell] & = \frac{1}{m}\sum_{i=1}^m \left(\langle \bs a_{i\setminus \left\{1, \ell\right\}}, \bs x_0\rangle  + a_i[1]x_0[1] + a_i[\ell]x_0[\ell]\right)^2 a_i[1]a_i[\ell]\label{estimatorOnSupport01}\\
& = \frac{1}{m}\sum_{i=1}^m \langle \bs a_{i\setminus \left\{1, \ell\right\}}, \bs x_0\rangle^2 a_i[1]a_i[\ell]\label{termBoundedThroughMassart1PRinsupport01}\\
&+ \frac{2}{m}\sum_{i=1}^m a_i^2[1] x_0[1] a_i[\ell] \langle \bs a_{i\setminus \left\{1, \ell\right\}}, \bs x_0\rangle\label{termBoundedThroughMassart1PRinsupport02}\\
&+ \frac{2}{m} \sum_{i=1}^m a_i^2[\ell] x_0[\ell] \langle \bs a_{i\setminus \left\{1, \ell\right\}}, \bs x_0\rangle a_i[1] \label{termBoundedThroughMassart1PRinsupport03}\\
&+ \frac{1}{m}\sum_{i=1}^m a_i^3[1] x_0^2[1] a_i[\ell]+ \frac{1}{m}\sum_{i=1}^m a_i^3[\ell] x_0^2[\ell] a_i[1]\label{termCube}\\
& + \frac{2}{m} \sum_{i=1}^m a_i^2[1] a_i^2[\ell] x_0[1]x_0[\ell] \label{termBoundedThroughBentkusPRinsupport}
\end{align}
For~\eqref{termBoundedThroughMassart1PRinsupport01} to~\eqref{termBoundedThroughMassart1PRinsupport03}, we turn to Lemma~\ref{lemmaLaurentMassart}.  We first define $Z_{1i}$ and $Z_{2i}$ as 
\begin{align}
Z_{1i}& = \frac{(a_i[1] + a_{i}[\ell])}{\sqrt{2}}, \quad Z_{2i} = \frac{(a_i[1] - a_{i}[\ell])}{\sqrt{2}}
\end{align}
\begin{align}
&P\left(\frac{1}{m}\sum_{i=1}^m Z_{1i}^2 \langle \bs a_{i\setminus \left\{1, \ell\right\}}, \bs x_0\rangle^2 > \frac{2}{m}\sqrt{\sum_{i=1}^m \langle \bs a_{i\setminus \left\{1, \ell\right\}}, \bs x_0\rangle^4 } \sqrt{t} + 2\max_i \left|\langle \bs a_{i\setminus \left\{1, \ell\right\}}, \bs x_0\rangle \right|^2 t\right)< e^{-t},
\end{align}
and similarly for $Z_{i2}$.  Combining this with Chebyshev's inequality and a union bound for the maximum, 
\begin{align}
P\left(\sum_{i=1}^m \frac{\langle \bs a_{i\setminus \left\{1, \ell\right\}}, \bs x_0\rangle^4 }{\|\bs x_0\|^4} > m +\sqrt{m}t\right)< t^{-2}\\
P\left(\max_i \langle \bs a_{i\setminus \left\{1, \ell\right\}}, \bs x_0\rangle^2 > t \right) < e^{-(t^2 \wedge t)/2}
\end{align}
Substituting those bounds in~\eqref{termBoundedThroughMassart1PRinsupport01} with $t = \log(m)$, we obtain 
\begin{align}
\left|\frac{1}{m}\sum_{i=1}^m \langle \bs a_{i\setminus \left\{1, \ell\right\}}, \bs x_0\rangle^2 a_i[1]a_i[\ell] \right|\lesssim 2\sqrt{\frac{\log(m)}{m}}\|\bs x_0\|^2 +  \|\bs x_0\|^2 \frac{\log(m)}{m}
\end{align}
with probability at least $1-o_m(1)$.  A similar reasoning can be applied to~\eqref{termBoundedThroughMassart1PRinsupport02} and~\eqref{termBoundedThroughMassart1PRinsupport03} with $Z_{1i} = (a_i[\ell] + \langle \bs a_{i\setminus \left\{1, \ell\right\}}, \bs x_0\rangle )/\sqrt{2}$ and $ Z_{2i} = \left(a_i[\ell] - \langle \bs a_{i\setminus \left\{1, \ell\right\}}, \bs x_0\rangle \right)/\sqrt{2}$ (resp. $Z_{1i} = (a_i[1] + \langle \bs a_{i\setminus \left\{1, \ell\right\}},\bs x_0\rangle )/\sqrt{2}$ and $Z_{2i} = (a_i[1] - \langle \bs a_{i\setminus \left\{1, \ell\right\}}\rangle )\sqrt{2}$). 
\begin{align}
\left|\frac{2}{m}\sum_{i=1}^m a_i^2[1] x_0[1] a_i[\ell] \langle \bs a_{i\setminus \left\{1, \ell\right\}}, \bs x_0\rangle \right| \lesssim \sqrt{\frac{\log(m)}{m}}\|\bs x_0\|^2 + \frac{\log(m)}{m}\|\bs x_0\|^2\label{boundSumOfsquareMultipliedByChiSquared01}
\end{align}
as well as 
\begin{align}
\left|\frac{2}{m}\sum_{i=1}^m a_i^2[\ell] x_0[\ell] \langle \bs a_{i\setminus \ell}, \bs x_0\rangle a_i[1] \right| < \sqrt{\frac{\log(m)}{m}} \|\bs x_0\|^2 + \frac{\log(m)}{m}\|\bs x_0\|^2\label{boundSumOfsquareMultipliedByChiSquared02}
\end{align}
For~\eqref{termBoundedThroughBentkusPRinsupport} we use Lemma~\ref{lemmaBentkus}, noting that $X_i \equiv 1-a_i^2[1]a_i^2[\ell]\leq 1$ hence
\begin{align}
\mathbb{E}X_i^2 = 1+ \mathbb{E}\left\{a_i^4[1]a_i^4[\ell]\right\} - 2\mathbb{E}\left\{a_i^2[1]a_i^2[\ell]\right\} = 8
\end{align}
we get 
\begin{align}
P\left(1-\frac{1}{m}\sum_{i=1}^m a_i^2[1]a_i^2[\ell] \geq t/m\right) < \exp(-t^2/2)
\end{align}
which implies with probability $1-o_m(1)$
\begin{align}
\frac{1}{m}\sum_{i=1}^m a_i^2[1]a_i^2[\ell]\geq 1- \log(m)/m
\end{align}
Substituting in~\eqref{termBoundedThroughBentkusPRinsupport}, we obtain 
\begin{align}
\left|\frac{2}{m}\sum_{i=1}^m a_i^2[1]a_i^2[\ell]x_0[1]x_0[\ell] \right|\geq \left|x_0[1]x_0[\ell] \right|\left(1- \frac{\log(m)}{m}\right)\label{boundSumOfSquareSquare}
\end{align}
Controling~\eqref{termCube} requires more advanced tools as the variables are not non-negative anymore and cannot be decoupled into a product of independent chi-squared.  The variables $a_i^3[1]a_i[\ell]$ and $a_i^3[\ell]a_i[1]$ have $\alpha$-exponential tails (sometimes known as sub-Weibull~\cite{vladimirova2020sub}), and we can thus turn to the Orlicz norms and Talagrand's inequality (see Theorem~\ref{talagrandTheorem} as well as Theorem 3 in~\cite{talagrand1989isoperimetry}).  For any $p>0$, the Orlicz $p$-norm of a random variable $X$ is defined as 
\begin{align}
\left\|X\right\|_{\psi_p} = \inf \left\{c>0\;|\; \mathbb{E}\left\{\exp\left(\frac{|X|}{c}\right)^p - 1\leq 1\right\}\right\}
\end{align}
In particular, for any gaussian variables $X_1, X_2$, taking $p=1/2$ gives 
\begin{align}
\inf\left\{c\;|\;\mathbb{E}\left\{\exp\left(\frac{|X_1^3X_2|^{1/2}}{c^{1/2}}\right)\right\}\leq 2\right\}& =\inf\left\{c\;|\;\mathbb{E}\left\{\exp\left(\frac{|X_1 (|X_1X_2|)^{1/2}|}{c^{1/2}}\right)\right\}\leq 2\right\}\\
& = \inf\left\{c'\;|\;\mathbb{E}\left\{\exp\left(\frac{|X_1 (|X_1X_2|)^{1/2}|}{c'}\right)\right\}\leq 2\right\}^{2}\label{OrliczNormBoundoneHalf}
\end{align}
Now the numerator in the exponential from the last line is a product of two subgaussian random variables. $X_1$ is gaussian and $|X_1X_2|^{1/2}$ follows a chi distribution.  From the fact that $|X_1X_2|^{1/2}$ has a chi distribution with two degrees of freedom (Simply use $X_1X_2 = (X_1+X_2)^2 - (X_1-X_2)^2$ where $X_1+X_2$ and $X_1 - X_2$ are independent),  using standard tail bounds for chi-squared random variables (e.g. Lemma~\ref{lemmaLaurentMassart}). 
\begin{align}
P\left(|X_1X_2|^{1/2}>8t\right) = P\left(|X_1X_2|>64 t^2\right)\leq \exp(-t^2) 
\end{align}
Now using Lemma 2.2.1 in~\cite{wellner2013weak} (see Lemma~\ref{LemmaWellner01}), we have 
\begin{align}
\left\||X_1X_2|^{1/2}\right\|_{\psi_2}\leq 8
\end{align}
Similarly we have (see for example~\cite{wainwright2019high}) $\|X_1\|_{\psi_2}\lesssim 1$.  Combining those results with Lemma~\ref{OrliczNormProductSubgaussianVariables} 
\begin{align}
\left\||X_1X_2|^{1/2}X_1\right\|_{\psi_1} = \inf\left\{c\;|\; \mathbb{E}\left\{\exp\left(\frac{|X_1X_2|^{1/2}|X_1|}{c}\right)\right\}\leq 2\right\} \lesssim 1
\end{align}
Combining this with~\eqref{OrliczNormBoundoneHalf} gives $\|X_1^3X_2\|_{\psi_{1/2}}\lesssim 1$ and similarly for $\|X_2^3X_1\|_{\psi_{1/2}}$.  To control the first term in~\eqref{TalagrandInequality01}, noting that $\mathbb{E}a_i^3[1]a_i[\ell] = 0$, and using Jensen's inequality
\begin{align}
\mathbb{E}\left\{\left|\sum_{i=1}^m a_i^3[1]a_i[\ell]\right|\right\} \leq \mathbb{E}^{1/2}\left\{\left|\sum_{i=1}^m a_i^3[1]a_i[\ell]\right|^2\right\}\leq (15m)^{1/2}\label{boundL1normTalagrand}
\end{align}
For the second term in~\eqref{TalagrandInequality01}, we use the following maximal inequality (see property (viii), section 4 in~\cite{chamakh2020orlicz} or Lemma 2.2.2 in~\cite{wellner2013weak}). For any $p\in \mathbb{R}^+$,
\begin{align}
\left\|\max_{j\in[m]} |X_j|\right\|_{\psi_{p}} \leq \max_{j\in [m]} \|X_j\|_{\psi_p} \left[\frac{\log(1+m)}{\log(3/2)}\right]^{1/p} 
\end{align}
Using $X_j = a_j^3[1]a_j[\ell]$ or $a_j^3[\ell]a_j[1]$ together with $p=1/2$ gives
\begin{align}
\left\|\max_{j\in[m]} |X_j|\right\|_{\psi_{p}} \lesssim \left[\frac{\log(1+m)}{\log(3/2)}\right]^{1/p} \label{maximalInequalityresult01}
\end{align}
Combining~\eqref{boundL1normTalagrand} and~\eqref{maximalInequalityresult01} and substituting in Theorem~\ref{talagrandTheorem} gives 
\begin{align}
\left\|\sum_{i=1}^m a_i^3[1]a_i[\ell]\right\|_{\psi_{1/2}}, \left\|\sum_{i=1}^m a_i^3[\ell]a_i[1]\right\|_{\psi_{1/2}}\lesssim \left(\sqrt{m} + \left(\frac{\log(1+m)}{\log(3/2)}\right)^2\right)\label{boundOrliczNormSumCube}
\end{align}
To derive a tail bound, we rely on the following deviation inequality (see Property (vi), section 4 in~\cite{chamakh2020orlicz}). For any $X\in L_{\psi}$, we have 
\begin{align}
P\left(|X|\geq c\right) \leq \frac{2}{\psi(c/\|X\|_{\psi})+1}
\end{align}
Taking $X = \sum_{i=1}^m a_i^3[1]a_i[\ell]$ or $\sum_{i=1}^m a_i^3[\ell]a_i[1]$ as well as $\psi = \psi_{1/2} = e^{x^{1/2}} - 1$ gives 
\begin{align}
P\left(\sum_{i=1}^m a_i^3[1]a_i[\ell] > c\right)\lesssim \exp\left(-\frac{c^{1/2}}{\|\sum_{i=1}^m a_i^3[1]a_i[\ell]\|_{\psi_{1/2}}^{1/2}})\right)
\end{align}
Using~\eqref{boundOrliczNormSumCube} and taking $c \gtrsim ( \sqrt{m}\vee \log^2(m))\log^2(m)\log^2(k)$ gives 
\begin{align}
\left|\frac{1}{m}\sum_{i=1}^m a_i^3[1]a_i[\ell]\right| < \sqrt{\frac{\log^4(m)\log^2(k)}{m}}\label{finalTailBoundCubeTalagrand}
\end{align}
with probability $1-o_m(1)$ for every $\ell\in S$.  Combining this with~\eqref{boundSumOfsquareMultipliedByChiSquared01}, ~\eqref{boundSumOfsquareMultipliedByChiSquared02} and~\eqref{boundSumOfSquareSquare} and substituting in~\eqref{estimatorOnSupport01}, we get 
\begin{align}
\begin{split}
\left|\frac{1}{m}\sum_{i=1}^m |\langle \bs a_i, \bs x \rangle |^2 a_i[1]a_i[\ell]\right|&\geq |x_0[1]x_0[\ell]| \left(1 - \frac{\log(m)}{m}\right) - \left(\sqrt{\frac{\log(m)}{m}} + \frac{\log(m)}{m}\right) \|\bs x_0\|^2\\
& - 2\sqrt{\frac{\log^4(m)\log^2(k)}{m}}
\end{split}\label{finalBoundTailPRinS}
\end{align}
with probability $1-o_m(1)$.  Note that alternatively, we could have used 
\begin{align}
\left|\frac{1}{m}\sum_{i=1}^m a_i^3[1]a_i[\ell]\right|\leq \sup_{i} \left|a_i[1]a_i[\ell]\right| \left|\frac{1}{m}\sum_{i=1}^m a_i^2[1]\right|\label{approach2Order4terms}
\end{align}
Then using the fact that 
\begin{align}
a_i[1]a_i[\ell] = (a_i[1] + a_i[\ell])^2/2 - (a_i[1] - a_i[\ell])^2/2
\end{align}
where $(a_i[1] + a_i[\ell])$ and $(a_i[1] - a_i[\ell])$ are independent, Proposition~\ref{scalarBernstein} then gives 
\begin{align}
P(|a_i[1]a_i[\ell]|>t)\lesssim e^{-t^2\wedge t}\label{boundOnSupremumApproach2}
\end{align}
Taking $t = 2\log(m)$ and combining this with a union bound then implies $\max_i |a_i[1]a_i[\ell]| < \log(m)$ with probability $1-o_m(1)$.  Applying Proposition~\ref{scalarBernstein} again to the sum in~\eqref{approach2Order4terms} and combining with~\eqref{boundOnSupremumApproach2} would then yield
\begin{align}
\left|\frac{1}{m}\sum_{i=1}^m a_i^3[\ell]a_i[1]\right| \leq \sqrt{\frac{\log^3(m)}{m}}
\end{align}
with probability $1-o_m(1)$. 

When $\ell\notin S$, the sum in~\eqref{supportEstimatorPR01} decomposes as 
\begin{align}
\sum_{i=1}^m |\langle \bs a_i, \bs x_0\rangle |^2 a_i[1]a_i[\ell]& = \sum_{i=1}^m \left(a_i[1]x_0[1] + \langle \bs a_{i\setminus 1}, \bs x_{0\setminus 1}\rangle \right)^2 a_i[1]a_i[\ell]\label{firstTwoTermsInNotInSupport}\\
& = \sum_{i=1}^m a_i^3[1]x_0^2[1] a_i[\ell] + \frac{1}{m}\sum_{i=1}^m \langle \bs a_{i\setminus 1}, \bs x_{0\setminus 1}\rangle^2 a_i[1]a_i[\ell]\\
&+ \frac{2}{m}\sum_{i=1}^m \langle \bs a_{i\setminus 1}, \bs x_{0 \setminus 1}\rangle a_i^2[1] a_i[\ell] x_0[1]  
\end{align}
For the last two terms above, using the independence we have 
\begin{align}
&\mathbb{E}\left\{\frac{1}{m}\sum_{i=1}^m \langle \bs a_{i\setminus 1}, \bs x_{0\setminus 1}\rangle^2 a_i[1] a_i[\ell]  \right\} = 0\\
& \mathbb{E}\left\{\frac{2}{m}\sum_{i=1}^m \langle \bs a_{i\setminus 1}, \bs x_{0\setminus 1} a_i^2[1] a_i[\ell] x_0[1]\rangle \right\} = 0
\end{align}
Now using Lemma~\ref{lemmaLaurentMassart} again, noting that for any two independent random gaussian random variables $X_1$ and $X_2$ we have $X_1X_2 = ((X_1+X_2)^2 - (X_1-X_2)^2)/2$, we get 
\begin{align}
P\left(\left|\frac{1}{m}\sum_{i=1}^m \langle \bs a_{i\setminus 1}, \bs x_{0\setminus 1}\rangle^2 a_i[1]a_i[\ell]\right|\geq 2\left(\sum_{i=1}^m \langle \bs a_{i\setminus 1}, \bs x_{0\setminus 1}\rangle^4 \right)^{1/2}\sqrt{t} + 2\max_{i\in [m]} \langle \bs a_{i\setminus 1}, \bs x_{0\setminus 1}\rangle^2 t  \right)< e^{-t}\\
P\left(\left|\frac{2}{m}\sum_{i=1}^m a_i^2[1] \langle \bs a_{i\setminus 1}, \bs x_{0\setminus 1}\rangle a_i[\ell]\right| > 2\left(\sum_{i=1}^m a_i^4[1]\right)^{1/2}\sqrt{t} + 2\max_{i\in [m]}a_i^2[1] t\right)<e^{-t}
\end{align}
Using Chebyshev's inequality as well as the union bound, and noting that
\begin{align}
\mathbb{E}\left\{\left( a_i^4\right)^2\right\} - \mathbb{E}\left\{ a_i^4\right\}^2 = m\sum_{i=1}^m a_i^8 = 105 m\\
\mathbb{E}\left\{\left(\sum_{i=1}^m \langle \bs a_{i\setminus 1}, \bs x_{0\setminus 1}\rangle^4 \right)^2\right\} - \mathbb{E}\left\{\sum_{i=1}^m \langle \bs a_{i\setminus 1}, \bs x_{0\setminus 1}\rangle^4 \right\}^2& = m \|\bs x_{0\setminus 1}\|^8
\end{align}
we can write 
\begin{align}
P\left(\sum_{i=1}^m a_i^4 - 3m > \sqrt{105 m} t\right) < t^{-2}\\
P\left(\sum_{i=1}^m \langle \bs a_{i\setminus 1}, \bs x_{0\setminus 1}\rangle^4 - \|\bs x_{0\setminus 1}\|^4 m > \sqrt{105 m} \|\bs x_{0\setminus 1}\|^4 t\right) < t^{-2}
\end{align}
On the other hand, using any standard tail bound on chi-squared random variables, we have 
\begin{align}
P\left(\max_{i\in [m]} a_i^2[1] > t\right)\leq m e^{-(t^2\wedge t)/2}\\
P\left(\max_{i\in [m]} \frac{\langle \bs a_{i\setminus 1}, \bs x_{0\setminus 1}\rangle }{\|\bs x_{0\setminus 1}\|^2}>t\right)\leq me^{-(t^2\wedge t)/2}
\end{align}
Grouping those bounds and using $t = \log(m)$ gives
\begin{align}
\left|\frac{1}{m}\sum_{i=1}^m \langle \bs a_{i\setminus 1}, \bs x_{0\setminus 1}\rangle^2 a_i[1]a_i[\ell] \right|\leq 2\left(\sqrt{105} + \|\bs x_{0\setminus 1}\|^2 \right) \sqrt{\frac{\log(m)}{m}} + 2\|\bs x_{0\setminus 1}\|^2 \frac{\log(m)}{m},\label{notInSupportMassart01}\\
\left|\frac{1}{m}\sum_{i=1}^m \langle \bs a_{i\setminus 1}, \bs x_{0\setminus 1}\rangle a_i[\ell] a_i^2[1] \right| \leq 2\left(\sqrt{105}+3\right)\sqrt{\frac{\log(m)}{m}} + 2\frac{\log(m)}{m}. \label{notInSupportMassart02}
\end{align}
For the first term in~\eqref{firstTwoTermsInNotInSupport}, we again turn to Talagrand's inequality on the $L_{\psi_{1/2}}$ space and following the reasoning that led to~\eqref{finalTailBoundCubeTalagrand}, we get
\begin{align}
\left|\frac{1}{m}\sum_{i=1}^m a_i^3[1]a_i[\ell]\right|< \sqrt{\frac{\log^4(m)\log^2(k)}{m}}\label{notInSupportCube}
\end{align}
with probability $1-o_m(1)$.  Grouping~\eqref{notInSupportMassart01}, ~\eqref{notInSupportMassart01} and~\eqref{notInSupportCube} gives for $\ell\notin S$, 
\begin{align}
\left|\frac{1}{m}\sum_{i=1}^m |\langle \bs a_i, \bs x_0\rangle |^2 a_i[1]a_i[\ell]\right|\lesssim \left(\sqrt{\frac{\log(m)}{m}} + \frac{\log(m)}{m}\right) (\|\bs x_0\|^2 + 1) + \sqrt{\frac{\log^4(m)\log^2(k)}{m}}\label{finalBoundPRNotInS}
\end{align}
From~\eqref{finalBoundPRNotInS} and~\eqref{finalBoundTailPRinS}, we see that for $m$ sufficiently large, the estimator~\eqref{supportEstimatorPR01} thus satisfies 
\begin{align}
&\left|\hat{v}[\ell]\right| \geq |x_0[1]x_0[\ell]| \left(1 - \frac{\log(m)}{m}\right) - C\sqrt{\frac{\log^4(m)\log^2(k)}{m}} \quad \text{if $\ell \in S$}\\
&\left|\hat{v}[\ell]\right| \leq c\sqrt{\frac{\log^4(m)\log^2(k)}{m}} \quad \text{if $\ell \in S^c$}
\end{align}
so that if we define $S_- = \left\{\ell\;|\; |x_0[\ell]x_1[\ell]| > 2C\sqrt{\frac{\log^4(m)\log^2(k)}{m}}\right\}$ and $\hat{S} = \left\{\ell\;|\; |\hat{v}[\ell]|> 2C\sqrt{\frac{\log^4(m)\log^2(k)}{m}}\right\}$
we clearly have $S_-\subseteq \hat{S}$. Moreover,  we can guarantee that
\begin{align}
\|(\bs v_0)_{\hat{S}}\| = \|\bs x_0[1](\bs x_0)_{\hat{S}^c}\|\leq \sqrt{k}2C\sqrt{\frac{\log^4(m)\log^2(k)}{m}}
\end{align} 
In particular
\begin{align}
\|(\bs x_0)_{\hat{S}^c}\|\leq  \sqrt{\frac{\mu_0^{-2} k \log^4(m)\log^2(k)}{m}}\label{boundPRx0OnShatcomplement}
\end{align}
To estimate $\bs x_0$ on $\hat{S}$, we follow~\cite{cai2016optimal, candes2015phase} and rely on the top eigenvector of the matrix $m^{-1}\sum_{i=1}^m b_i \bs a_i\bs a_i^\intercal  = m^{-1}\sum_{i=1}^m \langle \bs a_i, \bs x_0\rangle^2  \bs a_i\bs a_i^\intercal $.  Following the proof of lemma 6.3 in~\cite{cai2016optimal} (see Lemma A.6 in the same paper), we can write 
\begin{align}
&\left\|\frac{1}{m}\sum_{i=1}^m \langle \bs a_i, \bs x_0^\intercal\rangle^2 (\bs a_{i})_{\hat{S}} (\bs a_{i})_{\hat{S}}^\intercal - (\|\bs x_0\|^2 \bs I_{\hat{S}, \hat{S}} + 2(\bs x_0)_{\hat{S}}(\bs x_0)_{\hat{S}}^\intercal) \right\|\\
&\leq \left\|\frac{1}{m} \langle \bs a_i, \bs x_0\rangle^2 (\bs a_{i})_{S_0}(\bs a_i)_{S_0}^\intercal  - \left(\|\bs x_0\|^2\bs I_{S_0\times S_0} + 2\bs x_0\bs x_0^\intercal\right) \right\|\\
&\leq \delta \|\bs x_0\|_2^2
\end{align}
as soon as $m\gtrsim C(\delta) k\log(n)$ with probability $1-o_m(1)$.  Applying Davis-Kahan $\sin \theta$ Theorem (see e.g.\cite{yu2015useful} Theorem 1),  if we use $\bs v$ to denote the normalized leading eigenvector of the matrix $m^{-1} \sum_{i=1}^m \langle \bs a_i, \bs x_0\rangle^2 (\bs a_i)_{\hat{S}}(\bs a_i)_{\hat{S}}^\intercal $, we can write  
\begin{align}
\left\|\bs v\bs v^\intercal - \frac{(\bs x_0)_{\hat{S}}(\bs x_0)_{\hat{S}}^\intercal}{\|(\bs x_0)_{\hat{S}}\|^2}\right\|\leq \frac{\delta}{2-\delta}
\end{align}
In particular, noting that 
\begin{align}
\left\|\bs v\bs v^\intercal - \frac{(\bs x_0)_{\hat{S}}(\bs x_0)_{\hat{S}}^\intercal}{\|(\bs x_0)_{\hat{S}}\|^2}\right\|^2 = 2 - 2\langle \bs v, (\bs x_0)_{\hat{S}}/\|(\bs x_0)_{\hat{S}}\|\rangle = 2(1-\cos^2 \theta(\bs v, (\bs x_0)_{\hat{S}}/\|(\bs x_0)_{\hat{S}}\|))
\end{align}
while $\|\bs v \pm (\bs x_0)_{\hat{S}}/\|(\bs x_0)_{\hat{S}}\|\| = 2 \pm 2|\cos \theta(\bs v, (\bs x_0)_{\hat{S}}/\|(\bs x_0)_{\hat{S}}\||$, we have 
\begin{align}
\min\left\{\|\bs v + (\bs x_0)_{\hat{S}}/\|(\bs x_0)_{\hat{S}}\|\|, \|\bs v - (\bs x_0)_{\hat{S}}/\|(\bs x_0)_{\hat{S}}\|\|  \right\}&=  2\left(1 - |\cos \theta(\bs v, (\bs x_0)_{\hat{S}}/\|(\bs x_0)_{\hat{S}}\||\right)\\
& \leq 2\left(1 - \sqrt{1 - \frac{1}{2}\left(\frac{\delta}{2-\delta}\right)^2}\right)\\
&\leq  \frac{\sqrt{2}\delta}{2-\delta}
\end{align}
where the last line follows from $\sqrt{1-x^2}\geq 1-x$ for $x\leq 1$.  Now note that if we let $\phi^2 =  \frac{1}{m}\sum_{i=1}^m \langle \bs a_i, \bs x_0\rangle^2$, using Lemma 6.2 in~\cite{cai2016optimal} gives $(1-\delta)\leq \phi/\|\bs x_0\|\leq (1+\delta)$. On the other hand, using~\eqref{boundPRx0OnShatcomplement} we have 
\begin{align}
\|(\bs x_0)_{S}\|/\|\bs x_0\| \geq 1-\delta
\end{align}
which gives 
\begin{align}
1 \leq \frac{\phi}{\|(\bs x_0)_{\hat{S}}\|} \leq \frac{1+\delta}{1-\delta}
\end{align}
Those lines in particular imply
\begin{align}
\left\|\bs v\phi - \frac{(\bs x_0)_{\hat{S}}\phi}{\|(\bs x_0)_{\hat{S}}\|}\right\| &\geq \left\|\bs v \phi -  (\bs x_0)_{\hat{S}}\right\| - \frac{2\delta_1}{1-\delta_2} \|(\bs x_0)_{\hat{S}}\|\\
&\geq  \|\bs v\phi - \bs x_0\| - \delta \|\bs x_0\| \label{boundInitializationPR01}
\end{align}
as soon as $m\gtrsim \mu_0^{-2}k\delta^{-1}$ (up to log factors) with probability $1-o_m(1)$ and similarly for $\|\bs v \phi - (\bs x_0)_{\hat{S}}\phi/\|(\bs x_0)_{\hat{S}}\|\|$, i.e.
\begin{align}
\left\|\bs v \phi + \frac{(\bs x_0)_{\hat{S}}\phi}{\|(\bs x_0)_{\hat{S}}\|}\right\|&\geq \|\bs v \phi + \bs x_0\|  - \delta \|\bs x_0\|\label{boundInitializationPR02}
\end{align}
Grouping~\eqref{boundInitializationPR01} and~\eqref{boundInitializationPR02}, we recover the desired initialization bound
\begin{align}
\min\left\{\|\bs v \phi - \bs x_0\|, \|\bs v \phi + \bs x_0\|\right\} &\leq \frac{\sqrt{2}\delta}{2-\delta} \phi\\
&\leq  \frac{\sqrt{2}\delta}{2-\delta} \delta (1+\delta') \|\bs x_0\|
\end{align}
as soon as $m\gtrsim k\mu_0^{-2}\vee \mu_0^{-4}$ up to log factors, with probability $1-o_m(1)$. 

\section{\label{overlapGapPropertyProof}Overlap Gap Property}

We start by considering the following restricted version of problem~\eqref{generalProblem00} which corresponds to searching for a $k'$-sparse vector $\bs x$ that is as close as possible to $\langle \bs A_i, \bs x_0\bs x_0^\intercal\rangle $ but with the additional condition that $\bs x_0$ has a subset $S$ of its entries in common with $\bs x$. 
\begin{align}
\begin{split}
\min_{\|\bs x\|_0\leq k'} \quad & \hat{R}_m^{kk'}(\bs x) = \frac{1}{m}\sum_{i=1}^m \left(\langle \bs A_i, \bs x\bs x^\intercal\rangle - \langle \bs A_i, \bs x_0\bs x_0^\intercal \rangle +\varepsilon_i \right)^2\\
s.t. \quad& \bs x\in \left\{0,1\right\}^n, \quad \supp \bs x \cap \supp \bs x_0 = S 
\end{split}\tag{$\Phi(S)$}\label{problemwithfixedSupport01}
\end{align} 
If we let $\ell = |S|$, we can also define the following subproblem
\begin{align}
\begin{split}
\min_{\|\bs x\|_0\leq k'}\quad  & \hat{R}_m^{kk'}(\bs x)= \frac{1}{m} \sum_{i=1}^m  \left(\langle \bs A_i, \bs x\bs x^\intercal\rangle - \langle \bs A_i, \bs x_0\bs x_0^\intercal\rangle +\varepsilon_i  \right)^2\\
\text{s.t. }\quad &\bs x\in \left\{0,1\right\}^n, \langle \bs x, \bs x_0\rangle = \ell
\end{split}\tag{$\Phi(\ell)$}\label{Problemtwokdistance2}
\end{align}
Problem~\eqref{Problemtwokdistance2} is similar to~\eqref{problemwithfixedSupport01} except that we do not explicitly specify the subset of common entries but merely require a fixed number of ($\ell = |S|$) common entries between $\bs x$ and $\bs x_0$.  We will also follow the notations in~\cite{gamarnik2017high, arous2023free} and denote the square root of the objective value at the optimum for problem~\eqref{problemwithfixedSupport01} (resp. problem~\eqref{Problemtwokdistance2}) as $\varphi_{kk'}(S)$ (resp. $\varphi_{kk'}(\ell)$).  I.e.  if we define $\mathcal{S}_{\ell}$ as 
\begin{align}
\mathcal{S}_{\ell} = \left\{\bs x \in \left\{0,1\right\}^n\; |\; \|\bs x\|_0 = k', \langle \bs x, \bs x_0\rangle = \ell \right\},
\end{align}
we have
\begin{align}
\varphi_{kk'}(\ell) = \min_{\bs x \in \mathcal{S}_\ell}\sqrt{\frac{1}{m} \sum_{i=1}^m  \left(\langle \bs A_i, \bs x\bs x^\intercal\rangle - \langle \bs A_i, \bs x_0\bs x_0^\intercal\rangle +\varepsilon_i  \right)^2}
\end{align}
We will start by showing the following lower bound on the empirical risk $\hat{R}_m(\bs x)$
\begin{proposition}\label{upperBoundOGP01}
Let $\alpha_k$ be such that $\lim_{k, m\rightarrow \infty} \frac{\alpha_k}{\log(k)} = +\infty$.  With probability $1-o_k(1)$,  we have 
\begin{align}
 \varphi_{kk'}(\ell) \geq \sqrt{((k')^2 + k^2 - 2\ell^2) \left(1 - 2\sqrt{\frac{\log(N_{k'\ell})+ \alpha_k}{m} }\right)} \label{finalLowerBound01}
\end{align}
\end{proposition}
If we let $\tilde{\Gamma}(\ell)$ (we reserve the notation $\Gamma$ for the classical Gamma function) to denote the curve 
\begin{align}
\tilde{\Gamma}(\ell) = \sqrt{((k')^2 +k^2 - 2\ell^2) \left(1 - 2\sqrt{\frac{\log(N_{k'\ell})+ \alpha_k}{m} }\right)}\label{definitionFirstMomentsCurve01}
\end{align}

we can say the following
\begin{proposition}\label{analysisFirstMomentsCurve}
Let $\tilde{\Gamma}(\ell)$ be defined as in~\eqref{definitionFirstMomentsCurve01}. Let us assume $k = o(m)$, as well as $(k')^{3/2}<\sqrt{m}k/\sqrt{\log(m)}$.  Then there exists $\ell_c \in [0,k]$ with $\ell_c = \frac{(k')^2}{\sqrt{mk'}}\sqrt{\log(n)}$ such that $\tilde{\Gamma}(\ell)$ is increasing for all $\ell<\ell_c$ and decreasing for all $\ell>\ell_c$. 
More importantly we have
\begin{align}
\min\left\{\tilde{\Gamma}(\ell_c) - \tilde{\Gamma}(1), \tilde{\Gamma}(\ell_c) - \tilde{\Gamma}(0)\right\} \geq \frac{k^2}{k'} \wedge \frac{k'\ell_c}{\sqrt{mk'}}\sqrt{\log(n-k)}
\end{align}
\end{proposition}

Finally we will prove the following upper bound 

\begin{proposition}\label{upperBoundSecondMoments}
Let $k'<k^2/\log(k)$, $\max \left\{k',k\right\} = o(m)$. Provided that $\ell_c = \frac{(k')^{3/2}}{\sqrt{m}} < k$, with probability $1-o_k(1)$,  we have 
\begin{align}
\varphi_{kk'}(\ell) < \tilde{\Gamma}(\ell) +  \frac{k'\log^2(k')}{\sqrt{m}} \label{finalUpperBound01}
\end{align}
\end{proposition}

Combining Proposition~\ref{upperBoundOGP01} with Propositions~\ref{analysisFirstMomentsCurve} and~\ref{upperBoundSecondMoments}, we get 
\begin{align}
\tilde{\Gamma}(\ell_c)< \varphi_{kk'}(\ell_c) < \tilde{\Gamma}(\ell_c)  + \frac{k'\log^2(k')}{\sqrt{m}}
\end{align}
\begin{align}
\tilde{\Gamma}(1)< \varphi_{kk'}(1)  < \tilde{\Gamma}(1)  + \frac{k'\log^2(k')}{\sqrt{m}}
\end{align}
\begin{align}
\tilde{\Gamma}(0)< \varphi_{kk'}(0)  < \tilde{\Gamma}(0)  + \frac{k'\log^2(k')}{\sqrt{m}}
\end{align}
Now from the discussion of section~\ref{sectionLowerBoundOverparametrization},  $\tilde{\Gamma}(\ell_c)$ also satisfies
\begin{align}
\tilde{\Gamma}(\ell_c) - \max\left\{\tilde{\Gamma}(0), \tilde{\Gamma}(1) \right\}> O\left( \frac{k^2}{k'} \wedge \frac{k'\ell_c}{\sqrt{mk'}}\sqrt{\log(n-k)}\right)
\end{align}
From this, as soon as $m,n,k\rightarrow \infty$, we get 
\begin{align}
\varphi_{kk'}(\ell_c) &> \max\left\{\varphi_{kk'}(1), \varphi_{kk'}(0)\right\} + O\left( \frac{k^2}{k'} \wedge \frac{k'\ell_c}{\sqrt{mk'}}\sqrt{\log(n-k)}\right)\\
& - O\left( \frac{k'\log^2(k')}{\sqrt{m}}\right) 
\end{align}
which implies for $(k')^{3/2}\asymp \sqrt{m}k/\log^{1/2}(n/k^2)$
\begin{align}
\varphi_{kk'}(\ell_c) &> \max\left\{\varphi_{kk'}(1), \varphi_{kk'}(0)\right\} + O\left( \frac{k^2}{k'} \wedge \frac{k'\ell_c}{\sqrt{mk'}}\sqrt{\log(n-k)}\right)
\end{align}


Since the analysis of section~\ref{overparametrizationSection} does not depend on the constant $\ell_c$, provided that $\ell_c = O(k)$,  for any sufficiently small $\delta$ and for $k$ large enough, we have
\begin{align}
\max\left\{\varphi_{kk'}((1-\delta)\ell_c), \varphi_{kk'}((1+\delta )\ell_c)\right\} \leq  \min_{\ell \in ((1-\delta)\ell_c, (1+\delta)\ell_c)} \varphi_{kk'}(\ell)
\end{align}

The conclusion follows from an application of Proposition~\ref{propositionBenArousOGP} below. 
\begin{proposition}[\label{propositionBenArousOGP}see Proposition 4.6 in~\cite{arous2023free}]
Suppose that for some overlap sizes $0\leq \ell_1\leq z_1<z_2-1<z_2\leq \ell_2\leq k$ it holds
\begin{align}
\max\left\{\varphi_{kk'}(\ell_1), \varphi_{kk'}(\ell_2)\right\}< \min_{\ell \in (z_1, z_2)}\varphi_{kk'}(\ell)
\end{align}
with high probability. Then Problem~\eqref{overparametrizedProblem} exhibits the $k'$-OGP with $\zeta_{1n} = z_1$ and $\zeta_{2n} = z_2$. 
\end{proposition}

\subsection{\label{sectionLowerBoundOverparametrization}Lower bound (first moments method)}

{

%
%

The total number of vectors $\bs x$ having an overlap of size $\ell$ with the grountruth si given by
\begin{align}
N_{k, \ell} = {k\choose \ell} {n-k \choose k-\ell}
\end{align}
Recall that we defined $\tilde{\Gamma}$ as 
\begin{align}
\tilde{\Gamma}^2(\ell) &= 2(k^2 - \ell^2) \left(1 - 2\sqrt{\frac{1}{m}\left[\log\left({k\choose \ell} {n-k \choose k-\ell}\right) + \alpha_k\right]}\right)\\
& = 2(k^2 - \ell^2) \left(1 - 2\sqrt{\frac{1}{m}\left[\log\left(N_{k, \ell}\right) + \alpha_k\right]}\right)\label{definitionGammaEll}
\end{align}
We let $\Lambda = \left\{\supp \bs x \times \supp\bs x\right\} \setminus \left\{S\times S\right\}$ and $\Lambda_0 = \left\{\supp \bs x_0 \times \supp \bs x_0\right\} \setminus \left\{S \times S\right\}$.  For any fixed $\bs x$, we let
\begin{align}
Z_i&  =\langle \bs A_i, \bs x\bs x^\intercal\rangle - \langle \bs A_i ,\bs x_0\bs x_0^\intercal\rangle + \varepsilon_i \\
&=\sum_{(\ell_1, \ell_2) \in \Lambda} (\bs A_i)_{\ell_1, \ell_2} \bs x_{\ell_1}\bs x_{\ell_2} - \sum_{(\ell_1', \ell_2') \in \Lambda_0} (\bs A_i)_{\ell_1, \ell_2}(\bs x_0)_{\ell_1'}(\bs x_0)_{\ell_2'} + \varepsilon_i
\end{align}
The variable $Z_i$
\begin{align}
Z_i&=\sum_{(\ell_1, \ell_2) \in \Lambda} (\bs A_i)_{\ell_1, \ell_2}- \sum_{(\ell_1', \ell_2') \in \Lambda_0} (\bs A_i)_{\ell_1, \ell_2}+ \varepsilon_i
\end{align} 
is a  sum of $(k^2 - |S|^2)+ (k^2 - |S|^2) + 1$ independent random Gaussian variables from which we have, 
\begin{align}
\mathbb{E}\left\{\frac{1}{m}\sum_{i=1}^m Z_i^2\right\} = (k^2 - |S|^2)+ (k^2 - |S|^2) + \sigma^2
\end{align}
Letting $v_Z = \sqrt{ (k^2 - |S|^2)+ (k^2 - |S|^2)  + \sigma^2}$ as well as $\tilde{Z}_i = Z_i/v_Z$, the sum 
\begin{align}
m\tilde{Z} = \sum_{i=1}^m \tilde{Z}_i^2 
\end{align}
can thus be viewed as a chi-squared random variable with $m$ degrees of freedom.  Using Lemma 1 in~\cite{laurent2000adaptive} or equivalently a Chernoff bound, we have 
\begin{align}
P \left(\sum_{i=1}^m \tilde{Z}_i^2 < m - 2\sqrt{m}t\right) \leq e^{-t}
\end{align}
which implies that for any fixed solution $\bs x$ having an overlap of size $|S|$ with the planted signal $\bs x_0$, we can write 
\begin{align}
P \left(\hat{R}_m(\bs x) < mv_Z^2  - 2v_Z^2 \sqrt{mt}\right) \leq e^{-t}
\end{align}
This in turn implies 
\begin{align}
P\left(\varphi_{kk}^2(\ell) < mv_Z^2  - 2v_Z^2 \sqrt{mt}\right)& = P\left(\bigcup_{\substack{\|\bs x\|_0 = k' , \langle \bs x, \bs x_0\rangle = \ell }}\hat{R}_m(\bs x) < mv_Z^2  - 2v_Z^2 \sqrt{mt} \right)\\
&\leq {k \choose \ell} {n-k \choose k-\ell} P \left(\hat{R}_m(\bs x) < mv_Z^2  - 2v_Z^2 \sqrt{mt}\right) \\
&\leq {k \choose \ell} {n-k \choose k-\ell} e^{-t}
\end{align}
Taking $t = \log({n \choose \ell} {n-k \choose k-\ell}) + \alpha_k$, we get 
\begin{align}
P \left(\varphi_{kk}(\ell)< \tilde{\Gamma}(\ell)\right) &< N_{\ell,k} \exp\left( -\log N_{\ell,k} - \alpha_k\right)\\
& \leq e^{-\alpha_k}
\end{align}
so that taking $\alpha_k = \log(k)$ we recover 
\begin{align}
P \left(\varphi_{kk}(\ell)> \tilde{\Gamma}(\ell)\right)  > 1 -  o_k(1)
\end{align}
Using a similar reasonining for the misparametrized problem, i.e. $k'>k$, and letting $v_Z^2 =  (k')^2 + k^2 - 2\ell^2$, as well as 
\begin{align}
\tilde{\Gamma}^2(\ell) = \left((k')^2 + k^2 - 2\ell^2\right) \left(1 - 2\sqrt{\frac{1}{m}\left[\log\left({k\choose \ell} {n-k \choose k'-\ell}\right)  + \alpha_k\right]}\right)
\end{align}
we can write 
\begin{align}
P \left(\varphi_{kk'}(\ell)> \tilde{\Gamma}(\ell)\right)  > 1 -  o_k(1)
\end{align}

\subsection{Proof of Proposition~\ref{analysisFirstMomentsCurve}. Analysis of the first moments curve\label{analysisFirstMomentsCurveAnalysis}}

We start by providing a charaterization of the first moments furve $\tilde{\Gamma}(\ell)$ for the original problem.  We then study the misparametrized setting separately in section~\ref{overparametrizationSection}. Using~\eqref{definitionGammaEll}, we have
\begin{align}
\begin{split}
\tilde{\Gamma}^2(\ell+1)& = \left(2k^2 -2(\ell+1)^2\right) \left(1 - \sqrt{\frac{1}{m} \log N_{\ell+1, k}}\right)\\
\tilde{\Gamma}^2(\ell)& = (2k^2- 2\ell^2) \left(1 - \sqrt{\frac{1}{m} \log N_{\ell,k}}\right)
\end{split}\label{definitionGammaCurve01}
\end{align}
Taking the difference, we have
\begin{align}
\tilde{\Gamma}^2(\ell+1) - \tilde{\Gamma}^2(\ell) =& \left(2k^2 -2(\ell+1)^2 \right) \left(1 - 2\sqrt{\frac{1}{m} \log N_{\ell+1, k}}\right) \\
& -  (2k^2 - 2\ell^2) \left(1 - 2\sqrt{\frac{1}{m} \log N_{\ell+1, k}}\right)\\
 = &-2 - 4\ell  - 2\frac{2k^2}{\sqrt{m}} \left(\sqrt{\log N_{\ell+1, k}} - \sqrt{\log N_{\ell k}}\right) \label{decompositionTermDifferenceGamma01}\\
& + 4 \frac{\ell^2 }{\sqrt{m}} \left(\sqrt{\log N_{\ell+1, k}} - \sqrt{\log N_{\ell k}}\right)\label{decompositionTermDifferenceGamma02}\\
& + 4  \sqrt{\frac{1}{m} \log N_{\ell+1,k}}+ 8\ell \sqrt{\frac{1}{m} \log N_{\ell+1,k}}
\end{align}
For~\eqref{decompositionTermDifferenceGamma01} and~\eqref{decompositionTermDifferenceGamma02} we follow~\cite{arous2023free} (see section 5 in particular). 

\begin{align}
{k \choose \ell+1}/ {k \choose \ell}& = \frac{k!}{(k - \ell+1)! (\ell+1)!}  \frac{(k-\ell)!\ell!}{k!} = \frac{(k - \ell)}{(\ell+1)}\\
{n-k \choose k - \ell-1}/ {n-k \choose k - \ell} &= \frac{(n-k)!}{(n-2k+ \ell + 1)! (k - \ell-1)!} \frac{(k-\ell)! (n-2k+ \ell)!}{(n-k)!}\\
& = \frac{(k-\ell)}{n-2k+\ell+1}
\end{align}
As a result,  we can write 
\begin{align}
-\left(\sqrt{\log(N_{\ell+1,k})} - \sqrt{\log(N_{\ell,k})}\right)& = \frac{\log\frac{(\ell+1)(n-2k+\ell+1)}{(k-\ell)^2}}{\sqrt{\log N_{\ell+1,k}} + \sqrt{\log N_{\ell,k}}}\label{ratioLog001}
\end{align}
On the other hand, provided that $k = o(n)$, $\ell = o(\min\left\{k, k'\right\})$,  and $k=o(k')$, we have
\begin{align}
\log\left[{k\choose \ell} {n-k\choose k - \ell}\right] &= (1+o(1)) \left[\ell \log \left(\frac{k}{\ell}\right) + (k - \ell) \log\left(\frac{n-k}{k - \ell}\right)\right]\\
& = (1+o(1)) k \log\left(\frac{n}{k}\right).
\end{align}
Substituting this in~\eqref{ratioLog001} gives 
\begin{align}
&-\left(\sqrt{\log(N_{\ell+1,k})} - \sqrt{\log(N_{\ell,k})}\right)\\
& = (1+o(1)) \frac{1}{\sqrt{mk\log(n/k)}}\log\frac{(\ell+1)(n-2k+\ell+1)}{(k-\ell)^2}\label{ratioLog002}
\end{align}
Now note that for $k = o(n)$,  we have 
\begin{align}
\log \frac{(\ell+1)(n-2k+\ell+1)}{(k-\ell)^2} = \left[\log\left(\frac{(\ell+1)n }{k^2}\right) + \log\left(\frac{n-2k+\ell+1}{n}\right) - \log\left(\frac{(k-\ell)^2}{k^2}\right) \right]
\end{align}
where provided that $\ell=o(k)$
\begin{align}
\log\left(\frac{n-2k+\ell+1}{n}\right) - 2\log\left(\frac{k-\ell}{k}\right) = o(1)
\end{align}
Substituting this in~\eqref{ratioLog002} we get 
\begin{align}
-\left(\sqrt{\log(N_{\ell+1,k})} - \sqrt{\log(N_{\ell,k})}\right) = (1+o(1))\sqrt{\frac{1}{k m\log(n/k)}} \log\left(\frac{n\ell}{k^2}\right)
\end{align}
Finally combining this with~\eqref{decompositionTermDifferenceGamma01} and~\eqref{decompositionTermDifferenceGamma02} we obtain
\begin{align}
\begin{split}
\left(\tilde{\Gamma}^2(\ell+1) - \tilde{\Gamma}^2(\ell)\right)  = &-2 - 4\ell + \frac{2k^2}{\sqrt{m k \log(\frac{n}{k})}} \log\left(\frac{(\ell+1)n}{k^2}\right) \\
& -2 \frac{\ell^2}{\sqrt{m k  \log(\frac{n}{k})}} \log\left(\frac{(\ell+1)n}{k^2}\right)  \\
& + 4 (1+o(1)) \sqrt{\frac{k \log(n/k) }{m}} \\
&+ 8\ell (1+o(1))\sqrt{\frac{k\log(n/k)}{m}  }
\end{split}\label{analysisGamma0003}
\end{align}

When $k\ll m$ we can neglect the last two terms and write 
\begin{align}
\left[\tilde{\Gamma}^2(\ell+1) - \tilde{\Gamma}^2(\ell)\right]& = -2-4\ell + \frac{2k^2}{\sqrt{mk} \log(n/k)} \log\left(\frac{(\ell+1)n}{k^2}\right)\\
&- \frac{2\ell^2}{\sqrt{mk} \log(n/k)} \log\left(\frac{(\ell+1)n}{k^2}\right)
\end{align}
Solving for $\ell$, we get 
\begin{align}
-2 - 4\ell + \left(2k^2 - 2\ell^2\right)\beta = 0
\end{align}
where 
\begin{align}
\beta= \frac{\log(\frac{(\ell+1)n}{k^2})}{\sqrt{mk\log(n/k)}}
\end{align}
we get 
\begin{align}
\zeta_{1,2} = \frac{4\pm \sqrt{16 + 4 \left(-2 + 2k^2\beta\right)2\beta}}{-4\beta}
\end{align}
In particular, assuming $k^2 = o(k m)$ as well as $k^2 = O(\sqrt{mk})$,  taking the negative root, and using the Taylor expansion for the square root, we get
\begin{align}
\zeta_1 = \frac{4 - \sqrt{16 + 4(-2 + 2k^2 \beta) 2\beta}}{-4\beta} \asymp \frac{k^2}{\sqrt{km}}
\end{align}
which for the barrier to fall in $[0, k]$ requires $\frac{1}{m}<1$. The parabola has negative curvature and we focus on the largest root. This implies $\Gamma(\ell+1)>\Gamma(\ell)$ for $\ell<\ell_c$ and $\Gamma(\ell+1)<\Gamma(\ell)$ for $\ell>\ell_c$.

Let $\ell_c = c k$ with $0< c<1$, and note that we have 
\begin{align}
\tilde{\Gamma}(0) &= k \left(1 - 2\sqrt{\frac{ k\log(n)}{m}}\right)^{1/2}\label{ValuesFirstMomentCurve01}\\
\tilde{\Gamma}(1)& =0\\
\tilde{\Gamma}(\ell_c)& =\sqrt{(1-c)} k\left(1 - 2\sqrt{\frac{(k - c k)\log(n)}{m}  }\right)^{1/2}\label{ValuesFirstMomentCurve03}
\end{align}

%
On the other hand, we have
\begin{align}
\left(1 - 2\sqrt{\frac{(k- c k) \log(n)}{m}}\right)^{1/2} &= (1+o(1)) \left(1 - 2\sqrt{\frac{k \log(n)}{m}} + \frac{2ck\sqrt{\log(n)}}{\sqrt{m}\sqrt{k}}\right)^{1/2}\\
&=\left(1 - 2\sqrt{\frac{k \log(n)}{m}}\right)^{1/2} + (1+o(1))\frac{2ck\sqrt{\log(n)}}{\sqrt{m}\sqrt{k}}
\end{align}
Using $k^2<k m$ and $k^2>\sqrt{mk}$, we get 
\begin{align}
\tilde{\Gamma}(\ell_c)& = \frac{2kk_c}{\sqrt{m k}} 
\end{align}
Substituting this in~\eqref{ValuesFirstMomentCurve01} to~\eqref{ValuesFirstMomentCurve03}
\begin{align}
\min\left\{\tilde{\Gamma}(\ell_c) - \tilde{\Gamma}(1), \tilde{\Gamma}(\ell_c) - \tilde{\Gamma}(0)\right\}&  \geq  \frac{kk_c}{\sqrt{mk}} \wedge k
\end{align}
Taking $k_c = k^2/\sqrt{k'm} =  \sqrt{k}, k< m =  o( k^2)$, we get
\begin{align}
\min\left\{\tilde{\Gamma}(\ell_c) - \tilde{\Gamma}(1), \tilde{\Gamma}(\ell_c) - \tilde{\Gamma}(0)\right\} \geq \frac{k^3}{mk} \wedge k
\end{align}

\subsection{\label{overparametrizationSection}Overparametrization}

In the overparametrized regime,  going back to~\eqref{definitionGammaCurve01} and~\eqref{analysisGamma0003}, we have 
\begin{align}
\tilde{\Gamma}^2(\ell+1) & = \left((k')^2 + k^2 - 2(\ell+1)^2\right)\left(1 - \sqrt{\frac{N_{\ell+1, k'}}{m}}\right)\\
\tilde{\Gamma}^2(\ell) & = \left((k')^2 + k^2 - 2\ell^2\right)\left(1 - \sqrt{\frac{N_{\ell, k'}}{m}}\right)
\end{align}
Applying the same reasoning as above, 
\begin{align}
\left(\tilde{\Gamma}^2(\ell+1) - \tilde{\Gamma}^2(\ell)\right)  = &-2 - 4\ell + \frac{((k')^2 + k^2)}{\sqrt{m k' \log(\frac{n}{k'})}} \log\left(\frac{(\ell+1)n}{k^2}\right) \\
& -2 \frac{\ell^2}{\sqrt{m k'  \log(\frac{n}{k})}} \log\left(\frac{(\ell+1)n}{k^2}\right)  \\
& + 4 (1+o(1)) \sqrt{\frac{k' \log(n/k) }{m}} \\
&+ 8\ell (1+o(1))\sqrt{\frac{k'\log(n/k)}{m}  }
\end{align}
If we assume $k'<m$ with $(k')^2/\sqrt{mk'}<k$, we can neglect the last two terms.  Focusing on the first three terms, we get the solution
\begin{align}
\zeta_{1,2} = \frac{4\pm \sqrt{16 + 4 \left(-2 + 2((k')^2+k^2)\beta'\right)2\beta'}}{-4\beta'}
\end{align}
where $\beta' = \log^{1/2}(n/k^2)/\sqrt{mk'}$. In particular, assuming $(k')^2 = o(k' m)$ (which is always true for informative $k'$), taking the positive root, and using the Taylor expansion for the square root, we get
\begin{align}
\zeta_1 = \frac{4 - \sqrt{16 +4(-2 + ((k')^2 + k^2) \beta') 2\beta'}}{-4\beta} \asymp \frac{((k')^2 + k^2)}{\sqrt{mk'}}
\end{align}
which for the barrier to fall in $[0, k]$ requires $\frac{((k')^2 + k^2)\log^{1/2}(n/k^2)}{\sqrt{k'm}}<k$ or equivalently $(k')^{3/2}<\sqrt{m}k/\log^{1/2}(n/k^2)$

In particular, if we choose $(k')^{3/2} = \Theta\left( \sqrt{m}k/\log^{1/2}(n/k^2)\right)$, for $k'$ to be informative, we need
\begin{align}
k<\frac{m^{1/3} k^{2/3}}{\log^{1/3}(n/k')}<m
\end{align}

 The parabola has negative curvature and we focus on the largest root. This implies $\Gamma(\ell+1)>\Gamma(\ell)$ for $\ell<\ell_c$ and $\Gamma(\ell+1)<\Gamma(\ell)$ for $\ell>\ell_c$

From this, using 
\begin{align}
\sqrt{\frac{k' - \ell_c}{m}} \approx \sqrt{\frac{k'}{m}} - \frac{\ell_c}{\sqrt{ m k'}}\\
\sqrt{\frac{k' - \ell_c}{m}} \approx \sqrt{\frac{k'-k}{m}} + \frac{k -  \ell_c}{\sqrt{m k'}}\\
\end{align}

\begin{align}
&\sqrt{(k')^2 + (1-2(\zeta_c)^2)k^2} \left(1 - \sqrt{\frac{(k'- \ell_c)\log(n-k)}{m}}\right)^{1/2}\\
&\approx \sqrt{k^2 + (k')^2} \left(1 - \sqrt{\frac{(k'- \ell_c)\log(n)}{m}}\right)^{1/2}\\
& - 2(k') \frac{\zeta_c^2k^2}{(k')^2 + k^2} \sqrt{\log(n-k)}\left(1 - \sqrt{\frac{(k'- \ell_c)\log(n)}{m}}\right)^{1/2}\\
&\sqrt{(k')^2 + (1-2(\zeta_c)^2)k^2} \left(1 - \sqrt{\frac{(k'-\ell_c)\log(n-k)}{m}}\right)^{1/2}\\
&\approx \sqrt{(k')^2 - k^2} \left(1 - \sqrt{\frac{(k'- \ell_c)\log(n)}{m}}\right)^{1/2}\\
&+ 2(k') \frac{(1-\zeta_c^2)k^2}{(k')^2 + k^2} \left(1 - \sqrt{\frac{(k'- \ell_c)\log(n)}{m}}\right)^{1/2}
\end{align}
We have
\begin{align}
\min\left\{\tilde{\Gamma}(\ell_c) - \tilde{\Gamma}(0), \tilde{\Gamma}(\ell_c) - \tilde{\Gamma}(1)\right\} \geq \frac{k^2}{k'} \wedge \frac{k'\ell_c}{\sqrt{mk'}}\sqrt{\log(n-k)}
\end{align}

\subsection{Proof of Proposition~\ref{upperBoundSecondMoments} (Second moments method)}

We now proceed with the proof of Proposition~\ref{upperBoundSecondMoments}.  Recall that we have
\begin{align}
\mathcal{S}_{\ell} = \left\{\bs x \in \left\{0,1\right\}^n\; |\; \|\bs x\|_0 = k, \langle \bs x, \bs x_0\rangle = \ell \right\}
\end{align}
Recall that our problem reads as
\begin{align}
\begin{split}
\min\quad  & \hat{R}_m^{kk'}(\bs x)= \frac{1}{m} \sum_{i=1}^m  \left(\langle \bs A_i, \bs x\bs x^\intercal\rangle - \langle \bs A_i, \bs x_0\bs x_0^\intercal\rangle +\varepsilon_i  \right)^2\\
\text{s.t. }\quad &\|\bs x\|_0 = k', \;   \bs x\in \left\{0,1\right\}^n, \langle \bs x, \bs x_0\rangle = \ell
\end{split}\tag{$\Phi(\ell)$}\label{Problemtwokdistance2bis}
\end{align}
We first focus on the pure noise model, setting $\bs x_0 = 0$ (the translation to problem~\eqref{Problemtwokdistance2bis} can be done by removing the overlap between $\bs x$ and $\bs x_0$ and setting $(k')^2 \leftarrow (k')^2 - \ell^2$ together with $\sigma^2 = k^2 - \ell^2 \leq k^2$ as explained in~\cite{gamarnik2017high})
\begin{align}
\begin{split}
\min\quad  & \hat{R}_m^{kk'}(\bs x)= \frac{1}{m} \sum_{i=1}^m  \left(\langle \bs A_i, \bs x\bs x^\intercal\rangle +\varepsilon_i  \right)^2\\
\text{s.t. }\quad &\|\bs x\|_0 = k', \;   \bs x\in \left\{0,1\right\}^n,
\end{split}\label{Problemtwokdistance2bisPureNoise}
\end{align}
where $\varepsilon_i\sim \mathcal{N}(0, \sigma^2)$ is independent of $\left\{\bs A_i\right\}_{i=1}^m$.  

We consider the cardinality
\begin{align}
Z_{s} =\left| \left\{\bs x\;|\; \|\bs x\|_0= k',\frac{1}{m}\sum_{i=1}^m |\langle \bs A_i, \bs x\bs x^\intercal\rangle - Y_i|^2 < ((k')^2 + \sigma^2)\left(1 - 2\sqrt{\frac{1}{m}\log N_{\ell, k}}\right) \right\}\right|\label{definitionSetS}
\end{align}
Following~\cite{gamarnik2017high, arous2023free,gamarnik2019landscape} we will control this cardinality through a second moment method. We let $s$ to denote the bound on the RHS of the inequality in~\eqref{definitionSetS} and use $\bs Z, \tilde{\bs Z}$ where $\bs Z = (Z_1, Z_2, \ldots, Z_m)$, $\tilde{\bs Z} = (\tilde{Z}_1, \tilde{Z}_2, \ldots, \tilde{Z}_m)$ and $Z_i = |\langle \bs A_i, \bs x\bs x^\intercal\rangle - Y_i|^2, \tilde{Z}_i = |\langle \bs A_i, \tilde{\bs x}\tilde{\bs x}^\intercal \rangle- Y_i|^2 \sim \chi^2_2$ with correlation $(\ell^2 + \sigma^2)/((k')^2+ \sigma^2)$ where $\ell$ is used to denote the overlap between the supports of $\bs x$ in $Z_i$ and $\tilde{\bs x}$ in $\tilde{Z}_i$. We let
\begin{align}
\mathcal{S}_{k'} = \left\{\bs x \;|\; \bs x \in \left\{0,1\right\}^n, \;  \|\bs x\|_0 = k'\right\}
\end{align}
\begin{align}
Z_s^2 = \left|\left\{\bs x, \tilde{\bs x} \in \mathcal{S}_{k'} \;|\; \hat{R}_m(\bs x)\vee \hat{R}_m(\tilde{\bs x}) < ((k')^2 + \sigma^2)\left(1 - 2\sqrt{\frac{\log N_{\ell,k}}{m}}\right)\right\} \right|\label{definitionZsSquared}
\end{align}
Following~\cite{arous2023free, gamarnik2017high}, we decompose the set in~\eqref{definitionZsSquared} according to the correlation between the variables $\bs Z$ and $\tilde{\bs Z}$.  For any fixed $\bs x, \tilde{\bs x}$, we define the events
\begin{align}
E_{\ell}(\bs x, \tilde{\bs x}) &= \left\{\left.\hat{R}_m(\bs x)\vee \hat{R}_m(\tilde{\bs x}) < ((k')^2 + \sigma^2)\left(1 - 2\sqrt{\frac{\log N_{\ell,k}}{m}}\right)\right| \text{corr}\left(Z_i, \tilde{Z}_i\right) = \frac{\ell^2 +\sigma^2}{(k')^2 + \sigma^2} \right\}\\
E(\bs x)& = \left\{\hat{R}_m(\bs x) < ((k')^2 + \sigma^2)\left(1 - 2\sqrt{\frac{\log N_{\ell,k}}{m}} \right)\right\}
\end{align} 
\begin{align}
\frac{\mathbb{E}\left\{Z_{s}^2\right\}}{\mathbb{E}\left\{Z_{s} \right\}^2}  = \sum_{\ell = 0}^{k'} \frac{{n \choose k'-\ell, k'-\ell, \ell, n-2k'+\ell}}{{n\choose k'}^2} \frac{P\left(E_{\ell}(\bs x, \tilde{\bs x})\right)}{P^2\left(E(\bs  x)\right)}\label{tmpratioRho}
\end{align}
Using 
\begin{align}
{n \choose k'-\ell, k'-\ell, \ell, n-2k'+\ell}/{n \choose k'}^2 = \frac{{n-k' \choose k'-\ell} {k' \choose \ell}}{{n\choose k'}} &\leq \frac{{n-k' \choose k'-\ell} {k'\choose \ell} 4 k'!}{n^{k'}}\\
&\leq  \frac{(n-k')^{(k'-\ell)}}{(k'-\ell)!} \frac{(k')^\ell}{\ell !} \frac{1}{n^{k'}} 4(k')!\\
& \leq \frac{4(k')^{2\ell}}{n^{\ell}} \label{boundBinomial00}
\end{align}

We will decompose the sum in~\eqref{tmpratioRho} into a ``small correlation" contribution and a``large correlation" contribution.  When the variables are strongly correlated, we will use
\begin{align}
\frac{P(E_\ell(\bs x, \tilde{\bs x}))}{P^2(E(\bs x))} \leq P^{-1}(E(\bs x))
\end{align}
To control $P^{-1}(E(\bs x))$, we use a moderate deviation principle.  Consider Proposition~\ref{lowerBoundChiSquared01} below (see Corollary 3 in~\cite{zhang2020non})
\begin{proposition}\label{lowerBoundChiSquared01}
Suppose $Y \sim \chi_m^2$ and $X = Y-m$ for integer $m\geq 1$.  There exist uniform constants $C, c>0$ and a constant $C_\varepsilon>0$ that only depends on $\varepsilon$, such that
\begin{align}
P\left(X \geq x\right) \geq c\exp\left( - C x \wedge \frac{x^2}{m}\right), \quad \forall x>0\\
P\left(X\leq -x\right) \left\{\begin{array}{ll}
\geq c \exp\left(-\frac{C_\varepsilon x^2}{m}\right)& \forall 0<x<(1-\varepsilon) m\\
 = 0 & x\geq m 
\end{array}\right.\label{lowerBoundTailCepsilon}
\end{align}\label{nonasymptoticBoundChiSquaredLower}
\end{proposition}
Let us assume for now that we can take $C_\varepsilon = 1/2$ in~\eqref{lowerBoundTailCepsilon}. We will show later that this is true for $k', m$ large enough.  For $C_\varepsilon = 1/2$, note that one can write
\begin{align}
\begin{split}
\sum_{\ell= \left(1+C_\varepsilon  - 3/4\right)k'}^{k'} \frac{(n-k')^{(k' - \ell)} {k'}^{\ell}}{(k' - \ell)! \ell !} \frac{1}{n^{k'}} 4k'! e^{C_\varepsilon k'\log(n-k)}& \leq \sum_{\ell = \left(1+C_\varepsilon  - 3/4\right)k'}^{k'} \frac{(n-k')^{(k' - \ell)}(n-k)^{C_\varepsilon k'} (k')^{2k'}}{n^{k'}}\\
& \leq \sum_{\ell=\left(1+C_\varepsilon  - 3/4\right)k'}^{k'}\frac{{k'}^{2k'}}{n^{k'/4}} < (3/4 - C_\varepsilon)k'
\end{split}\label{largeValuesL}
\end{align}

We now show why the choice $C_\varepsilon = 1/2$ can be used provided that $k'$ and $m$ are sufficiently large.  Note that if we let $b_m = \sqrt{k' m \log(n)}$, for any $k'=o(m)$ we have $\sqrt{k'm \log(m)}/m\rightarrow 0$ but $\sqrt{k'm\log(n)}/\sqrt{m} \rightarrow \infty$.  This precisely satisfies the definition of a moderate deviation principle (see Definition~\ref{definitionMDP} below)

\begin{definition}[\label{definitionMDP}MDP]
Given a sequence of random variables $\left\{X_m\right\}_{m=1}^M$ with $\left\{b_m\right\}$ a positive sequence such that $b_m/m^{1/2}\rightarrow \infty$ as well as $b_m/m \rightarrow 0$, we say that the sum $S_m = \sum_{i=1}^m X_m$ satisfies a moderate deviation principle (MDP) with good rate function $I(x)$ such that $I(x)>0$ for all $x\neq 0$ if 
\begin{align}
\limsup\limits_{n\rightarrow\infty}\frac{n}{b_n^2} \log P\left\{S_n/b_n \in F\right\} \leq -\inf_{x\in F} I(x)\quad \text{for $F$ closed}\label{boundMDPupper}\\
\liminf \limits_{n\rightarrow\infty}\frac{n}{b_n^2} \log P\left\{S_n/b_n \in G\right\} \geq -\inf_{x\in G} I(x)\quad \text{for $G$ open}\label{boundMDPlower}
\end{align}
\end{definition}

As shown by Theorem~\ref{necessarySufficientMDP} below (see Theorem 2.2. in~\cite{eichelsbacher2003moderate}. Also see the work of de Acosta~\cite{de1992moderate})  for the sum $\frac{1}{b_n}\sum_{i=1}^n X_i$ to satisfy an MDP, it is necessary and sufficient for the ratio $\frac{n}{b_n^2} \log(n P(|X_1|>b_n))$ to diverge to $-\infty$. 

\begin{theorem}[\label{necessarySufficientMDP}Theorem 2.2. in~\cite{eichelsbacher2003moderate}]
Let $X_1, \ldots,\ldots$ be i.i.d. real valued random variables and $b_m$ be an increasing sequence of numbers with $b_m/\sqrt{m} \rightarrow \infty$ and $b_m/m \rightarrow 0$. The following are equivalent
\begin{enumerate}
\item The random variables $X_i$ satisfy $\mathbb{E}X_i = 0$ and 
\begin{align}
\limsup\limits_{n\rightarrow\infty} \frac{m}{b_m^2} \log\left[m P(|X_1|>b_m)\right] = -\infty
\end{align}
\item The sequence $\frac{1}{b_m}\sum_{i=1}^m X_i$ satisfy a moderate deviation principle with good rate function $I$ such that $I(x)>0$ for all $x\neq 0$ and $\lim_{x\rightarrow \infty} I(x) = \lim_{x\rightarrow -\infty} I(x) = \infty$. 
\end{enumerate}
\end{theorem}

In the case of a chi-squared random variable $X_1' = X_1 - \mathbb{E}X_1$, noting that $X_1$ is subexponential with parameters $(\nu, \alpha) = (2, 4)$ (see Example 2.8 in~\cite{wainwright2019high}) applying Proposition~\ref{subexponentialTailBound}, we get 
\begin{align}
P\left(|X_1'|> \sqrt{k' m\log(n)}\right)\leq e^{-\sqrt{k'm\log(n)}/4} 
\end{align}
Substituting in the necessary and sufficient condition of Theorem~\ref{necessarySufficientMDP}, we get 
\begin{align}
\lim_{m\rightarrow \infty}\frac{m}{k'm} \log(m P\left(|X_1'|>\sqrt{k'm\log(n)}\right)) \leq \lim_{m\rightarrow \infty}\left( \frac{\log(m)}{k'} - \frac{\sqrt{mk' \log(n)}}{k'} \right)= -\infty
\end{align}
as soon as $m>k'\log^2(m)$.

Applying this with $x = \sqrt{k'm \log(n-k)}$ and using the rate function $I(t) = \frac{t^2}{2\mathbb{E}X_i^2}$ (again see~\cite{eichelsbacher2003moderate}), for sufficiently large $k', m$, shows that we can use $C_\varepsilon = 1/2$ in~\eqref{largeValuesL}.

For the smaller values of $\ell$, we use the joint pdf of two correlated chi-squared with $m$ degrees of freedom~\cite{joarder2009moments}. Let $\left\{X_{1i}\right\}_{i=1}^m$, $\left\{X_{2,i}\right\}_{i=1}^m$ denote two sets of independent Gaussian random variables with correlation $\rho$ and variance $\sigma^2 = 1$. We define the variables $U_1$ and $U_2$ as 
\begin{align}
U_k& = \sum_{j=1}^m X_{kj}^2 
\end{align}
We further let $\rho = \ell/k$ to denote the correlation coefficient between $X_{1j}$ and $X_{2j}$.  Then the joint distribution of $U_1$ and $U_2$ can read as 
\begin{align}
f_\rho(u_1, u_2) = \frac{2^{-(m+1)}(u_1u_2)^{(m-2)/2}e^{-\frac{(u_1 + u_2)}{2(1-\rho^2)}}}{\sqrt{\pi} \Gamma\left(\frac{m}{2}\right)(1-\rho^2)^{m/2}} \sum_{k=0}^\infty \left[1+(-1)^k\right] \left(\frac{\rho \sqrt{u_1u_2}}{1-\rho^2}\right)^k \frac{\Gamma\left(\frac{k+1}{2}\right)}{k! \Gamma\left(\frac{k+m}{2}\right)} \label{densityCorrelatedChiSquared}
\end{align}
For the infinite series, 
\begin{align}
\sum_{k=0}^\infty [1+(-1)^k] \left(\frac{\rho\sqrt{u_1u_2}}{1-\rho^2}\right)^k \frac{\Gamma(\frac{k+1}{2})}{k!\Gamma(\frac{k+m}{2})}\label{correlatedChiSquared00}
\end{align}
Using the upper and lower bounds on the factorial and $\Gamma$ function, we have
\begin{align}
\frac{\Gamma(\frac{k+1}{2})}{k!} &\leq \frac{1}{e }\left(\frac{e}{k}\right)^k \left(\frac{k+1}{2}\right)^{\frac{k-1}{2}} \\
&\leq \frac{1}{e}\left( \frac{1}{2}\right)^{k/2} \left( \frac{k+1}{2}\right)^{-1/2} \left(\frac{e^2}{k}\right)^{k/2} \left(\frac{k+1}{ k}\right)^{k/2}\\
&\leq \left( \frac{1}{2}\right)^{k/2} e^1 \frac{1}{e} \left(\frac{k+1}{2}\right)^{-1/2} \left(\frac{e^2}{k}\right)^{k/2} \label{approximationFactorial01}
\end{align}
where we used the logarithm identity $1/(1+x)\leq \log(1+1/x)\leq 1/x$.

We can in fact get a tighter bound using Stirling's approximation. We have 
\begin{align}
\frac{\Gamma\left(\frac{k+1}{2}\right)}{k!} \sim \left(\frac{e}{2}\right)^{k/2} \left(\frac{1}{k}\right)^{k/2}
\end{align}

On the other hand, using bounds on the ratio of $\Gamma$ functions, for $m$ sufficiently large, one can write (using Stirling's approximation)
\begin{align}
\frac{\Gamma(\frac{k+m}{2})}{\Gamma(m/2)} \sim \left( \frac{m}{2}\right)^{k/2} \label{ratioGammaFunctions02}
\end{align}
substituting~\eqref{ratioGammaFunctions02} and~\eqref{approximationFactorial01} in~\eqref{correlatedChiSquared00} gives for any $\rho<1$,
\begin{align}
&\sum_{k=0}^\infty [1+(-1)^k] \left(\frac{\rho\sqrt{u_1u_2}}{1-\rho^2}\right)^k \frac{\Gamma(\frac{k+1}{2})}{k!\Gamma(\frac{k+m}{2})}\\
 &\leq \frac{1}{\Gamma\left(m/2\right)}\sum_{k=0}^\infty \left[1 + (-1)^k\right] \left(\frac{\rho\sqrt{u_1u_2}}{1-\rho^2}\right)^k \left(\frac{e}{2k}\right)^{k/2} \left(\frac{2}{m}\right)^{k/2}\\
&\leq  \frac{1}{\Gamma\left(m/2\right)}\sum_{k=0}^\infty \left[1 + (-1)^k\right] \left(\frac{\rho\sqrt{u_1u_2}}{1-\rho^2}\right)^k \left(\frac{e}{km}\right)^{k/2} 
\end{align}
We will consider the cases $k>2e m$ and $k<2e m$ separately. When $k>2e m$ we have 
\begin{align}
&\sum_{k=2em}^\infty [1+(-1)^k] \left(\frac{\rho\sqrt{u_1u_2}}{1-\rho^2}\right)^k \frac{\Gamma(\frac{k+1}{2})}{k!\Gamma(\frac{k+m}{2})}\\
&\leq \frac{1}{\Gamma\left(m/2\right)}\sum_{k=2em}^\infty[1+(-1)^k] \left(\frac{\rho}{1-\rho^2}\right)^k \left(\frac{m^2e}{km}\right)^{k/2}\\
&\leq \frac{1}{\Gamma\left(m/2\right)}\sum_{k=2em}^\infty[1+(-1)^k] \left(\frac{\rho}{1-\rho^2}\right)^k \\
&\lesssim \frac{1}{\Gamma\left(m/2\right)} \label{boundRemainderInfiniteSum}
\end{align}
For any $k = Cm$, note that by definition of the Gamma function
\begin{align}
&\iint_{0}^\infty\frac{2^{-(m+1)}(u_1u_2)^{(m-2)/2}e^{-\frac{(u_1 + u_2)}{2(1-\rho^2)}}}{\sqrt{\pi} \Gamma\left(\frac{m}{2}\right)(1-\rho^2)^{m/2}}  \left[1+(-1)^k\right] \left(\frac{\rho \sqrt{u_1u_2}}{1-\rho^2}\right)^k \frac{\Gamma\left(\frac{k+1}{2}\right)}{k! \Gamma\left(\frac{k+m}{2}\right)}\\
& = \iint_{0}^\infty \frac{1}{2^m} \frac{(u_1u_2)^{(m+k)/2 - 1} e^{-\frac{(u_1 + u_2)}{2(1-\rho^2)}} } {\Gamma(m/2)}\left(\frac{\rho}{1-\rho^2}\right)^k \left(\frac{1}{1-\rho^2}\right)^{m/2} \frac{\Gamma\left(\frac{k+1}{2}\right)}{k! \Gamma\left(\frac{k+m}{2}\right)}\\
&\leq  \left(\frac{\rho}{1-\rho^2}\right)^k \left(\frac{1}{1-\rho^2}\right)^{m/2} \frac{\Gamma\left(\frac{k+1}{2}\right)\Gamma\left(\frac{k+m}{2}\right)}{2^{m} k! \Gamma(m/2)}
\end{align}
Noting that
\begin{align}
\frac{\Gamma\left(\frac{k+1}{2}\right)\Gamma\left(\frac{k+m}{2}\right)}{2^{m} k! \Gamma(m/2)} = \frac{\left[\frac{k+1}{2} \left(\frac{k+1}{2}-1\right) \ldots 1 \right] \left[\frac{k+m}{2} (\frac{k+m}{2}-1 )\ldots \right]}{\left[k (k-1) \ldots 1\right] \left[\frac{m}{2} \ldots 1\right]}\frac{1}{2^m}
\end{align}
For any $k = Cm$, with $C\geq 1$ we have 
\begin{align}
\frac{\Gamma\left(\frac{k+1}{2}\right)\Gamma\left(\frac{k+m}{2}\right)}{2^{m} k! \Gamma(m/2)} =  \frac{\left[\frac{Cm+1}{2} \left(\frac{Cm+1}{2}-1\right) \ldots 1 \right] \left[\frac{(C+1)m}{2} (\frac{(C+1)m}{2}-1 )\ldots \right]}{\left[Cm (Cm-1) \ldots 1\right] \left[\frac{m}{2} \ldots 1\right]}\frac{1}{2^m}\leq 2^{-m}\label{largeKjoitChiSquared}
\end{align}
Similarly for $k<m$, we can write 
\begin{align}
&\iint_{0}^\infty\frac{2^{-(m+1)}(u_1u_2)^{(m-2)/2}e^{-\frac{(u_1 + u_2)}{2(1-\rho^2)}}}{\sqrt{\pi} \Gamma\left(\frac{m}{2}\right)(1-\rho^2)^{m/2}}  \left[1+(-1)^k\right] \left(\frac{\rho \sqrt{u_1u_2}}{1-\rho^2}\right)^k \frac{\Gamma\left(\frac{k+1}{2}\right)}{k! \Gamma\left(\frac{k+m}{2}\right)}\\
&\leq \left(\frac{\rho}{1-\rho^2}\right)^k \left(\frac{1}{1-\rho^2}\right)^{m/2} \frac{\Gamma\left(\frac{k+1}{2}\right)\Gamma\left(\frac{k+m}{2}\right)}{2^{m} k! \Gamma(m/2)}
\end{align}
Letting $k = Cm$ for $C\leq 1$, and using Stirling's approximation and neglecting the polynomial terms yields 
\begin{align}
\frac{\Gamma\left(\frac{k+1}{2}\right)\Gamma\left(\frac{k+m}{2}\right)}{k! \Gamma\left(\frac{m}{2}\right)}&\sim \frac{\left(\frac{Cm+1}{2e}\right)^{\frac{Cm+1}{2}} \left(\frac{(C+1)m}{2e}\right)^{\frac{(C+1)m}{2}}}{(Cm)! \left(\frac{m}{2e}\right)^{m/2}} \\
&\sim \frac{\left(\frac{Cm + 1}{2e}\right)^{\frac{Cm+1}{2}} \left(\frac{(C+1)m}{2e}\right)^{\frac{(C+1)m}{2}} }{\left(\frac{Cm}{e}\right)^{Cm} \left(\frac{m}{2e}\right)^{m/2}}\\
&\sim \left(\frac{\frac{Cm+1}{2e}  \frac{(C+1)m}{2e}}{\left(\frac{Cm}{e}\right)^2 \left(\frac{m}{2e}\right)^{1/C}}\right)^{\frac{Cm}{2}} \left(\frac{(C+1)m}{2e}\right)^{m/2}\\
&\sim \left(\frac{C^2 m^2 + Cm^2 + Cm + m}{C^2 m^2 m^{1/C}}\right)^{\frac{Cm}{2}} \left(\frac{(C+1)m}{2e}\right)^{m/2}\\
&\sim \left(\frac{C^2+C}{C^2 m^{1/C}}\right)^{\frac{Cm}{2}} \left(\frac{(C+1)m}{2e}\right)^{m/2}\\
&\sim \left(\left(\frac{C^2+C}{C^2}\right)^{C} \frac{1}{m}\right)^{m/2} \left(\frac{C+1}{2e}m\right)^{m/2} \\
&\sim \left(\left(\frac{C^2+C}{C^2}\right)^C \frac{C+1}{2e}\right)^{m/2} \label{smallKjoitChiSquared}
\end{align}
where the last line follows $\left(1+ \frac{1}{C}\right)^C \leq e$ and $\lim_{C\rightarrow 0}
 \left(1+ \frac{1}{C}\right)^{C} = e$.
Combining~\eqref{largeKjoitChiSquared} and~\eqref{smallKjoitChiSquared} with~\eqref{boundRemainderInfiniteSum}, we can thus write 
\begin{align}
&\iint_{0}^{m-\sqrt{k'm\log(n)}} f_\rho(u_1, u_2) du_1 du_2 \lesssim O(2^{-m}) \\
&+  \iint_{0}^{m-\sqrt{k'm\log(n)}}\frac{2^{-(m+1)}(u_1u_2)^{(m-2)/2}e^{-\frac{(u_1 + u_2)}{2(1-\rho^2)}}}{\sqrt{\pi} \Gamma\left(\frac{m}{2}\right)(1-\rho^2)^{m/2}} \sum_{k=2em}^\infty \left[1+(-1)^k\right] \left(\frac{\rho \sqrt{u_1u_2}}{1-\rho^2}\right)^k \frac{\Gamma\left(\frac{k+1}{2}\right)}{k! \Gamma\left(\frac{k+m}{2}\right)} du_1du_2\\
&\lesssim \iint_{0}^{m-\sqrt{k'm\log(n)}}\frac{2^{-(m+1)}(u_1u_2)^{(m-2)/2}e^{-\frac{(u_1 + u_2)}{2(1-\rho^2)}}}{\sqrt{\pi} \Gamma^2\left(\frac{m}{2}\right)(1-\rho^2)^{m/2}} du_1du_2 +O(2^{-m}).\label{decompositionInfiniteSumPlusO2minusm}
\end{align}
Following this, we let
\begin{align}
 \overline{f}_\rho(u_1, u_2) \equiv \frac{2^{-(m+1)}(u_1u_2)^{(m-2)/2}e^{-\frac{(u_1 + u_2)}{2(1-\rho^2)}}}{\sqrt{\pi} \Gamma^{2}\left(\frac{m}{2}\right)(1-\rho^2)^{m/2}} \label{simplification}
\end{align}
Now using $s= m - \sqrt{mk'\log(n)}$ and letting $\zeta = \left(1 - \frac{1}{2\left(C_\varepsilon + 1\right)}\right)$, 
\begin{align}
\sum_{\rho = 0}^{(\zeta)^2/(k')^2}\frac{P(\hat{R}_m^2(\bs x), \hat{R}^2_m(\hat{\bs x})<s|\rho)}{P(\hat{R}_m^2(\bs x)<s)^2} \leq  1+\sum_{\rho = 1/(k')^2}^{(\zeta)^2/(k')^2}\frac{\displaystyle \iint_{0}^{s} f_{\rho}(u_1, u_2) \; du_1 du_2 }{\left(\displaystyle \int_{0}^{s} f(u)\;  du\right)^2}\label{ratioProbabilitiescorrelated}
\end{align}
Using the upper bound~\eqref{simplification} and applying the change of variable $u' \leftarrow u/(1-\rho^2)$ to the numerator in~\eqref{ratioProbabilitiescorrelated} gives 
\begin{align}
\iint_{0}^s f_{\rho}(u_1,u_2)\; du_1du_2 &\leq (1-\rho^2)^{m/2}\iint_{0}^{s/(1-\rho^2)} \frac{2^{-(m+1)} (u_1u_2)^{(m-2)/2} e^{ - \frac{(u_1 + u_2)}{2}}}{\sqrt{\pi} \Gamma^2(\frac{m}{2})} du_1\; du_2\\
& \leq (1-\rho^2)^m \left(\int_{0}^{s/(1-\rho^2)} f(u)\; du\right)^2
\end{align}
Substituting in~\eqref{ratioProbabilitiescorrelated} we finally obtain
\begin{align}
\sum_{\rho = 0}^{(\zeta)^2/(k')^2}\frac{P(\hat{R}_m^2(\bs x), \hat{R}^2_m(\hat{\bs x})<s|\rho)}{P(\hat{R}_m^2(\bs x)<s)^2} \leq \sum_{\rho = 0}^{(\zeta)^2/(k')^2}\left(\frac{\int_{0}^{s/(1-\rho^2)} f(u)\; du}{\int_{0}^s f(u)\; du}\right)^2\label{ratioPcorrelatedOverP}
\end{align}

To control the ratio~\eqref{ratioPcorrelatedOverP}, first note that if we let $w_k = \sqrt{k'm \log(n)}$,  we have 
\begin{align}
\iint_{0}^{s/(1-\rho^2)} f(u_1, u_2)\; du_1 du_2& = \iint_{0}^{m - w_k/(1-\rho^2) + m\rho^2} f(u_1, u_2)\; du_1 du_2 \\
&\leq \iint_{0}^{m - w_k/(1-\rho^2)}  f(u_1, u_2)\; du_1 du_2\\
& + \iint_{m - w_k/(1-\rho^2)}^{m - w_k/(1-\rho^2) + m\rho^2} f(u_1, u_2)\; du_1 du_2\\
&+2 \int_{0}^{m - w_k/(1-\rho^2)} \int_{m - w_k/(1-\rho^2)}^{m - w_k/(1-\rho^2) + m\rho^2} f(u_1, u_2)\; du_1 du_2
\end{align}
For all the terms for which $\rho^2 m < \sqrt{mk'}$,  the three integrals are of the same order.  We thus start by focusing on the second integral for $\rho^2 m = \omega(\sqrt{k'm})$. Provided that $w_k\lll m$ (which can be achieved for $m$ sufficiently larger than $k'\log(n)$), we can write 
\begin{align}
(1-\rho^2)^{m/2}\int_{m - w_k/(1-\rho^2)}^{m - w_k/(1-\rho^2) + m\rho^2} f(u)\; du  \leq (1-\rho^2)^{m/2 }  \int_{0}^\infty f_0(u_1, u_2)du_1du_2 \leq (1-\rho^2)^m 
\end{align}
Using the non-asymptotic bound from Proposition~\ref{nonasymptoticBoundChiSquaredLower} on the denominator, we can then write
\begin{align}
\frac{\int_{m - w_k(1-\rho^2)}^{m/(1-\rho^2) - w_k/(1-\rho^2)} f(u)\; du}{\int_{0}^s f(u)\; du} \lesssim e^{Ck'}  \left(1-\rho^2\right)^{m/2} 
\end{align}
Noting that for $k'<m$, $\rho^2 m = \omega(\sqrt{k'm})$ which implies $\rho^2=\omega(\sqrt{k'/m})$, we have
\begin{align}
\lim_{\substack{k',m\rightarrow \infty \\ \sqrt{k'm}/m \rightarrow 0}} C k' - m\rho^2 = -\infty
\end{align}
we can write 
\begin{align}
\lim_{\substack{k',m\rightarrow \infty \\ \sqrt{k'm}/m \rightarrow 0}}\sum_{\ell=1}^{k'}(1-\rho^2)^m\frac{\int_{m - w_k(1-\rho^2)}^{m/(1-\rho^2) - w_k/(1-\rho^2)} f(u)\; du}{\int_{0}^s f(u)\; du} \rightarrow 0\label{boundPerturbationRhom}
\end{align}
Now, going back to~\eqref{ratioPcorrelatedOverP}
\begin{align}
&\sum_{\ell=1}^{(1-\delta)k'}\frac{P(\hat{R}_m^2(\bs x), \hat{R}^2_m(\hat{\bs x})<s|\rho)}{P(\hat{R}_m^2(\bs x)<s)^2} \\
&\lesssim 1 + \sum_{\ell=1}^{(1-\delta)k'}\left((1-\rho^2)^{m/2}\frac{\int_{0}^{m - w_k/(1-\rho^2)} f(u)\; du}{\int_{0}^{m - w_k} f(u)\; du}\right)^2\label{firstRatioDecompositionMDP01}\\
& + \sum_{\ell=1}^{(1-\delta)k'}\left((1-\rho^2)^{m/2}\frac{\int_{m - w_k(1-\rho^2)}^{m/(1-\rho^2) - w_k/(1-\rho^2)} f(u)\; du}{\int_{0}^s f(u)\; du} \right)^2\\
& + \sum_{\ell=1}^{(1-\delta)k'}2(1-\rho^2)^{m/2}\frac{\int_{0}^{m - w_k/(1-\rho^2)} \int_{m - w_k(1-\rho^2)}^{m/(1-\rho^2) - w_k/(1-\rho^2)} f(u_1,u_2)\; du_1du_2}{\left(\int_{0}^s f(u)\; du\right)^2} 
\end{align}

To control the first term, we again turn to moderate deviation theory.  
Using the bounds~\eqref{boundMDPupper} and~\eqref{boundMDPlower},  and substituting in~\eqref{firstRatioDecompositionMDP01}, we get 
\begin{align}
&\sum_{\ell=1}^{(1-\delta)k'} (1-\rho^2)^{m}\left(\frac{(k')^2}{n}\right)^{\ell} e^{\frac{1}{2}\left(\frac{\ell^2 + \sigma^2}{(k')^2 + \sigma^2} \right)^2 k'\log(n)}
\end{align}

%
%
Applying the logarithm to each of the terms in the sum, we obtain
\begin{align}
\log\left((1-\rho^2)^{m}\left(\frac{(k')^2}{n}\right)^{\ell} e^{\frac{1}{2}\left(\frac{\ell^2 + \sigma^2}{(k')^2 + \sigma^2} \right)^2 k'\log(n)}\right) \asymp \rho^2 k' -\ell \log(n/(k')^2) - m\rho^2
\end{align}
which is negative provided that $m$ is sufficiently larger than $k'$ and $\ell \neq 0$.  Combining this with~\eqref{largeValuesL}, ~\eqref{decompositionInfiniteSumPlusO2minusm}and~\eqref{boundPerturbationRhom}, we can write for $m,k'$ large enough as well as $k' = o(m)$, 
\begin{align}
\frac{\mathbb{E}\left\{Z_{s}^2\right\}}{\mathbb{E}\left\{Z_{s} \right\}^2}  = \sum_{\ell = 0}^{k'} \frac{{n \choose k'-\ell, k'-\ell, \ell, n-2k'+\ell}}{{n\choose k'}^2} \frac{P\left(E_{\ell}(\bs x, \tilde{\bs x})\right)}{P^2\left(E(\bs  x)\right)}\leq O(k')
\end{align}

Now recalling that for $\|\bs x\|_0 = k'$, we have 
\begin{align}
\mathbb{E}\left[Z_{s}\right] &= {n \choose k} {n-k\choose k'-k}  P\left(\hat{R}^{1/2}_m(\bs x)< \sqrt{k^2 + \sigma_Y^2}\left(1 - \sqrt{\frac{k'\log(n)}{m}}\right)^{1/2}\right)\\
& = {n \choose k} {n-k\choose k'-k}  P\left(\frac{\hat{R}_m(\bs x)}{(k')^2 + \sigma_Y^2}<\left(1-\sqrt{k'\log(n)/m}\right)\right)\\
& \geq {n \choose k} {n-k\choose k'-k} ce^{-C_\varepsilon k'}>0
\end{align}
and using the Paley-Zygmund inequality (see e.g.~\cite{steele2004paley}), we can then derive the lower bound 
\begin{align}
P\left(\varphi_{kk'}(0)\leq  s\right) \geq  P\left(Z_{s, \infty} \geq 1 \right)&\geq P\left(Z_{s, \infty} \geq \mathbb{E}Z_{s, \infty}/2 \right)\\
&\geq \frac{1}{4}\frac{\mathbb{E}\left\{Z_{s, \infty}\right\}^2}{\mathbb{E}\left\{Z_{s, \infty}^2\right\}}\\
&\gtrsim 1/k'\label{lowerBound001}
\end{align}

To conclude, note that the function $\varphi(0)$ can be seen as a Lipschitz function of the $\left\{\bs A_i\right\}_{i=1}^m$. I.e.  for any $Y_i$ independent of the variables $\langle \bs A_i, \bs x\bs x^\intercal\rangle $, we have 
\begin{align}
\max_{\|\bs x\|_0 \leq k'}\sqrt{\hat{R}_m(\bs x)} = \max_{\|\bs x\|_0 \leq k'}\sqrt{\frac{1}{m}\sum_{i=1}^m \left(\langle \bs A_i , \bs x\bs x^\intercal \rangle - Y_i\right)^2}.
\end{align}
Using a reverse triangle inequality, we first get 
\begin{align}
&\left|\sqrt{\frac{1}{m}\sum_{i=1}^m \left(\langle \bs A_i , \bs x\bs x^\intercal \rangle - Y_i\right)^2} - \sqrt{\frac{1}{m}\sum_{i=1}^m \left(\langle \bs A'_i , \bs x\bs x^\intercal \rangle - Y_i\right)^2}\right|\\
& \leq \sqrt{\frac{1}{m}\sum_{i=1}^m \left(\langle \bs A_i , \bs x\bs x^\intercal \rangle - \langle \bs A'_i , \bs x\bs x^\intercal \rangle \right)^2}\\
&\leq \frac{k'}{\sqrt{m}}\sqrt{\sum_{i=1}^m \|\bs A_i - \bs A_i'\|^2}. 
\end{align}
The claim then follows by noting that 
\begin{align}
\left|\max_{\|\bs x\|_0\leq k'} f(\bs A, \bs x) - \max_{\|\bs x\|_0\leq k'} f(\bs A', \bs x)\right| \leq \max_{\|\bs x\|_0\leq k'} \left|f(\bs A, \bs x) - f(\bs A', \bs x)\right|
\end{align}
Using the Gaussian concentration inequality for Lipschitz functions (see e.g.~\cite{fresen2023variations,pisier2006probabilistic}), we can thus write 
\begin{align}
P\left(\left|\varphi_{kk'}(0)  - \mathbb{E}\varphi_{kk'}(0)\right|>t\right)<\exp\left(-\frac{mt^2}{(k')^2}\right)\label{gaussianConcentrationLipschitz01}
\end{align}
Taking $t \gtrsim \frac{k'\log^2(k')}{\sqrt{m}}$ in particular implies 
\begin{align}
P\left(\varphi_{kk'}(0) < \mathbb{E}\varphi_{kk'}(0) - t\right) &\leq \frac{1}{(k')^2} \label{subexponentialConcentrationUpperBound002}
\end{align}
Which, when comparing with~\eqref{lowerBound001} must imply
\begin{align}
\mathbb{E}\varphi_{kk'}(0) &\leq  s +  \frac{k'\log^2(k')}{\sqrt{m}} \\
\end{align}
Substituting this back into~\eqref{gaussianConcentrationLipschitz01} finally gives
\begin{align}
P\left(\varphi_{kk'}(0)> s + \frac{k'\log^2(k')}{\sqrt{m}}\right) &\leq o_k(1)
\end{align}
provided that $k' = o(k^2)$ and for $k$ large enough.  For the deviation $ \frac{k'\log^2(k')}{\sqrt{m}}$ to be smaller than the lower bound on the Gap $\min\left\{\tilde{\Gamma}(\ell_c) - \tilde{\Gamma}(1), \tilde{\Gamma}(\ell_c) - \tilde{\Gamma}(0)\right\}$ derived in section~\ref{analysisFirstMomentsCurveAnalysis}, we need 
\begin{align}
  \frac{k'\log^2(k')}{\sqrt{m}} < \frac{k^2}{k'} \wedge \frac{k'\ell_c}{\sqrt{mk'}}\sqrt{\log(n-k)}
\end{align} 
Taking $\ell_c = \frac{(k')^{3/2}\sqrt{\log(n)}}{\sqrt{m}} = O(k)$ which can be achieved for $(k')^{3/2} \asymp \frac{\sqrt{m}k}{\log^{1/2}(n/k^2)}$ (see the discussion in section~\eqref{overparametrizationSection}), for $n$ large enough, we get the condition
\begin{align}
 \frac{k'\log^2(k')}{\sqrt{m}}< \frac{(k')^2}{m}\log(n) \wedge \frac{k^2}{k'}
\end{align}
Combining this with the discussion of section~\ref{overparametrizationSection}, we get the overparametrization range 
\begin{align}
\sqrt{m} \log^2(k')/\log(n) \leq k' \leq m^{1/3}k^{2/3}{\log^{1/3}(n)} \wedge k m^{1/4} /\log(n)
\end{align}

To conclude,  we relate the bound on the objective of problem~\eqref{Problemtwokdistance2bisPureNoise} and problem~\eqref{Problemtwokdistance2} by following~\cite{gamarnik2017high}, section 4.  Simply note that since $\bs x,\bs x_0 \in \left\{0,1\right\}^n$, we have
\begin{align}
\hat{R}_m(\bs x, \bs x_0) &= \frac{1}{m}\sum_{i=1}^m \left(\langle \bs A_i, \bs x\bs x^\intercal \rangle  - \langle \bs A_i, \bs x_0\bs x_0^\intercal\rangle +\varepsilon_i \right)^2\\
& = \frac{1}{m}\sum_{i=1}^m \left(\langle \tilde{\bs A}_i, \bs x\bs x^\intercal \rangle   +\varepsilon_i' \right)^2
\end{align}
Where $\tilde{\bs A}_i$ is a gaussian i.i.d. matrix such that $\tilde{\bs A}_i[k,k] =0$ for $k\in \supp(\bs x_0)$. The variables $Z_i = \langle \tilde{\bs A}_i, \bs x\bs x^\intercal \rangle $ thus have variance bounded by $(k')^2$ and always sufficiently larger than the variance of $\langle \bs A_i, \bs x_0\bs x_0\rangle$ which is no more than $k^2$ (i.e. $\sigma^2\leq k^2$). Moreover the noise $\varepsilon_i'$ is independent of $\left\{\tilde{\bs A}_i\right\}_{i=1}^m$ and has variance $2(k-\ell)k + k^2\lesssim k^2$.  
Since we only assume that the variance was bounded by $(k')^2$,  for $k'$ sufficiently larger than $\omega(k)\vee \sigma^2$, all the results derived above still hold. In particular, we can thus write
\begin{align}
\tilde{\Gamma}(\ell)< \varphi_{kk'}(\ell) < \tilde{\Gamma}(\ell) + \frac{k'\log^2(k')\log^{1/2}(n)}{\sqrt{m}}
\end{align}

\section{\label{auxilliaryResults}Auxilliary results}

\begin{proposition}[\label{Corollary74Solta}see Corollary 7.4 in~\cite{soltanolkotabi2019structured}]
Let $\bs d\in \mathbb{R}^n$ be a fixed vector with nonzero entries and let $\mathcal{T}\subset \mathbb{B}^n$. Furthermore,  assume 
\begin{align}
\left(\sum_{i=1}^m d_r^2\right) \geq \max\left(20 \|\bs d\|_\infty^2 , \frac{\omega^2(\mathcal{T})}{\delta^2}, \frac{3}{2\delta} - 1\right)
\end{align}
Then for all $\bs u\in \mathcal{T}$, 
\begin{align}
\left|\frac{\sum_{i=1}^m d_i^2 \langle \bs a_i, \bs u\rangle^2 }{\sum_{i=1}^m d_i^2} - \|\bs u\|^2_2\right|\leq \delta
\end{align}
with probability at least $1-6\exp\left(-\delta^2\left(\sum_{i=1}^m d_i^2\right)/1440\right)$

\end{proposition}

\begin{proposition}[\label{subexponentialTailBound}Sub-exponential tail bound, see Proposition 2.9 in~\cite{wainwright2019high}]
Suppose $X$ is subexponential with parameters $(\nu, \alpha)$. Then 
\begin{align}
P\left(X - \mu\geq  t\right) \leq \left\{\begin{array}{ll}
e^{-\frac{t^2}{2\nu^2}} & \text{if $0\leq t \leq \nu^2/\alpha$}\\
e^{-t/2\alpha}& \text{for $t>\nu^2/\alpha$}
\end{array}\right.
\end{align}
\end{proposition}

\begin{proposition}[\label{BernsteinProposition}subexponential Bernstein inequality (matrix version)~\cite{koltchinskii2011nuclear}]
    Let $Z, Z_1, \ldots, Z_n$ be i.i.d. random matrices with dimensions $m_1\times m_2$ that satisfy $\mathbb{E}Z = 0$. Suppose that $U_Z^{\alpha}<\infty$ for some $\alpha\geq 1$. Let $U_Z^{\alpha}$ denote any upper bound on the Orlicz norm of $Z, Z_1, \ldots, Z_n$. I.e.
    \begin{align}
    U_Z^{(\alpha)}& = \inf\left\{u>0\;|\; \mathbb{E} \exp\left(\|Z\|^{\alpha}/u^{\alpha}\leq 2\right)\right\}, \quad \alpha \geq 1     
    \end{align}
    and let 
    \begin{align}
        \sigma_Z = \max\left\{\left\|\frac{1}{m}\sum_{i=1}^m \mathbb{E} Z_iZ_i^\intercal\right\|^{1/2}, \left\|\frac{1}{m}\sum_{i=1}^m \mathbb{E} Z_i^\intercal Z_i\right\|^{1/2}\right\}
    \end{align}
    Then there exists a constant $C>0$ such that, for all $t>0$, with probability at least $1-e^{-t}$, 
    \begin{align}
        \left\|\frac{Z_1+\ldots+ Z_n}{n}\right\|\leq C\max \left\{\sigma_Z \sqrt{\frac{t+\log(m)}{n}}, U_Z^{(\alpha)}\left(\log\left(\frac{U_Z^{\alpha}}{\sigma_Z}\right)^{1/\alpha}\right)\frac{t+\log(m)}{n}\right\}
    \end{align}
    where $m = m_1+m_2$. 
\end{proposition}

\begin{proposition}[\label{scalarBernstein}subexponential Bernstein inequality (scalar version), also see Theorem 2.8.1 in~\cite{vershynin2010introduction}]
   Let $X_1, \ldots, X_N$ be independent zero mean, subexponential variables. Then for every $t\geq 0$ we have 
   \begin{align}
       P\left(\left|\sum_{i=1}^m X_i\right|>t\right)\leq 2\exp \left[-c \min \left(\frac{t^2}{\sum_{i=1}^N \|X_i\|_{\psi_1}^2},\frac{t}{\max_i \|X_i\|_{\psi_1}}\right)\right]
   \end{align}
   where $c$ is an absolute constant. 
\end{proposition}

\begin{lemma}[\label{lemmaLaurentMassart}Lemma 1 in~\cite{laurent2000adaptive}]
Let $(Y_1,\ldots, Y_D)$ be i.i.d. Gaussian variables with mean $0$ and variance $1$. Let $a_1, \ldots, a_D$ be non negative. We set
\begin{align}
\|\bs a\|_\infty = \sup_{i=1, \ldots, D} |a_i|, \quad \|\bs a\|_2^2 = \sum_{i=1}^D a_i^2
\end{align}
Let 
\begin{align}
Z = \sum_{i=1}^D a_i (Y_i^2 - 1)
\end{align}
Then the following inequalities hold for any positive $t$
\begin{align}
P(Z\geq 2\|\bs a\|_2\sqrt{t} + 2\|\bs a\|_\infty t)&\leq \exp(-t)\\
P(Z\leq -2\|\bs a\|_2\sqrt{t})&\leq \exp(-t)
\end{align}
\end{lemma}

We will also need the following extension whose proof can be easily derived from the proof of Lemma~\ref{lemmaLaurentMassart}

\begin{lemma}\label{extensionLemmeMassart}
Let $U_1, \ldots, U_D$, $V_1, \ldots, V_D$ be i.i.d. Gaussian variables with mean $0$ and variance $1$. Let $a_1, \ldots, a_D$ be non negative. Then for $Z = \sum_{i=1}^D a_i U_i V_i$
\begin{align}
P\left(Z \geq 2\|\bs a\|_2\sqrt{t} + 2\|\bs a\|_{\infty} t\right)\leq \exp(-t)
\end{align}
\end{lemma}

\begin{lemma}[Theorem 1 in~\cite{bentkus2003inequality}]\label{lemmaBentkus}
Let $M_n = X_1+\ldots + X_n$ be a martingale with $X_j$ bounded from above by some non-random $\varepsilon\geq 0$ such that $P(X_i \leq \varepsilon)=1$ for $i=1, \ldots, n$.  Assume that the variance $s_i^2 = \mathbb{E}\left\{X_i^2\right\}$ is bounded from above, that is, that $P(s_i^2\leq \sigma^2) = 1$ for some $\sigma$.  Write $b =  \sigma\vee \varepsilon$
Let $I(x) = 1-\Phi(x)$ be the survival function of the standard normal distribution.  Introduce $D(t)=1$ for $t\leq 0$ and $D(t) = \exp(-t^2/2)\wedge cI(t)$.  Let $2\leq c \leq c_0 = 1/I(\sqrt{3})$. Then, for $t \in \mathbb{R}$, we have 
\begin{align}
P\left(\sum_{i=1}^m X_i \geq t\right)\leq \min\left\{\exp\left\{-t^2/2\right\}, cI(x)\right\}, \quad \text{for $t\geq 0$}
\end{align}
\end{lemma}

\begin{lemma}[\label{lemmaPhi}See Lemma 6.2. in~\cite{cai2016optimal}]
Let $\bs x_0$ denote any $k$-sparse vector from $\mathbb{R}^n$ and $\bs a_i\in \mathbb{R}^n$ denote $m$ gaussian i.i.d vectors with $a_i[\ell]\sim \mathcal{N}(0,1)$, then with probability $1-o_m(1)$,  as soon as $m\gtrsim \log(m)\delta^{-1}$, we have 
\begin{align}
(1-\delta)\|\bs x_0\|^2 \leq \frac{1}{m}\sum_{i=1}^m \left|\langle \bs a_i, \bs x_0\rangle\right|^2 \leq (1+\delta)\|\bs x_0\|^2 
\end{align}
\end{lemma}

\begin{theorem}[\label{talagrandTheorem}Talagrand. see Theorem 3 in~\cite{talagrand1989isoperimetry}]
Consider a sequence $(X_i)_{i\geq 1}$ of independent random variables valued in a Banach space $B$. We assume that those variables are integrable with $\mathbb{E}X_i = 0$. For $0\leq \alpha\leq 1$, we have 
\begin{align}
\left\|\sum_{i=1}^m X_i\right\|_{\psi_{\alpha}} \leq K(\alpha) \left(\left\|\sum_{i=1}^m X_i\right\|_1 + \left\|\max_{i\in [m]} |X_i| \right\|_{\psi_{\alpha}}\right)\label{TalagrandInequality01}
\end{align}
where $K(\alpha)$ depends on $\alpha$ only. 
\end{theorem}

\begin{lemma}[\label{LemmaWellner01}See Lemma 2.2.1 in~\cite{wellner2013weak}]
Let $X$ be a random variable with $P(|X|>x)\leq Ke^{-Cx^p}$ for every $x$, for constant $K$ and $C$, and for $p\geq 1$. Then its Orlicz norm satisfies $\|X\|_{\psi_p}\leq ((1+K)/C)^{1/p}$
\end{lemma}

\begin{lemma}[\label{OrliczNormProductSubgaussianVariables}see e.g. Lemma 7 in~\cite{ahmed2014compressive}]
Let $X_1$ and $X_2$ be two subgaussian random variables, i.e. ,$\|X_1\|_{\psi_2}<\infty$ and $\|X_2\|_{\psi_2}<\infty$. Then the product $X_1X_2$ is a subexponential random variable with 
\begin{align}
\|X_1X_2\|_{\psi_1} \leq c\|X_1\|_{\psi_2} \|X_2\|_{\psi_2}
\end{align}
\end{lemma}

\begin{proposition}[\label{candesSoltanolkotabiRegularity}Regularity condition (see~\cite{candes2015phase})]
Let $\bs x_0$ is any solution to the noiseless version of problem~\eqref{generalProblemPR00}.  Let $S\subseteq \supp(\bs x_0)$ with $|S|\leq k$. Then with probability $1-13e^{-\gamma k} - \frac{8}{k^2} - me^{-1.5m}$,  for all $\bs h$ satisfying $\|\bs h\|\leq \|\bs x_0\|/8$, we have 
\begin{align}
\langle \left(\nabla f(\bs x)\right)_{S}, \bs h\rangle \geq \frac{1}{8} \|\bs h\|^2 + \frac{1}{3k + 550} \left\|\left(\nabla f(\bs x)\right)_{S}\right\|^2 
\end{align}
as soon as $m\gtrsim k\log(n)$
\end{proposition}

\begin{definition}\label{definitionSubWeibull}
A random variable $X$ satisfying 
\begin{align}
P\left(|X|\geq t\right)\leq a\exp\left(-b t^{1/\theta}\right), \quad \text{for all $t>0$, for some $\theta, a, b>0$}
\end{align}
is called a sub-Weibull random variable with tail parameter $\theta$. 
\end{definition}

\begin{proposition}[\label{propositionSubWeibull}see Corollary 3.1. in~\cite{vladimirova2020sub}]
Let $X_1, \ldots, X_n$ be identically distributed sub-Weibull (in the sense of Definition~\ref{definitionSubWeibull}) random variables with tail parameter $\theta$. Then, for all $t\geq n K_{\theta}$, we have
\begin{align}
P\left(\left|\sum_{i=1}^m X_i\right|\geq t\right)\leq \exp\left( - \left(\frac{t}{nK_\theta}\right)^{1/\theta}\right)
\end{align}
for some constant $K_\theta$ that depends on $\theta$. 
\end{proposition}


\begin{lemma}[\label{lemmaRIP2}see Lemma B.1 in~\cite{lee2017near}]
Let $F(\bs y)$ and $\bs G(\bs x)$ be defined as in~\eqref{definitionFandG}. Then 
\begin{align}
\left\|\Pi_{\tilde{J}_1}\left(F^\intercal(\bs y) F(\bs \zeta) - \langle \bs \zeta, \bs y\rangle \bs I_{n_1}  \right)\Pi_{\tilde{J}_1}\right\|\leq \delta \|\bs y\| \|\bs \zeta\|, \quad \forall \tilde{J}_1\subset [n_1], |\tilde{J}_1|\leq 3 k
\end{align}
for all $\bs y, \bs \zeta \in \mathbb{C}^{n_2}$ with $\|\bs y\|_0, \|\bs \zeta\|_0 \leq 3k$ and 
\begin{align}
\left\|\Pi_{\tilde{J}_2}\left(G^\intercal(\bs x)G(\bs \xi) - \langle \bs \xi, \bs x\rangle \bs I_{n_2} \right) \Pi_{\tilde{J}_2}\right\|\leq \delta \|\bs x\|\|\bs \xi\|, \quad \forall \tilde{J}_2\subset [n_2], |\tilde{J}_2|\leq 3k
\end{align}
for all $\bs x, \bs \xi\in \mathbb{C}^n$ with $\|\bs x\|_0+ \|\bs \xi\|_0\leq 3k$. 
\end{lemma}

\begin{proposition}[\label{approximateOrthogonality}see Proposition 3.2. in~\cite{needell2009cosamp}]
Suppose that $\bs \Phi$ has restricted isometry constant $\delta_r$. Let $S$ and $T$ be disjoint subsets of indices whose combined cardinality does not exceed $r$. Then 
\begin{align}
\left\|\bs \Phi_S^* \bs \Phi_T\right\|\leq \delta_r
\end{align} 
\end{proposition}

\begin{corollary}[\label{corollaryBlumensath}Corollary 1 in~\cite{blumensath2009iterative}]
Given a noisy observation $\bs x = \bs \Phi \bs y^s +\varepsilon$ where $\bs y^s$ is $s$-sparse, if $\Phi$ has the resttricted isometry property with $\beta_{3s}<1/8$, then, at iteration $k$, $\left\{IHT\right\}$ will recover an approximation $\bs y^k$ satisfying 
\begin{align}
\|\bs y^s - \bs y^k\|_2 \leq 2^{-k} \|\bs y^s\| + 4\|\varepsilon\|
\end{align}
\end{corollary}


\bibliographystyle{ieeetr}
\bibliography{sample}

\end{document}